\documentclass[12pt]{article}

\usepackage{jheppub}
\usepackage{graphicx} 
\usepackage{amsmath}
\usepackage{amsfonts}
\usepackage{hyperref}
\usepackage{comment} 
\usepackage{tikz}
\usetikzlibrary{arrows.meta, positioning, decorations.pathreplacing, calc}

\newcommand\secref[1]{{\S\ref{#1}}}
\newcommand\appref[1]{{Appendix~\ref{#1}}}
\newcommand\figref[1]{{Figure~\ref{#1}}}
\newcommand\tabref[1]{{Table~\ref{#1}}}
\DeclareMathOperator{\tr}{tr} 
\DeclareMathOperator{\Sym}{Sym} 

\title{\ \vspace{1.5in} \\ Analytic Torsion from Chern-Simons theory via the $(2,0)$-theory on Dicyclic Orbifolds of $S^3$}

\author{Emil Albrychiewicz, Andr\'{e}s Franco Valiente, and Ori Ganor}

\affiliation{
Leinweber Institute for Theoretical Physics and Department of Physics, \\
University of California,
Berkeley, CA 94720-7300, USA \\
and
Theoretical Physics Group, Lawrence Berkeley National Laboratory, \\
Berkeley, CA 94720-8162, USA}
\emailAdd{ealbrych@berkeley.edu}
\emailAdd{andresfranco@berkeley.edu}
\emailAdd{ganor@berkeley.edu}

\abstract{
The Witten index of the $(2,0)$-theory compactified on spaces of the form $S^3/\Gamma\times S^2$, with a freely acting group $\Gamma$, and with external string sources implemented via timelike surface operator insertions, is expressed in terms of Ray-Singer torsion of $S^3/\Gamma$ and characters of irreducible representations of $\Gamma$. We compute it explicitly for the Dicyclic groups $\Gamma=\text{Dic}_k$. The torsion and characters are generally irrational numbers, but they nicely combine to an integer index. Alternatively, the Witten index can be computed from Chern-Simons theory on $S^2$, and Ray-Singer torsion on $S^3/\text{Dic}_k$ is thus computable from Chern-Simons theory. The matching of the Witten index calculated by these dual approaches reveals new details about the partition function of the $(2,0)$-theory with surface operators.
}

\begin{document}

\def\be{\begin{equation}}
\def\ee{\end{equation}}
\def\bear{\begin{eqnarray}}
\def\eear{\end{eqnarray}}
\def\nn{\nonumber}
\makeatletter
\def\@fpheader{\relax}
\makeatother
\newcommand\bra[1]{{\left\langle{#1}\right\rvert}} 
\newcommand\ket[1]{{\left\lvert{#1}\right\rangle}} 

\newcommand{\C}{\mathbb{C}}
\newcommand{\R}{\mathbb{R}}
\newcommand{\Q}{\mathbb{Q}}
\newcommand{\Z}{\mathbb{Z}}
\newcommand{\CP}{\mathbb{CP}}
\newcommand{\RP}{\mathbb{RP}}
\newcommand{\Zq}{\mathbb{Z}_q}
\newcommand{\triv}{\mathbf{1}}
\newcommand{\Dic}{\text{Dic}}
\newcommand{\DP}{\mathcal{D}}
\newcommand{\ClassS}{\text{Class}\;\mathcal{S}}
\newcommand{\vol}{\text{vol}}

\def\iRG{{\mathfrak{i}}}
\def\nRG{{\mathfrak{n}}}
\def\NS{\mathcal{N}}

\def\rGen{{\mathbf{r}}} 
\def\sGen{{\mathbf{s}}} 
\def\tGen{{\mathbf{t}}} 

\newcommand\rep[1]{{\bf {{#1}}}}

\def\Hilb{{\mathcal H}}

\newcommand\eqdef{{\stackrel{\tiny\rm def}{=}}}

\newcommand\AbRep[1]{{\mathbf{1}_{{{#1}}}}} 
\newcommand\AbRepCC[1]{{\overline{\mathbf{1}}_{{{#1}}}}} 
\newcommand\SUTwoRep[1]{{\widetilde{\mathbf{2}}_{{{#1}}}}} 
\newcommand\NonAbRep[1]{{\mathbf{2}_{{#1}}}} 

\def\nC{{\mathfrak{n}}}
\def\RSurf{{\widetilde{R}}}

\def\Coq{{C}}

\def\PolyX{{X}}
\def\aRL{{a}}
\def\PolyDeg{{d}}
\def\PolyP{{\mathcal{P}}}
\def\PolyT{{t}}
\def\OpS{{\mathcal{W}}}

\def\ChebU{{U}}
\def\ChebT{{T}}

\def\vF{{\mathfrak{v}}}
\def\sF{{\mathfrak{s}}}
\def\wF{{\mathfrak{w}}}
\def\PFZ{{\mathcal{Z}}}
\newcommand\innerP[2]{{\left\langle{#1},{#2}\right\rangle}}

\def\antip{{\tilde{p}}}
\def\Disc{{\mathbf{D}}}
\def\HO{{\Omega}}
\def\Id{{\mathbf{I}}}

\def\PauliR{{\sigma}} 
\def\PauliL{{\widetilde{\sigma}}} 
\def\GGPauli{{\mathbf{\sigma}}} 

\def\rhoConst{{\mathbf{b}}} 
\def\chiConst{{\mathbf{d}}} 
\def\AreaS{{\varepsilon}} 

\def\mF{{\mathfrak{m}}} 

\def\AreaFormS{{\varepsilon}}

\def\Hchi{{\varphi}} 
\def\Hphi{{\phi}} 

\def\DiracVI{{\Gamma}} 
\def\DiracV{{\gamma}} 

\def\Xrho{{\tilde{\rho}}}
\def\Xpsi{{\tilde{\psi}}}
\def\Xchi{{\tilde{\chi}}}
\def\AdjField{{\Phi}}
\def\HopfRadius{{\mathfrak{b}}}

\def\Va{{\alpha}}
\def\Vb{{\beta}}
\def\Vc{{\mu}}
\def\Wa{{\mathcal W}}

\def\xi{{{i}}}
\def\xj{{{j}}}
\def\xk{{{k}}}
\def\yi{{\mathbf{i}}}
\def\yj{{\mathbf{j}}}
\def\yk{{\mathbf{k}}}
\def\yl{{\mathbf{l}}}
\def\ym{{\mathbf{m}}}
\def\yn{{\mathbf{n}}}

\def\EpChS{{{\upsilon}}}
\def\EpChV{{\zeta}}

\def\psiF{{\mathfrak{P}}}
\def\chiF{{\mathfrak{C}}}
\def\rhoS{{\kappa}}

\def\yCo{{\mathbf{y}}}
\def\xCo{{\mathbf{x}}}

\def\rhoP{{\widetilde{\widetilde{\rho}}}}
\def\psiP{{\widetilde{\widetilde{\psi}}}}
\def\chiP{{\widetilde{\widetilde{\chi}}}}

\def\psiEX{{\overline{\varpi}}}

\def\Xt{{\varpi}}
\def\Xpm{{\varrho}}
\def\ZGen{{\gamma}}
\def\ProjZ{{\mathcal{P}}}

\def\aG{{\mathfrak{a}}}
\def\aGp{{\mathfrak{a}'}}

\def\Yg{{\Upsilon}} %
\def\Pg{{\Pi}}

\def\tHm{{\nu}} 
\def\tHmS{{\mathfrak{m}}} 
\def\tHPg{{\mu}}
\def\tHX{{\lambda}}

\def\Norm{{\mathcal{C}}}

\def\qXrho{{\rho}}
\def\qXchi{{\chi}}

\maketitle

\section{Introduction}
\label{sec:intro}

Our goal is to describe a connection between analytic (Ray-Singer) torsion of certain free orbifolds of $S^3$ and elements of the Verlinde matrix of $\text{SU}(2)$ Chern-Simons theory \cite{Verlinde:1988sn, Witten:1988hf}. The connection is revealed by calculating the dimension of the Hilbert space of a certain topological compactification of the 6d $(2,0)$-theory \cite{Witten:1995zh} in two different ways.
We consider the $(2,0)$-theory associated with the Lie group SU$(2)$ ($A_1$) (i.e., the low energy limit of the worldvolume theory of two coincident M5 branes after the center of mass is decoupled \cite{Strominger:1995ac}) along a Lorentzian manifold $\R\times X_5$, with a topological twist to preserve supersymmetry, where $X_5$ is a generic 5-dimensional smooth Riemannian manifold with a positive scalar curvature. We will later set $X_5$ to a quotient of either $S^5$ or $S^3\times S^2$ by a freely acting finite group $\Gamma$. Compactifying on $S^3/\Gamma \times S^2$ leads to $\NS=2$ supersymmetry in 0+1d, while compactifying on $S^5/\Gamma$ leads to $\NS=1$. Our main result is \eqref{eqn:NeedsToBeInZn}, which is an expression for the Witten index of the theory on $S^3/\Gamma \times S^2$, with $\Gamma$ set to the Dicyclic group $\text{Dic}_k$ and with $\nC$ external string-like sources introduced via surface operators.

There is an ongoing effort to propose a fundamental formulation of the $(2,0)$-theory, and we list several approaches that have been suggested. (This list is by no means comprehensive and extensive reviews of the $(2,0)$-theory can be found in \cite{Moore2012} and \cite{Heckman:2018jxk}.) The matrix model formulation of the $(2,0)$-theory on spaces with an $\R^{1,1}$ factor, based on instanton quantum mechanics, was proposed in \cite{Aharony:1997th}, following an approach closely related to the BFSS matrix model \cite{Banks:1996vh}. Another approach, known as deconstruction, was introduced in \cite{Arkani-Hamed:2001wsh}. The conjecture is that in the large $N$ limit, 4d $\NS=2$ quiver gauge theories on the Higgs branch reproduce the $(2,0)$-theory. A third proposal involves studying the $(2,0)$-theory compactified on an $S^1$ circle as a 5d SYM theory \cite{Douglas:2010iu, Lambert:2010iw,Lambert:2012qy}. This idea has its origins in observations made in \cite{Seiberg:1997ax}. Another approach involves the use of 3-algebras to obtain an action for multiple M5 branes, as discussed in \cite{Lambert:2010wm}. This approach was motivated by the success of BLG models \cite{Bagger:2006sk, Gustavsson:2007vu} for multiple M2 branes. Finally, we mention a twistor approach to non-abelian higher gauge fields introduced in \cite{Saemann:2011nb}.

Independently of the full UV description of the $(2,0)$-theory, the fundamental observation \cite{Seiberg:1997ax} that 5d SYM describes the low-energy limit of the 6d theory compactified on $S^1$, was employed to extract remarkable exact results, using supersymmetric localization \cite{Kallen:2012cs, Kim:2012ava, Kim:2013nva}, and it will be our main tool as well.

In addition to the compactification on $S^1$ to 5d SYM mentioned above, several other low-energy limits of compactifications of the $(2,0)$-theory to lower dimensions on various manifolds, and dualities thereof, have uncovered some information about the nature of the theory. This includes compactification on $T^2$ \cite{Witten:1995zh}, which leads to 4d $\NS=4$ SYM theory, with S-duality emerging from the mapping class group of $T^2$. Compactifications on $T^2$ with R-symmetry twists have been studied in \cite{Cheung:1998te,Cheung:1998wj}. In \cite{Gaiotto:2009we, Gaiotto:2009hg}, the $(2,0)$-theory was compactified on a general Riemann surface $\Sigma$ with an R-symmetry background connection that reduces the supersymmetry to $\NS=2$, leading to 4d $\ClassS$ theories. 
Various properties of the the quantum theory can be determined from the geometry of $\Sigma$. For example, the moduli space of vacua is connected to the moduli space of flat gauge bundles on $\Sigma$, and the mapping class group of $\Sigma$ leads to dualities that generalize the modular transformations in the $T^2$ case.
The AGT correspondence \cite{Alday:2009aq} then allows to express the Nekrasov partition function \cite{Nekrasov:2002qd} of $\ClassS$ theories in terms of 2d Liouville/Toda CFT on $\Sigma$.

Ccompactification on three manifolds \cite{Dimofte:2011ju} results in 3d $\NS=2$ theories referred to as Class $\mathcal{R}$. These theories exhibit a rich duality structure including mirror symmetry. In \cite{Gukov:2015sna} compactification on three dimensional Seifert manifolds was studied and the resulting twisted partition function was connected with Chern-Simons theory on the base of the manifold. The last compactification we would like to mention, in the increasing dimension of compactified space, is the compactification on four manifolds $S^3\times S^1$. This results in a zero area limit of 2d q-deformed SYM theory \cite{Gadde:2009kb}, where the q-deformation introduces a quantum group structure for the gauge group. This theory can actually be used to compute superconformal indices of $\ClassS$ theories \cite{Gadde:2010te, Gadde:2011ik}. 

In this paper, we compute the Witten index \cite{Witten:1982df} of the space of ground states of the 6d $\NS=(2,0)$ theory with SU$(2)$ gauge group, compactified on a few particular examples of $X_5$, with a supersymmetry-preserving R-symmetry twist.
We carry out the calculations by compactifying the time direction into a Euclidean circle $S^1$ and shrinking the circle to small radius. In this limit \cite{Seiberg:1997ax}, the 6d theory is described by an effective 5d SYM with gauge group SU$(2)$, and one finds that the circle's radius is proportional to a square of the gauge coupling constant of the resulting 5d SYM theory. The R-symmetry twist becomes a topological twist. Hence, we can compute the Witten index by calculating the partition function of the 5d SYM theory on $X_5$. As explained in \cite{Aharony:1998qu,Henningson:2010rc,Tachikawa:2013hya}, additional discrete data in the form of an element $\vF\in H^3(X_5,\Z_2)$ needs to be specified in order to uniquely specify what is meant by ``the Hilbert space'' of the $(2,0)$-theory on $X_5$, and in 5d SYM $\vF$ specifies which $\text{SU}(2)/\Z_2$ gauge bundles are to be included in the path integral.

We can probe further by inserting supersymmetric Wilson surface operators of the $(2,0)$-theory \cite{Ganor:1996nf,Maldacena:1998im,Drukker:1999zq} into the $X_5\times S^1$ geometry. This corresponds to asking for the Witten index of a modified Hilbert space of the system with defects along a collection of loops $C_i$ in $X_5$ (the surfaces are $C_i\times S^1$), and the Witten index depends only on the homotopy class of the loops. Insertion of such one-dimensional defects can be thought of as the 6d analog of inserting ``heavy quark'' sources in Yang-Mills theory. The Witten index calculation reduces to the calculation of a correlation function of Wilson loops in topologically twisted 5d SYM on $X_5$. 

Similarly to what was originally found for 4d SYM \cite{Witten:1994cg, Vafa:1994tf}, the partition function of the 5d SYM theory and the correlation functions of SUSY Wilson loops are independent of the coupling constant and the metric because of the topological twist, which in 5d inserts a background gauge field for the $\mathfrak{so}(5)_R$ R-symmetry algebra that matches the geometrical spin connection of $X_5$. 
It was shown in \cite{Bershadsky:1995qy} that the worldvolume theories that describe the dynamics of branes wrapping analytic submanifolds of a Calabi-Yau space naturally become topologically twisted. In our case, we can view the topological twist as the result of wrapping M5-branes on the zero section of the 10d space $T^\star X_5$ (the cotangent space of $X_5$). The spatial slices of the brane will reside along $X_5\subset T^\star X_5$, and the respective topological twist identifies the $\mathfrak{so}(5)$ holonomy algebra with the $\mathfrak{so}(5)_R$ R-symmetry algebra; both are subalgebras of the real superalgebra $\mathfrak{osp}(8^*|2)$ symmetry of the 6d $\NS=(2,0)$ theory.
This particular topological twist of the 5d SYM theory complexifies the gauge field and also results in an additional 5d Chern Simons theory with Grassmann odd sections. This twist was discussed first in \cite{Geyer:2002gy} and then in \cite{Bak:2015hba, Bak:2017jwd}. 

Once the zero modes are addressed, the partition function can be computed perturbatively using Ray-Singer analytic torsion \cite{ray1970reidemeister, RAY1971145}, and whether zero modes exist or not depends on the topology of $X_5$. We will argue that unlike the cases studied in \cite{Bak:2015hba, Bak:2017jwd}, if the quotient group $\Gamma$ is non-abelian, fermionic zero modes are absent around some of the classical solutions that contribute to the partition function, thus leading to a non-vanishing result. In this paper, we will restrict the class of geometries we consider based on those observations. 

In what follows, we choose $X_5$ to be $S^3/\Gamma\times S^2$. Here, $\Gamma$ is one of the finite subgroups of SU$(2)$ of A or D type, which are constructed by the ADE McKay correspondence \cite{mckay1981graphs}. For A-type, $\Gamma$ is a cyclic group $\mathbb{Z}_k$, and $S^3/\Gamma$ is a {\it Lens space} denoted by $L(k,1)$.\footnote{In general one can add integers $l_1,\dots,l_n$ coprime with $k$, where $n$ is the complex dimension of orbifold but in this paper we set them all to 1.} For D-type, $\Gamma$ is a dicyclic group $\text{Dic}_k$ of order $4k$, and the resulting orbifold is referred to as a {\it prism manifold.} We will also briefly discuss $S^5/\Gamma$ with $\Gamma$ a cyclic group $\Z_k$ of order $k$. We review these geometries and possible extensions in detail in the first part of this paper. 

We are able to compute the partition function on these geometries by applying supersymmetric localization methods. The localization equations result in BPS solutions that are given by flat connections. The $X_5$ spaces that we consider will have nontrivial fundamental groups and consequently, there will be nontrivial flat connections. After the path integral localization, we will need to compute functional determinants of twisted Hodge Laplacians on orbifolds. We will find that we can take a computationally simpler approach by using the Ray-Singer torsion which allows us to extract the answer by simply studying the group theoretic properties of representations of $\Gamma$. 

The main result of this paper is formula \eqref{eqn:NeedsToBeInZn} for the Witten index of  space of the $(2,0)$-theory on (non-singular) orbifolds $S^3/\Gamma\times S^2$ with external sources along a collection of $1$-cycles, and a confirmation that the index is indeed an integer. Since the Witten index is calculated indirectly as a correlation function of Wilson loops in an effective 5d SYM, it is not immediately obvious that it would come out as an integer, but we will verify it.

Moreover, we can compute the Witten index in a completely different way by exploiting the fact that $S^3/\text{Dic}_k$ is a circle fibration. This leads to a linear combination of dimensions of Hilbert spaces of Chern-Simons theory with quarks on $S^2$, and various gauge groups and levels. We will argue that the agreement of the two methods explains the simplicity of the Ray-Singer torsion of $S^3/\text{Dic}_k$, and expresses it in terms of Verlinde matrix \cite{Verlinde:1988sn,Witten:1988hf} elements.

The outline of this paper is as follows. In \secref{sec:geometry}, we provide and review basic facts about the geometrical and topological properties of the manifolds we consider in this paper (for the role of $X_5)$. In \secref{sec:physics}, we describe the topological twist and the emergence of fermionic Chern-Simons theory (FCS). We also review how the partition function of FCS can be computed using Ray-Singer torsion. In \secref{sec:ray}, we provide an explicit calculation of torsion for $\Gamma=\Z_k$ and $\Gamma=\text{Dic}_k$: in the latter case we derive a novel closed form expression. We use this calculation to compute and list the Witten index in \secref{sec:HilbNoS} for a few values of $k$. We then introduce timelike surface operators into the Witten index and prove that the resulting correlation function is always an integer. In \secref{sec:DerCS}, we discuss an alternative way of computing the Witten index using a combination of Chern Simons theories on $\R\times S^2$ with gauge groups SU$(2)$ at level $(k-2)$ and $\text{U}(1)$ at level $2k$. They appear once one compactifies on the orbifold $S^3/\text{Dic}_k$. There we explain the match between Verlinde's formula for the dimension of the Hilbert space of Chern-Simons theory (with quarks) \cite{Verlinde:1988sn} and the Witten index computed using the Ray-Singer torsion in previous sections. In section \secref{sec:SU2Z2}, we discuss the case where the gauge group of the analytic torsion side is SU$(2)/\Z_2$ and the gauge group of the dual Chern-Simons theory is SU$(2)$. We conclude our results in section \secref{sec:conclusions},  with a discussion of possible extensions. 
\appref{app:LevelShift} explains the level shift from $k$ to $(k-2)$ with a detailed analysis of the fermionic modes.

\section{Geometry of Spherical Orbifolds}
\label{sec:geometry}

We begin with a discussion of the various manifolds $X_5$ on which we will study the (twisted) $A_1$ 6d $\NS=(2,0)$ theory, and we review some important facts about their homology groups. Recall that after compactification on a small $S^1$, we get a twisted version of 5d Super Yang-Mills theory with gauge group $G=\text{SU}(2)$ on the manifold $X_5$, which is a topological field theory that we will review below. To compute the partition function for a topological field theory that arises in this compactification, we will need to add the contributions coming from nontrivial flat connections of the SU$(2)$ gauge field. To get a nontrivial result, we will need nontrivial flat connections, which require nontrivial elements of the fundamental group $\pi_1(X_5)$. Therefore, we want to pick manifolds $X_5$ that have a nontrivial fundamental group $\pi_1(X_5)$. We will need to calculate the Ray-Singer torsion of the manifolds and to obtain non-vanishing results we will need to address the issue of fermionic zero modes. Ideal candidates for $X_5$ are quotients of spheres or products of spheres by a finite subgroup $\Gamma$ of the spin group.

In the rest of this section we present some choices of $X_5$ in the form $S^5/\Gamma$ and $S^3/\Gamma\times S^2$, where $\Gamma$ is chosen to act freely. For $S^5/\Gamma$, there are several good choices of $\Gamma$ \cite{wolf1972spaces} so that the resulting orbifold is smooth, but in this paper we only perform the calculations for $\Gamma=\Z_k$, leaving the non-abelian $\Gamma$'s for a future work \cite{WIP}.

For $S^3/\Gamma\times S^2$, we take $\Gamma$ to be a subgroup of the spin group $\text{Spin}(3)\times\text{Spin}(4)\simeq\text{SU}(2)_1\times \text{SU}(2)_2\times\text{SU}(2)_3$, where we added subscripts $1,2,3$ to distinguish the three SU$(2)$ factors, and where SU$(2)_1$ acts on $S^2$, and $\text{SU}(2)_2\times\text{SU}(2)_3$ acts on $S^3$. In general, one can consider two choices that act freely: either to embed $\Gamma$ as a (finite) subgroup in SU$(2)_2$ [or, equivalently, in SU$(2)_3$], or to embed as a diagonal subgroup in $\text{SU}(2)_1\times\text{SU}(2)_2$, thus acting nontrivially both on $S^2$ and $S^3$. Collectively we can denote both choices as $(S^3\times S^2)/\Gamma$. In either case, $\Gamma$ will be associated with an A, D, or E type Dynkin diagram, by the McKay correspondence. In the present paper, we focus on the first case where $\Gamma$ acts nontrivially on $S^3$. We perform the calculations for two choices of $\Gamma$: $\Z_k$ and Dic$_k$, which by the McKay correspondence have Dynkin diagrams of $A_{k-1}$ and $D_{k+2}$ type. Both groups act freely, leading to a smooth $X_5$.

Both $S^5$ and $S^3\times S^2$ are simply connected spaces, so the fundamental groups of the respective obifolds are:
\begin{align}
    \pi_1(S^5/\Gamma)=\Gamma,\qquad\pi_1((S^3\times S^2)/\Gamma)=\Gamma,
\end{align}
where (slightly abusing notation) we denoted by the same $\Gamma$ two different groups.
The holonomies of flat connections on these spaces are classified by group homomorphisms $\Gamma\rightarrow G$, up to conjugation, and for $G=\text{SU}(2)$, flat connections are given by two dimensional representations (not necessarily irreducible) of $\Gamma$, with unit determinant. 

We start with $\Gamma=\Z_k\rightarrow \text{SU}(2)$, and consider the holonomies $g_m$ (of the flat connection) that satisfy the conditions $g_m^k=1$ and $\det{g_m}=1$. Up to conjugation, we can set
\begin{align}
\label{eqn:GDefSU2}
    g_m=\begin{pmatrix}
        e^{2\pi i m/k} & 0\\
        0 & e^{-2\pi i m/k}
    \end{pmatrix},
\end{align}
with $m$ in the range
\begin{align}
\label{eqn:mRange}
    m=0,1,2,\dots,\left[\frac{k}{2}\right],
\end{align}
where by the bracket $\left[k/2\right]$, we mean the largest integer not bigger than $k/2$. We keep only this range of $m$, because replacing $m\rightarrow -m$ in \eqref{eqn:GDefSU2} gives a (Weyl) conjugate matrix.

Next, we consider the dicyclic group $\Gamma=\text{Dic}_k\rightarrow \text{SU}(2)$, with $k>1$\footnote{Dic${}_1$ is isomorphic to $\Z_4$, which is already covered by the previous case.}. The dicyclic group is a non-abelian group that is an extension of the cyclic group of order $2k$. It is generated by two elements $(r,s)$ that geometrically act as rotation and inversion respectively, and they satisfy
\begin{equation}
\label{eqn:GenDic}
    r^{2k}=1, \quad s^2=r^k, \quad s^{-1}rs=r^{-1}.
\end{equation}
We construct irreducible representations $\text{Dic}_k\rightarrow \text{SU}(2)$ (see also previous work \cite{GanorJu} and the follow-ups \cite{Ju:2023umb, Albrychiewicz:2024fkr} that used a similar construction), which satisfy these relations for each holonomy and are given by two-dimensional matrices with unit determinant. From \eqref{eqn:GenDic} one observes that there are abelian and non-abelian representations, which makes the manifold quotient by the dicyclic group more interesting than a quotient by a cyclic group only. We analyze these representations separately for even and odd $k$.
\begin{table}[ht]
\centering
$\begin{array}{|c|c|c|}
    \hline
    \text{Representation reference} \ & g(r) & g(s)  \\
    \hline 
    \AbRep{1} & 1 & 1 \\ 
    \hline 
    \AbRep{2} & 1 & -1 \\ 
    \hline  
    \AbRep{3} & -1 & 1 \\  
    \hline  
    \AbRep{4} & -1 & -1 \\ \hline
\end{array}$
\caption{Inequivalent abelian irreducible representations of Dic$_k$ for even $k$. $g(r)$ and $g(s)$ are the phases that represent the generators $r,s$.}
\label{tab:IRRepsDicEven}
\end{table}

For any even $k$, there are four one-dimensional representations, listed in \tabref{tab:IRRepsDicEven} together with the notation we use to refer to them ($\AbRep{1},\dots,\AbRep{4}$). Associated to them are four inequivalent embeddings of $\text{Dic}_k$ in SU$(2)$:
\be\label{eqn:RepsDicEven}
\SUTwoRep{j} =\AbRep{j}\oplus\AbRep{j},\qquad
j=1,2,3,4.
\ee
For non-abelian irreducible representations, still for even $k$, we take
\begin{align} 
    \label{eqn:NonARepsDic}
    g_m(r)=\begin{pmatrix}
        e^{\pi i m/k} & 0 \\
        0 & e^{-\pi i m/k}
    \end{pmatrix}, \quad
    g_m(s)=\begin{pmatrix}
        0 & 1 \\
        (-1)^m & 0 \\
    \end{pmatrix},
\end{align}
where $m=1,2,\dots,k-1$, and we refer to the representation associated with a particular $m$ as $\NonAbRep{m}$. For odd $m$, both matrices have determinant $1$, and the representation forms an embedding of Dic${}_k$ in SU$(2)$.

For odd $k$, the abelian irreducible representations are listed in \tabref{tab:AbRepsDicOdd}, and the inequivalent embeddings $\text{Dic}_k\rightarrow\text{SU}(2)$ are listed in \tabref{tab:RepsDicOdd}; there are $3$ of them.
As for the non-abelian representations, the only difference from \eqref{eqn:NonARepsDic} is, that for odd $k$, $m$ now runs up to $k-2$.

We will also need the decomposition of the symmetric part of tensor products of two-dimensional representations into irreducible ones:
\be\label{eqn:symtensorp}
\Sym^2\NonAbRep{m} = 
\left(\NonAbRep{m}\otimes\NonAbRep{m}\right)_S\cong
\left\{
\begin{array}{ll}
\NonAbRep{2m}\oplus\AbRep{1}, & \; \text{for $m\in 2\Z$;} \\
\NonAbRep{2m}\oplus\AbRep{2}, & \; \text{for $m\in 2\Z+1$.}
\end{array}
\right.
\ee
For odd $1\le m\le k-1$ the representation $\NonAbRep{m}$ is (pseudoreal) complex, while for even $m$ a reality condition can be imposed. Thus, imposing a reality condition on \eqref{eqn:symtensorp} we obtain the real $3$-dimensional representation of $\Dic_k$ that corresponds to mapping an element $g\in\text{Dic}_k$ to an $\text{SU}(2)$ matrix by the representation $\NonAbRep{m}$ and then mapping the $\text{SU}(2)$ matrix to $SO(3)$ via the adjoint representation of $\text{SU}(2)$.

\begin{table}[ht]
\centering
$\begin{array}{|c|c|c|}
    \hline
    \text{Representation name} \ & g(r) & g(s)  \\
    \hline 
    \AbRep{1} &  1 &  1 \\ 
    \hline 
    \AbRep{2}  & 1  & -1 \\ 
    \hline  
    \AbRep{3}  & -1 & i \\  
    \hline
    \AbRepCC{3}  & -1 & -i \\  
    \hline
\end{array}$
\caption{Inequivalent abelian representations of Dic$_k$ for odd $k$. 
Here $g(r)$ and $g(s)$ are the phases that represent the generators $r,s$.
}
\label{tab:AbRepsDicOdd}
\end{table}

\begin{table}[ht]
\centering
$\begin{array}{|c|c|}
    \hline
    \text{Representation name} \ & \text{Decomposition into irreducible representations}  \\
    \hline
    \SUTwoRep{1} & \AbRep{1}\oplus\AbRep{1}  \\ 
    \hline
    \SUTwoRep{2}  & \AbRep{2}\oplus\AbRep{2} \\ 
    \hline
    \SUTwoRep{3}  & \AbRep{3}\oplus\AbRepCC{3} \\  
    \hline  
\end{array}$
\caption{
For odd $k$, embeddings of Dic$_k$ in SU$(2)$ are constructed as direct sums of pairs of abelian representations from \tabref{tab:AbRepsDicOdd}.
}
\label{tab:RepsDicOdd}
\end{table}
This concludes the discussion of irreducible representations of $\Gamma$ and we proceed to a discussion of topological properties of quotiented spheres. 

We begin with the case of $S^5$ which has a spin isometry group Spin$(6)$, which is isomorphic to SU$(4)$. The various possible choices of $\Gamma$ that act freely on $S^5$, and hence lead to a smooth quotient space $S^5/\Gamma$ are listed in \cite{Wolf:1974}. The abelian $\Gamma$'s are $\Z_n$ and $\Z_n\times\Z_m$. There is also a non-abelian group whose generators are subject to relations $A^m=B^{3n}=\mathbb{I}$ and $BAB^{-1}=A^r$, and in terms of SO$(6)$ matrices they are 
\begin{align}
\label{eqn:BobbyGroup}
    A = \begin{pmatrix}
    R\left(\frac{2\pi}{m}\right) & 0 & 0 \\
    0 & R\left(\frac{2\pi r}{m}\right) & 0 \\
    0 & 0 & R\left(\frac{2\pi r^2}{m}\right) 
    \end{pmatrix}
    \; \text{and} \;
    B=\begin{pmatrix}
        0 & \mathbb{I} & 0 \\
        0 & 0 & \mathbb{I} \\
        R\left(\frac{2\pi s}{n}\right) & 0 & 0
    \end{pmatrix},
\end{align}
where $\mathbb{I}$ is a 2-dimensional identity matrix and $R$ is a rotation matrix:
\begin{align}
    R(\theta)=\begin{pmatrix}
        \cos{\theta} & \sin{\theta} \\
        -\sin{\theta} & \cos{\theta}
    \end{pmatrix}.
\end{align}
The integers $m, n, r, s$ are subject to following conditions: $(3(r-1)n,m)=(s,n)=1$ and $r\not\equiv r^3\equiv 1 \mod m$. 
This subgroup was utilized in \cite{Acharya:1998db} to construct conical singularities for D$3$-brane probes and discussed recently in \cite{Tsimpis:2022orc}. In this paper we will only analyze the $\Gamma=\Z_k$ case of $S^5/\Gamma$ (equivalent to a Lens space), and the non-abelian cases will be discussed in \cite{WIP}. The integer homology groups of $S^5/\Z_k$ differ from those of $S^5$ only in dimension $1$ and $H_1(S^5/\Z_k,\Z)=\Z_k$.

As we discussed at the top of this section, for $S^3\times S^2$ we have two choices for the embedding of $\Gamma$ in the spin isometry group: either let $\Gamma$ act trivially on $S^2$, or embed $\Gamma$ diagonally in two of the three $SU(2)$ factors of $\text{Spin}(3)\times\text{Spin}(4)$. The second option makes $(S^2\times S^3)/\Gamma$ a nontrivial $S^2$ fiber bundle over $S^3/\Gamma$, but the homology groups $H_\star((S^2\times S^3)/\Gamma,\Z)$ are the same for both cases (listed in \tabref{tab:HomologyXY}), and easily derived from those of $S^3/\Gamma$ (listed in \tabref{tab:HomologyX}) by the K\"unneth formula (or a spectral sequence in the second case) and Poincar\'e duality. 
\begin{table}[ht]
    \centering
    \[
    \begin{array}{|c|c|c|}
   \hline
         X & S^3/\Z_k & S^3/\text{Dic}_k \\ \hline
         H_0(X) & \Z & \Z \\ \hline
         H_1(X) & \Z_k & 
         \begin{cases} 
         \Z_2\times \Z_2, & k\; \text{even} \\
         \Z_4, & k\; \text{odd}
         \end{cases} \\ \hline   
         H_2(X) & 0 & 0 \\ \hline
         H_3(X) & \Z & \Z \\ \hline
    \end{array}
    \]
    \caption{Homology groups for $S^3/\Z_k$ and $S^3/\text{Dic}_k$ orbifolds.}
    \label{tab:HomologyX}
\end{table}
\begin{table}[ht]
    \centering
    \[
    \begin{array}{|c|c|c|c|}
   \hline
         X\times Y & S^3/\Z_k\times S^2 & S^3/\text{Dic}_k\times S^2 & \text{Betti number} \\ \hline
         H_0(X\times Y) & \Z & \Z & b_0=1 \\ \hline
         H_1(X\times Y) & \Z_k & 
         \begin{cases} 
         \Z_2\times \Z_2, & k\; \text{even} \\
         \Z_4, & k\; \text{odd}
         \end{cases} & b_1=0 \\ \hline   
         H_2(X\times Y) & \Z & \Z & b_2=1 \\ \hline
         H_3(X\times Y) & \Z\oplus\Z_k & \begin{cases} 
         (\Z_2\times \Z_2)\oplus\Z, & k\; \text{even} \\
         \Z_4\oplus\Z, & k\; \text{odd}
         \end{cases}  & b_3=1 \\ \hline
         H_4(X\times Y) & 0 & 0 & b_4=0 \\ \hline
         H_5(X\times Y) & \Z & \Z & b_5=1 \\ \hline
    \end{array}
    \]
    \caption{Homology groups and Betti numbers for product manifolds $S^3/\Z_k\times S^2$ and $S^3/\text{Dic}_k\times S^2$. Betti numbers are condensed into one column since they are equal for both manifolds.}
    \label{tab:HomologyXY}
\end{table}

The $(2,0)$-theory on $X_5$ defines several inequivalent Hilbert spaces, and as explained in \cite{Tachikawa:2013hya}, in order to specify a concrete Hilbert space one has to provide (in addition to a choice of spin structure) an element $\vF$ in the cohomology group $H^3(X_5,\Z_2)$. Here, the cohomology coefficient ring $\Z_2$ is identified with the center of the gauge group SU$(2)$. Compactified on $S^1$, the theory reduces to 5d SYM on $X_5$ with gauge group $\text{SO}(3)\simeq\text{SU}(2)/\Z_2$, and $\vF$ encodes the phases with which to weight, in a path integral, the contributions of topologically inequivalent $\text{SO}(3)$ gauge bundles. Specifically, the obstruction to lifting an $\text{SO}(3)$ gauge bundle on $X_5$ to an $\text{SU}(2)$ bundle is given by a Stiefel-Whitney class $\wF\in H^2(X_5,\Z_2)$, which captures the `t Hooft magnetic flux \cite{tHooft:1977nqb}. Since $X_5$ is assumed orientable, we have a cohomology pairing $\innerP{\vF}{\wF}\in\Z_2$, and the partition function of the theory labeled by $\vF$ is given, in terms of the path integrals $\PFZ_\wF$ that are restricted to gauge bundles with Stiefel-Whitney class $\wF$, by \cite{Tachikawa:2013hya} (see also \cite{Witten:1996hc,Freed:2006yc,Henningson:2010rc} for prior discussions related to the non uniqueness of  space of the $(2,0)$ theory):
\be\label{eqn:wFZ}
\sum_{\wF\in H^2(X_5,\Z_2)} (-1)^{\innerP{\vF}{\wF}}\PFZ_\wF.
\ee
In particular, $\vF=0$ means the 5d SYM theory has an $\text{SO}(3)$ gauge group.

We wish to insert surface operators along the time direction, which reduce to Wilson loops along $1$-cycles of $X_5$. If the path integral includes all gauge bundles with $\wF\neq 0$ then we cannot uniquely define Wilson loops in the fundamental representation of $\text{SU}(2)$, or in any representation where the center $\Z_2\subset\text{SU}(2)$ is nontrivial. We can, however, define Wilson loops in the adjoint representation. For $g\in\text{SU}(2)$ we have a relation between the character in the adjoint representation and the character in the fundamental representation of $g^2$:
$$
\tr_{\scriptstyle{adj}} g=
1+\tr_{\scriptstyle{fund}} (g^2),
$$
and so Wilson loops along loops that wind an even number of times are always well-defined. For $S^3/\text{Dic}_k\times S^2$, Wilson loops in the fundamental representation around cycles associated with $r^{2j}$, for example, are always well defined.

Now consider principal $\text{SO}(3)$ bundles on $S^3/\Gamma$. We can pull back the bundle under the projection $S^3\rightarrow S^3/\Gamma$ to obtain a principal $\text{SO}(3)$ bundle on $S^3$, but such bundles are topologically characterized by $\pi_2(\text{SO}(3))$, which is trivial, so the $\text{SO}(3)$ bundle can be trivialized on $S^3$. Fix a point $p\in S^3$, and take $\gamma\in\Gamma$. Because the fibration is trivial, we can then compare the fibers at $p$ and at $\gamma(p)$, and thus associate an element of $\text{SO}(3)$ with $\gamma$. Therefore, $\text{SO}(3)$ gauge bundles on $S^3/\Gamma$ are associated with group homomorphisms $\Gamma\rightarrow\text{SO}(3)\cong\text{SU}(2)/\Z_2$, and the obstruction to lifting an $\text{SO}(3)$ bundle to an $\text{SU}(2)$ bundle is given by the obstruction to lifting such a homomorphism to a homomorphism $\Gamma\rightarrow\text{SU}(2)$. For $\Gamma=\text{Dic}_k$, such a homomorphism is given by a solution to \eqref{eqn:GenDic} with the generators $r,s$ represented by $2\times 2$ matrices of unit determinant. A homomorphism $\Gamma\rightarrow\text{SO}(3)\cong\text{SU}(2)/\Z_2$, on the other hand, is represented by $2\times 2$ matrix solutions with unit determinant to the modified relations
\begin{equation}
\label{eqn:GenDicVF}
    r^{2k}=(-1)^{k\sF_1}, \quad s^2=(-1)^{\sF_2}r^k, \quad s^{-1}rs=(-1)^{\sF_1}r^{-1}.
\end{equation}
where $\sF_1,\sF_2\in\Z_2$, and the sign of $r^{2k}$ in the leftmost condition is required for consistency with the other two relations.
Thus, the pair $(\sF_1,\sF_2)\in\Z_2^2$ characterizes the obstruction to lifting an $\text{SO}(3)$ bundle over $S^3/\text{Dic}_k$ to an $\text{SU}(2)$ bundle. But since $(\pm r,\pm s)$, with all $4$ sign combinations, describe identical $\text{SO}(3)$ bundles, we can set $\sF_2=0$ for odd $k$, but we must keep it for even $k$.
Therefore, for even $k$, there must be an injective map $H^2(S^3/\text{Dic}_k,\Z_2)\rightarrow\Z_2^2$ that maps $\vF$ to $(\sF_1,\sF_2)$, and for odd $k$ we must have an injective map $H^2(S^3/\text{Dic}_k,\Z_2)\rightarrow\Z_2$ that maps $\vF$ to $\sF_1$. Indeed, from \tabref{tab:HomologyX} and the universal coefficient theorem we have an isomorphism
\begin{align}
\label{eqn:H2Z2}
H^2(S^3/\text{Dic}_k,\Z_2)\cong
\text{Ext}_\Z^1(H_1(S^3/\text{Dic}_k,\Z),\Z_2)=
\left\{\begin{array}{ll}
\Z_2^2 & \text{for even $k$;} \\
\Z_2 & \text{for odd $k$.} \\
\end{array}
\right.
\end{align}
We now analyze the solutions of \eqref{eqn:GenDicVF}.

If $\sF_2=\sF_1=0$ we get the solutions that can be lifted to $\text{SU}(2)$ and discussed above. Now set $\sF_2=0$ and $\sF_1=1$. Then, we diagonalize $g(r)$ and from $s^{-1}rs=-r^{-1}$ we see that either the eigenvalues of $g(r)$ are $\pm i$, or their product is $-1$, in which case $g(r)\notin\text{SU}(2)$. Thus, up to conjugation, there is one solution [in $\text{SU}(2)/\Z_2$]:
\be\label{eqn:sF1}
g(r) = \begin{pmatrix} i & 0 \\ 0 & -i \\ \end{pmatrix},\qquad
g(s) = \pm\begin{pmatrix} e^{\frac{k\pi i}{4}} & 0 \\ 0 & e^{-\frac{k\pi i}{4}} \\ \end{pmatrix}\qquad
(\sF_1=1).
\ee

For even $k$, we can also set $\sF_2=1$, and then we get additional solutions with $\sF_1=0$ that are similar to \eqref{eqn:NonARepsDic} and are given by
\begin{align} 
    \label{eqn:NonARepsDicSO}
    g_m(r)=\begin{pmatrix}
        e^{\pi i m/k} & 0 \\
        0 & e^{-\pi i m/k}
    \end{pmatrix}, \quad
    g_m(s)=\begin{pmatrix}
        0 & 1 \\
        -1 & 0 
    \end{pmatrix},
\end{align}
where $m=0,2,4,\dots,k-1$ this time.

To conclude, we list the third cohomology group with $\Z_2$ coefficients, which labels different ways to define  space of the $(2,0)$-theory on $S^3/\text{Dic}_k\times S^2$:
\begin{align}
    H^3(S^3/\text{Dic}_k\times S^2, \Z_2)=\begin{cases}
        \Z_2\oplus \Z_2 \oplus \Z_2= (\Z_2)^3 \quad k \; \text{even,} \\
        \Z_2\oplus \Z_2=(\Z_2)^2 \quad k \; \text{odd.}
    \end{cases}
\end{align}
The $S^2$ introduced another $\Z_2$ factor, but we note that a nontrivial Stiefel-Whitney class on $S^2$ does not support a flat connection, so the component of $\vF$ along $S^2$ has no effect on the topological partition function on $S^3/\text{Dic}_k\times S^2$.

\section{Fermionic Chern-Simons and the Ray-Singer Torsion}
\label{sec:physics}

Our setup, as discussed in \secref{sec:intro}, is the $(2,0)$-theory compactified on one of the 5d manifolds of \secref{sec:geometry}, denoted by $X_5$, with a (partial) topological twist that is equivalent to identifying the $\mathfrak{so}(5)_H$ holonomy algebra with the $\mathfrak{so}(5)_R$ R-symmetry algebra. After compactification of the remaining (``time'') direction on $S^1$, we obtain a twisted 5d SYM theory, and we now present its field contents. In our notation, $\mu=1,\dots,5$ labels local coordinates on $X_5$.
Due to the topological twist, the scalar fields $\Phi^I$ (which describe the perpendicular fluctuations of the M$5$-branes in the M-theory realization, with $I=1,\dots,5$ labeling R-symmetry vector representation indices before the topological twist) get replaced by a 1-form $\Phi=\Phi_{\mu}dx^\mu$. The spinor fields $\psi^a_\alpha$ (with $\alpha=1,\dots,4$ a spinor index and $a=1,\dots,4$ labeling a spinor representation of the R-symmetry group), which transform in the representation $(4,4)$ of $\mathfrak{so}(5)_H\oplus\mathfrak{so}(5)_R$, decompose under the following rule:
\begin{align}
4\otimes 4 = 1\oplus 5\oplus 10,
\end{align}
and, consequently, decompose into a scalar $\rho$, a 1-form $\psi=\psi_\mu dx^\mu$, and a 2-form $\chi=\frac{1}{2}\chi_{\mu\nu}dx^{\mu}\wedge dx^{\nu}$, all Grassmann odd. The fields before and after performing the topological twist are summarized in \tabref{tab:FieldTwist}.
\begin{table}[ht]
\centering
\begin{tabular}{|l|c|c|c|}
\hline
Field & Before twist & $\mathfrak{so}(5)_H\oplus\mathfrak{so}(5)_R$ & After twist \\
\hline
Gauge field & $A_\mu$ & $(5,1)$ & $A_\mu$ \\
\hline
Gluino & $\psi_\alpha^a$ & $(4,4)$ & $\rho,\psi_\mu,\chi_{\mu\nu}$
\\
\hline
Scalars & $\Phi^I$ & $(1,5)$ & $\Phi_\mu$ \\
\hline
\end{tabular}
\caption{Field contents before and after the topological twist.}
\label{tab:FieldTwist}
\end{table} 

The fermionic kinetic term of the resulting 5d SYM theory can be identified with a 5d ``fermionic Chern Simons theory'' (FCS). In the terminology of \cite{Elliott:2020ecf} (which presents a classification of topological twists of super Yang Mills theories up to 10 dimensions), the twist we are utilizing is a ``$\NS=2$ topological B twist with a supercharge of rank 4''. Unlike ordinary 3d Chern Simons theories which are constructed from connection 1-forms that are Grassmann even, the 5d FCS theories are constructed from Grassmann odd (fermionic) two-forms $\Psi$ with the action
\begin{align}
    \int_{X_5} \Psi\wedge d\Psi.
\end{align}
Note that this action would vanish if $\Psi$ were Grassmann even since the integrand would then be a total derivative. A comprehensive summary of FCS theories can be found in \cite{Bak:2017jwd}.

We now present the Lagrangian and the topological BRST symmetry of the twisted 5d SYM with gauge group SU$(2)$, following \cite{Bak:2017jwd}. Our fields are those of \tabref{tab:FieldTwist}, with all fields transforming in the adjoint representation of SU$(2)$. Using a point-wise linear combination of the adjoint fields, we can define twisted connections as
\begin{align}
    \nonumber
    \mathcal{A}&=A-\Phi, \\ \nonumber
    \overline{\mathcal{A}}&=A+\Phi, 
\end{align}
and the corresponding twisted covariant derivatives as
\begin{align}
    \nonumber
    \DP_\mu &= \nabla_\mu-i\mathcal{A}_{\mu}, \\ \nonumber
    \overline{\DP}_\mu &= \nabla_\mu-i\overline{\mathcal{A}}_{\mu}.
\end{align}
We can now write the 5d FCS action for the Grassmann odd two-form $\chi$  as
\begin{align}
\label{eqn:5dCSLag}
    \mathcal{S}_{CS}= -\frac{i}{2}\int_{X_5}  \operatorname{tr} \chi \wedge \DP \chi = -\frac{i}{8}\int_{X_5} \epsilon^{\mu\nu\rho\sigma\tau}\tr\chi_{\mu\nu}\DP_{\rho}\chi_{\sigma\tau} \hspace{0.1cm} d^5x.
\end{align}
The trace is induced by the Killing form on the Lie algebra $\mathfrak{su}(2)$. The Lagrangian density $\mathcal{L}_{CS}$ can be derived by applying the topological twist to the kinetic term of the Dirac fermion of 5d SYM, and expressing the fermion in terms of $\chi$. As a cohomological TQFT \cite{Witten:1990bs}, the theory obtained by twisting 5d SYM has a topological BRST symmetry generated by the component of the SUSY transformation that becomes a scalar after the twist. The topological BRST differential $\delta$ acts on the fields as
\begin{align*}
    \delta \rho &= -\epsilon \DP^{\dagger} \Phi = -\epsilon\DP^\mu\Phi_\mu, \\
    \delta \psi &= 0, \\
    \delta \chi&=\frac{1}{2}\epsilon\mathcal{F}, \\
    \delta \mathcal{A} &= 0,\\
    \delta \overline{\mathcal{A}} &= -2i\epsilon\psi,
\end{align*}
where the field strength $\mathcal{F}$ is defined with respect to the new gauge field via $\mathcal{F}=d \mathcal{A}-i\mathcal{A} \wedge \mathcal{A}$, and $\epsilon$ is a Grassmann odd parameter. $\mathcal{L}_0$ is invariant under $\delta$, up to a total derivative, and the full Lagrangian of the twisted 5d SYM theory can be obtained by adding to $\mathcal{L}_{CS}$ a $\delta$-exact term \cite{Bak:2017jwd},
\be\label{eqn:BRSTLag}
{\mathcal L}={\mathcal L}_{CS}+\delta V,
\ee
where
\begin{align}
    V = \tr\left(\frac{1}{2}\overline{\mathcal{F}}_{\mu\nu}\chi^{\mu\nu}-\frac{1}{2}\DP^\mu\Phi_\mu\rho-\Phi^{\mu}\DP_{\mu}\rho\right).
\end{align}
On flat space, ${\mathcal L}$ is the same as the Lagrangian of 5d SYM after a field redefinition is applied to the twisted fields $\chi$, $\psi$, $\rho$, $\Phi$, $\mathcal{A}$, $\overline{\mathcal{A}}$. The Lagrangian ${\mathcal L}$ possesses a SU$(2)$ gauge symmetry associated to the gauge field ${\mathcal A}$, which is gauge fixed by a standard BRST gauge fixing procedure (introducing additional Faddeev-Popov ghosts) separate from the topological BRST symmetry above.

The localization equations for this cohomological field theory can be found by looking for fixed points of the BRST differential. The solutions to the localization equations are known as BPS solutions. The BPS equations are then given by
\begin{align*}
    \mathcal{F}=F-i\DP\Phi+i\Phi\wedge\Phi=0, \quad \DP^\dagger\Phi=0, \quad \DP\Phi=0,
\end{align*}
where $\mathcal{F}$ is the Euclideanized SU$(2)$ field strength. As we discussed in \secref{sec:geometry}, all the five-manifolds $X_5$ that we will consider in this paper have $H^1(X_5,\Z)=0$, and consequently the covariant constancy condition $\mathcal{D} \Phi=0$ forces $\Phi$ to vanish since there are no nontrivial globally constant 1-forms on a manifold with trivial $H^1(X_5,\Z)$. In summary, the path integral localizes onto the moduli space of flat connections ($F=0$) with $\Phi$ vanishing.

For the manifolds $X_5$ that we are considering, the fundamental group is nontrivial and given by $\pi_1(X_5)=\Gamma$. Therefore, there exist a finite number of nontrivial flat connections, which are classified by group homomorphisms $\Gamma \rightarrow \text{SU}(2)$ [or $\Gamma \rightarrow \text{SO}(3)$], up to conjugation. One can compute the partition function by expanding around the saddle points, which correspond to these flat connections. This approach in the context of 3d Chern-Simons theory on a general manifold is developed in \cite{Witten:1988hf, Freed:1991wd, Rozansky:1993zx, Adams:1995np, Adams:1996hi, Marino:2002fk} and reviewed in \cite{Marino:2011nm}. In 3d, the renormalization of loop corrections introduces a metric-dependent phase in the partition function \cite{Witten:1988hf}, but this phase is absent in 5d, and the theory remains exactly topological \cite{Bak:2017jwd}.

For each connection, we can define a covariant derivative
\begin{align}
\label{eqn:ExtDerTwist}
    d_A=d-i[A,\cdot].
\end{align}
The saddle point conditions imply the flatness of the connection $F=d_A^2=0$. We can canonically define an inner product on the space of adjoint-valued $p$-forms on $X_5$ using the Hodge $\star$ operator induced by the metric and the non-degenerate bilinear Killing form on the Lie algebra associated to the gauge group:
\begin{align}
    \langle a, b\rangle = \int_{X_5} \tr(a\wedge\star b).
\end{align}
With respect to this inner product, one can define the adjoint operator of \eqref{eqn:ExtDerTwist} acting on a $p$-form, known as the codifferential:
\begin{align}
    \delta_A=(-1)^{5(1+p)+1}\star d_A\star.
\end{align}
The Hodge Laplacian, acting on $p$-forms, is defined in the standard way as
\begin{align}
\label{eqn:HodgeLap}
    \Delta^p_A=\delta_A d_A+d_A\delta_A.
\end{align}
After localization, we find that all leftover quadratic terms in the action can be expressed in terms of Hodge Laplacians, and the corresponding partition function factorizes as a product of their functional determinants. The 1-loop approximation becomes exact in this limit. 

We will now provide more details on the localization procedure \cite{Witten:1991mk}. The details for fermionic Chern-Simons theory is described in \cite{Bak:2017jwd} based on results of \cite{Kapustin:2009kz}. Formally speaking, we deform the action by adding a BRST exact term, $\mathcal{L}=\mathcal{L}_{CS}+t\delta V$. Consequently, the path integrand remains in the same BRST cohomology class. We take an asymptotic limit $t \rightarrow \infty$ which localizes the path integral on the BPS solutions $\delta V=0$.

As for the fluctuations around the BPS solutions, one can show that in the limit $t\rightarrow\infty$,  the action has to be expanded only up to quadratic order, and the one-loop determinant reduces to a product of functional determinants of Hodge Laplacians on $0$, $1$ and $2$-forms. Note that not all quadratic terms in the action are proportional to $t$, since the $\chi\wedge d\chi$ coupling in $\mathcal{L}_{CS}$ [see 
\eqref{eqn:5dCSLag}] is independent of $t$. However, we can follow an argument presented in \cite{Bak:2017jwd} to show that \eqref{eqn:5dCSLag} does not pose a problem. We first rescale all field fluctuations (both fermionic and bosonic) by $\frac{1}{\sqrt{t}}$, which results in the rescaling of $\mathcal{L}_{CS}$ by $\frac{1}{t}$. This rescaling also affects the measure of the path integral, and the factor by which the measure is rescaled can be computed by zeta-function regularization. If we expand the $p$-form fields in eigenmodes of the twisted (in the background field $A$) Hodge Laplacian $\Delta^p_A$, and consequently write the path integral measure as a product over the eigenmodes while removing the zero modes from the product, then the measure gets multiplied by a product of factors of $\sqrt{t}$ for each mode, which is regularized to 
$$
 \left( \frac{1}{\sqrt{t}} \right)^{\pm\zeta_{\Delta^p_A}(0)},
$$
where the sign $(\pm)$ is $+$ for bosons and $-$ for fermions, and
$$
 \zeta_{\Delta^p_A}(s)=\sum\limits_n\phantom{}^{'}\frac{1}{\lambda_n^s}
$$
is the Minakshisundaram–Pleijel $\zeta$-function defined as a sum of $(-s)$-powers of the nonzero eigenvalues $\lambda_n$ of $\Delta^p_A$. Next, we use a result \cite{rosenberg1997laplacian} that relates the value at zero of this regularized $\zeta$-function to the $p^{\text{th}}$ Betti number, which computes the rank of the twisted $p$-th cohomology group (i.e., the cohomology group obtained by replacing the exterior derivative of de-Rham cohomology with the covariant derivative with respect to the flat $\mathfrak{su}(2)$ connection $A$):
\begin{align}
\label{eqn:LocZeroModes}
-\zeta_{\Delta^p_A}(0)=b_p(A).
\end{align}
Our theory has fields that are 0-, 1-, and 2-forms, so we are only interested in $p=0,1,2$, and for the manifolds $X_5$ that we consider $b_1(A)=0$.\footnote{This statement is independent of the flat connection $A$, because with a gauge transformation we can convert a zero mode of the twisted Hodge Laplacian on $X_5$ to a zero mode of the standard Laplacian (with $A=0$) on a simply connected cover of $X_5$, and all the manifolds that we consider have finite simply connected covers.} Thus, the total measure gets multiplied by
\begin{align}
\label{eqn:MeasureTfactor}
t^{-(b_0(A)+b_2(A))/2}    
\end{align}
under the $t$-rescaling. Shortly, in \eqref{eqn:TfactorDet} we will see that this factor cancels the effect of $t$ rescaling on the determinant. Assuming that, we find that the path integral over the fluctuations over the bosonic and fermionic fields reduces to
\begin{align}
    \label{eqn:PartitionFunction}
    Z&=Z_\Phi Z_{YM} Z_F, \\ \label{eqn:PartitionFunctionBos}
    Z_\Phi &=\frac{1}{\det^{\frac{1}{2}}\Delta_A^1}, \\ \label{eqn:PartitionFunctionYM}
    Z_{YM} &=\frac{\det{\Delta_A^0}}{\det^{\frac{1}{2}}\Delta_A^1}, \\ \label{eqn:PartitionFunctionFer}
    Z_F&=\text{det}^\frac{1}{4}\Delta_A^2 \text{det}^\frac{1}{4}\Delta_A^1 \text{det}^\frac{1}{4}\Delta_A^0.
\end{align}
Here we mention another technical detail \cite{Bak:2017jwd} involved in deriving the factor $\det^\frac{1}{4}\Delta_A^2$ in \eqref{eqn:PartitionFunctionFer}. After rescaling by $t$, the relevant differential operator is not quite $\Delta_A^2$, but rather $\frac{1}{t^2}d_A^\dagger d_A + d_A d_A^\dagger$. This is because of the extra factor of $1/t$ in $\mathcal{L}_{CS}$. However, an argument of \cite{Bak:2017jwd}, which uses the fact that $(d_A^\dagger d_A)(d_A d_A^\dagger)=0$ and $(d_A d_A^\dagger)(d_A^\dagger d_A)=0$, shows that 
\begin{align}
\label{eqn:TfactorDet}
\det\left(\frac{1}{t^2}d_A^\dagger d_A + d_A d_A^\dagger\right)
=\left(\frac{1}{t^2}\right)^{-b_0(A)-b_2(A)}\det\left(d_A^\dagger d_A + d_A d_A^\dagger\right).
\end{align}
Since this determinant contributes to the partition function with the power $(1/4)$ [see \eqref{eqn:PartitionFunctionFer}], the prefactor is exactly canceled by the factor \eqref{eqn:MeasureTfactor} we obtained by removing zero modes from the measure. Indeed, in \cite{Bak:2017jwd}, the zero modes were removed by hand from the measure, but in this paper we must keep the zero modes since the gauge field $A$ is dynamical, and as we will see, different flat connections might have different numbers of zero modes. Thus, in this paper flat connections with (fermionic) zero modes will not contribute to the overall path integral.

We now address the issue of zero modes in more detail, focusing on two types relevant to our analysis. First, we consider the zero modes of the Faddeev-Popov ghosts $c,\bar{c}$ (modes whose existence would naively imply that the path integral is zero). Those modes are in the kernel of $\Delta_A^0 = d_A^\dagger d_A$, as appears in \eqref{eqn:PartitionFunctionYM}, and arise due to incomplete gauge fixing. For flat connections on $X_5$, the full group of gauge transformations $\mathcal{G}$ contains a stabilizer subgroup $\mathcal{H}_A$, referred to as the ``isotropy group'', which preserves the flat connection under gauge transformations. The Lie algebra of $\mathcal{H}_A$ is determined by the kernel of $d_A$ acting on 0-forms (in the adjoint representation), i.e., $\text{Lie}(\mathcal{H}_A) = \text{Ker}\,d_A$. This correspondence reflects the fact that $d_A$ generates the infinitesimal action of the gauge group on the flat connection.
These Faddeev-Popov zero modes can be addressed using the procedure outlined in \cite{Rozansky:1993zx, Adams:1996hi} and reviewed in \cite{Marino:2011nm}. The prescription involves splitting the integration over $c,\bar{c}$ modes into contributions from the isotropy group $\mathcal{H}_A$ and the orthogonal complement of $\text{Ker}\,d_A$. Using the nilpotency of $d_A$, the space of adjoint-valued 0-forms $\Omega^0(X_5, \mathfrak{su}(2))$ can be Hodge decomposed as \begin{align} 
\Omega^0(X_5, \mathfrak{su}(2)) = \text{Ker}\,d_A \oplus \text{Im} \,d_A^\dagger = \text{Lie}(\mathcal{H}_A) \oplus (\text{Ker}\,d_A)^\perp, \end{align}
where the second equality uses the fact that $(\text{Ker}\,d_A)^\perp = \text{Im}\,d_A^\dagger$. The integral over $\text{Lie}(\mathcal{H}_A)$, where the integrand is independent of these modes, yields the volume of the isotropy group $\vol(\mathcal{H}_A)$. The remaining integral is then restricted to the orthogonal component $(\text{Ker}\,d_A)^\perp$, effectively removing the zero modes from the determinant. In summary, we replace $\det(\Delta_A^0)$ with: 
\begin{align} 
\label{eqn:IsoGroup}
\frac{1}{\vol(\mathcal{H}_A)}\det\left(d_A^\dagger d_A \big|_{(\text{Ker}\, d_A)^\perp}\right). \end{align}
We still need to compute $\vol(\mathcal{H}_A)$, which is the volume of the subgroup of constant gauge transformations $\mathcal{H}_A$ that preserve $A$. $\mathcal{H}_A$ is isomorphic to a subgroup $H_A$ of the gauge group SU$(2)$, but its volume should be calculated using a different metric from that of SU$(2)$, and as a result it depends on the volume of the manifold $\vol(X_5)$. To see this, we will study how $\vol(\mathcal{H}_A)$ varies under rescaling of $\vol(X_5)$. $\vol(\mathcal{H}_A)$ is calculated using a metric on the tangent space of $\mathcal{H}_A$, which is the space of covariantly constant adjoint-valued $0$-forms $\varepsilon$ on $X_5$, whose norm is given by
$$
\int_{X_5}\|\varepsilon(x)\|^2 \sqrt{g} d^5x,
$$
where $\|\varepsilon\|^2$ is the local Killing form on $\mathfrak{su}(2)$, and $\sqrt{g}d^5x$ is the volume form on $X_5$.
Thus, the norm of $\varepsilon$ scales like a factor of $\vol(X_5)^{1/2}$. Consequently, $\vol(\mathcal{H}_A)$ can be expressed as 
\begin{align} 
\label{eqn:IsoVol}
\vol(\mathcal{H}_A) = \vol(X_5)^{\dim H_A / 2} \vol(H_A). 
\end{align} 
For the trivial flat connection, the isotropy group is SU$(2)$, as all gauge transformations preserve the connection. For nontrivial abelian flat connections, the isotropy group is the maximal abelian subgroup of SU$(2)$, which corresponds to the maximal torus U$(1)$. For non-abelian flat connections, such as those associated with the dihedral group Dic$_k$, the isotropy group reduces to the center of SU$(2)$, which is $\mathbb{Z}_2$.

The second type of zero modes we must address are the zero modes of physical (non-ghost) fields. We have seen near \eqref{eqn:MeasureTfactor}, from the study of the nontrivial cohomology groups of the universal cover of $X_5$, that only the $0$-form and $2$-form fields can have zero modes. Specifically, in the case of abelian flat connections we have fermionic zero modes associated with the 0-form $\rho$ and the 2-form $\chi$. The standard approach to address such zero modes is to insert BRST-invariant operators into the path integral, which absorb the zero modes, leading to a nonzero correlation function. However, the BRST transformations of $\rho$ and $\chi$ do not allow for simple BRST invariant operators that can soak up these modes. More importantly, there is no clear physical intuition about what types of more complicated BRST-invariant observables might be constructed to achieve this. 
Without any operator insertions, the partition function vanishes identically for all abelian flat connections, since those have fermionic zero modes. For example, when $X_5 = S^3 / \mathbb{Z}_k \times S^2$, all flat connections are abelian, as discussed in \secref{sec:geometry}. It is not hard to check that loop effects cannot remove these zero modes. In the case where $X_5 = S^3 / \text{Dic}_k \times S^2$, however, non-abelian flat connections do exist. For those non-abelian flat connections, the zero modes are lifted because there are no covariantly constant 0-forms or 2-forms. For example, if a covariantly constant $\rho$ satisfying $d_A\rho=0$ existed, its value $\rho(p)$ at some point $p\in X_5$ would have to commute with the holonomy group of $A$, which is isomorphic to a Dic$_k$ subgroup of SU$(2)$, and no elements of the Lie algebra $\mathfrak{su}(2)$ are invariant under the entire group. With the zero modes removed, we can proceed to compute the partition function.

As first observed by Schwarz in \cite{Schwarz:1978cn} and later extended by Witten in \cite{Witten:1991we}, the partition function for certain topological gauge theories can be computed using the Ray-Singer analytic torsion. This method applies to cases where the gauge fields define flat connections $A$ on a bundle $E$ over the manifold $X_5$. In particular, the Ray-Singer torsion is related to the volume of the moduli space of flat connections. The torsion $\tau(X_5, A)$ on a smooth $d$-dimensional manifold $M$ is a topological invariant that is defined as
\begin{align} 
\label{eqn:RSTDef}
\tau(M, A) = \prod_{p=0}^d (\text{det}' \Delta_A^p)^{-(-1)^p \frac{p}{2}}, 
\end{align} 
where $\det'$ indicates that zero modes of the twisted Laplacians were removed.  However, this approach in computing the partition function holds only when flat connections are both irreducible and isolated. In our case, the non-abelian flat connections are indeed isolated and irreducible, but the abelian flat connections are isolated but not irreducible, since their isotropy group is not in the center of the gauge group as discussed above. In the latter case, even though it does not contribute to the partition function because of the zero modes discussed above, in order to compute correlation functions (with the zero modes absorbed) one has to introduce a metric-dependent factor, leading to the modified torsion \cite{Friedmann:2002ty}
\begin{align} 
\label{eqn:TorsioMet} 
\tilde{\tau}(X_5, A) = (\text{vol}(X_5))^{\dim H_A} \tau(X_5, A), 
\end{align} 
where $\tau(X_5, A)$ depends solely on the topology of $X_5$.

Simplifying the expression for the partition function \eqref{eqn:PartitionFunction}, we find: 
\begin{align}
\label{eqn:PartX5}
Z(X_5) = \Norm\sum_A \frac{1}{\vol(\mathcal{H}_A)} \frac{\det'^{\frac{5}{4}} \Delta_A^0 \det'^{\frac{1}{4}} \Delta_A^2}{\det'^{\frac{3}{4}} \Delta_A^1},
\end{align} 
where $\Norm$ is a normalization constant which usually is fixed by imposing that trivial flat connection gives partition function equal to unity. (See also \cite{Murthy:2025ioh} for a recent discussion of normalization of the measure in the context of localization.) Since for us trivial flat connection does not contribute, we fix this normalization to be $\Norm=1$ using a different argument that will be presented in \secref{sec:HilbNoS}. In our case, only non-abelian flat connections $A$ are of interest, $\dim H_A = 0$ in \eqref{eqn:IsoVol} (since $H_A$ is finite), and there are no zero modes, so $\det'$ can be replaced with $\det$.
Comparing \eqref{eqn:PartX5} and \eqref{eqn:RSTDef}, we finally arrive at:
\begin{align} 
Z(X_5) = \Norm \sum_{\text{nonab $A$}}\frac{\sqrt{\tau(X_5, A)}}{|H_A|}, 
\end{align}
where the (finite) sum is taken over all allowed (non-abelian) inequivalent flat connections. The metric dependence, in the torsion definition for non-irreducible flat connections, cancels with the corresponding factor from \eqref{eqn:IsoVol}, leaving the partition function expressed in terms of topological invariants.

When $X_5$ is a product manifold with a rank 1 gauge bundle defined over it, this result can be further simplified. If $M$ and $N$ are oriented compact manifolds without boundary, and $N$ is simply connected, the torsion of the product manifold satisfies \cite{RAY1971145, Blau:2022odi}: 
\begin{align} 
\tau(M \times N,A) = \tau(M,A)^{\chi(N)}\tau(N)^{\chi(M)}, 
\end{align} where $\chi$ is the Euler characteristic. Applying this to the case $X_5 = S^3 / \Gamma \times S^2$, we find that the partition function evaluates to 
\begin{align} 
\label{eqn:PartTorsion}
Z(S^3 / \Gamma \times S^2) = \Norm \sum_{\text{nonab $A$}} \frac{\tau(S^3 / \Gamma, A)}{|H_A|}, 
\end{align} 
since $\chi(S^2) = 2$ and the Euler characteristic of the odd-dimensional manifold $S^3/\Gamma $ vanishes. This represents a significant simplification: to study the partition function in five dimensions, we only need to compute the torsion of the three-dimensional manifold $S^3 / \Gamma$ over all admissible choices of flat SU$(2)$ connections $A$. In particular, we only need to study the finite number of flat connections which are classified by group homomorphisms $\Gamma \rightarrow $ SU$(2)$ [or $\Gamma \rightarrow \text{SO}(3)$], up to conjugation. In the next section, we compute these torsions explicitly.

\section{Ray-Singer Torsion for Spherical Orbifolds}
\label{sec:ray}

In \cite{cheeger1979analytic} and \cite{muller1978analytic}, it was shown that for a closed manifold, the Ray-Singer analytic torsion is equal to Reidemeister torsion, the latter being computed using functional determinants of boundary operators on a co-chain complex. The Reidemeister torsion for the quotient space $S^{2n+1}/\Gamma$ was computed in \cite{ray1970reidemeister} and later used for topological field theories on quotient spaces in \cite{Nash:1992sf}. 

We start with a formula for Reidemeister torsion on a $d$-dimensional manifold $M$ given in \cite{cheeger1979analytic}:
\begin{align}
\label{eqn:CheegerForm}
    \log\tau(M, \mathbf{1})=\sum\limits_{p=0}^d(-1)^{p+1}\left(\log V_p (M)+\log O_p \right),
\end{align}
where $\mathbf{1}$ indicates that we compute the torsion for a trivial representation of the fundamental group, or equivalently a trivial flat connection. In this formula $O_p$ denotes
\begin{align}
    O_p=|\text{Tor}\,H^p(M,\mathbb{Z})|,
\end{align}
which is the cardinality of the torsion part of the integer cohomology group. The $\log V_p$ factors vanish if there are no zero modes, which is the only case we are interested in.  They are related to the volume dependent part that appears in \eqref{eqn:TorsioMet}, and we kept them for completeness. We will now present their values for a specific example, which is a quotient space $M=S^d/\Gamma$ (for any dimension $d$, but for us only $d=3$ is relevant). The only non zero Betti numbers are $b_0=b_d=1$ and we find $V_0=\frac{1}{\sqrt{\vol(M)}}$ and $V_d=\sqrt{\vol(M)}$ (which are the norms of a 0-form and a harmonic top form).

\subsection{\texorpdfstring{Ray-Singer Torsion for $\Gamma=\Z_k$}{Ray-Singer Torsion for Gamma=Zk}}
\label{subsec:RSTGammaZk}
We begin by presenting results for the Ray-Singer torsion for $\Gamma=\Z_k$ since these results will later be used to compute the torsion that we actually need for $\Gamma=\text{Dic}_k$. Choosing $\Gamma=\Z_k$, for an odd dimensional Lens space $O_p=k$ for $0<p<d$, since
\begin{align}
    \text{Tor}\,H^p(M,\mathbb{Z})=\mathbb{Z}_k, \; \text{for} \; 0<p<d.
\end{align}
$O_p$ vanishes for an even dimensional Lens space, where we used the fact that the general Lens space cohomology is
\begin{align}
\label{eqn:cohomLensSpace}
H^i(S^{2n+1}/\mathbb{Z}_k,\mathbb{Z})=\left\{
\begin{array}{ll}
\mathbb{Z} & \text{for $i=0,2n+1$,} \\
\mathbb{Z}_k & \text{for $i=2,4,\dots,2n$,} \\
0 & \text{for $i=1,3,\dots,2n-1$}. \\
\end{array}
\right.
\end{align}
Therefore, we find from \eqref{eqn:CheegerForm}:
\begin{align}
\label{eqn:TorsionSZk}
    \log\tau(S^{2n+1}/\mathbb{Z}_k, \mathbf{1})=\log{\left\lbrack V(S^{2n+1}/\mathbb{Z}_k)\right\rbrack}-n \log{k}.
\end{align}
In general, one can use representation theory to argue \cite{Friedmann:2002ty} that
\begin{align}
    \label{eqn:TorsionIrrep}
    \log{\tau}(S^3)=\log{\tau}(S^3/\Gamma, \triv)+ \sum_{\rho\notin \triv}\dim(\rho)\log\tau(S^3/\Gamma,\rho),
\end{align}
where $\rho$ counts inequivalent nontrivial representations of $\Gamma$.\footnote{The factor $\dim(\rho)$ does not appear in \cite{Friedmann:2002ty}, because only one-dimensional representations are needed there, for $\Gamma=\mathbb{Z}_k$.} The factor $\log\tau(S^3/\Gamma,\rho)$ has been originally computed by Ray in \cite{ray1970reidemeister}. Ray gave a formula for the torsion for any orbifold $S^3/\Gamma$ as long as $\Gamma$ acts freely. An improved approach was used in \cite{Dowker:2009in}, where an effective tau function is used to simplify Ray's derivation. To describe it, first define the representation $\rho_m:\, \Z_k \rightarrow \text{U}(1)$ where the generator $\mu$ of $\Z_k$ is represented by
\be\label{eqn:rhom}
\rho_m(\mu)=
e^{\frac{2\pi i m}{k}},
\ee
with representations $\rho_m$ labeled by an integer $m$. In the range $m=1,\dots,k-1$, we get nontrivial inequivalent representations of $\Z_k$. Using the approach of \cite{Dowker:2009in}, the formula for the analytic torsion on a Lens space with a nontrivial representation 
 is then
\begin{align}
\label{eqn:DowkerTorsion}
    \log\tau(S^3/\Z_k,\rho_m) = \frac{2}{k}\sum\limits_{p=0}^{k-1}\sum\limits_{l=1}^{k-1}e^{2\pi i m p/k}e^{-2\pi i l p/k}\log \left\lbrack 2\sin\left(\frac{\pi l}{k}\right)\right\rbrack.
\end{align}
Here, $p$ labels conjugacy classes of $\Gamma$, and $l$ runs over nontrivial ones. For a complete derivation we refer to \cite{Dowker:2009in}. Summing over $p$ fixes $l\equiv m \mod k$, leaving the formula for torsion for $\rho_m$:
\begin{align}
\label{eqn:LensSpaceTorRep}
    \log\tau(S^3/\Z_k,\rho_m)=2\log \left\lbrack 2\sin\left(\frac{\pi m}{k}\right)\right\rbrack.
\end{align}
As a quick check on formula \eqref{eqn:TorsionIrrep}, we can sum over all nontrivial representations to get 
\begin{align}
    \sum_{\rho\notin \triv}\log\tau(S^3/\Z_k,\rho)=2\sum\limits_{m=1}^{k-1}\log\left\lbrack 2\sin\left(\frac{\pi m}{k}\right)\right\rbrack=2\log\prod\limits_{m=1}^{k-1}\left\lbrack2\sin\left(\frac{\pi m}{k}\right)\right\rbrack=2\log{k}.
\end{align}
This can be seen to match \eqref{eqn:TorsionIrrep}, using \eqref{eqn:TorsionSZk} and $\tau(S^3)=2\pi^2$, which we get by setting $k=1$ in \eqref{eqn:TorsionSZk}, for example.

All our fields are in the adjoint representation $\lambda$ of the gauge group SU$(2)$, and we need to decompose $\lambda$ into irreducible (one-dimensional) representations of the subgroup $\Gamma=\Z_k$:
\begin{align}
    \lambda \rightarrow \lambda_m=\rho_{2m}\oplus\triv\oplus\rho_{-2m},
\end{align}
where $\rho_{m}$ (for any integer $m$) is the representation defined in \eqref{eqn:rhom}.
Therefore, for the $\Z_k$ representation $\lambda_m$ we get
\begin{align}
\label{eqn:RSTS3Zk}
    \tau(S^3/\Z_k,\lambda_m)=\tau(S^3/\Z_k,\rho_{2m})\tau(S^3/\Z_k,\triv)\tau(S^3/\Z_k,\rho_{-2m})=\frac{16}{k}\sin^4\left(\frac{2\pi m}{k}\right),
\end{align}
where $m$ is in range given by \eqref{eqn:mRange} 
[and we omitted an overall constant volume term $V(S^3)$]. Equation \eqref{eqn:RSTS3Zk} was derived in \cite{Freed:1991wd, Rozansky:1993zx}.

\subsection{\texorpdfstring{Ray-Singer Torsion for $\Gamma=\text{Dic}_k$}{Ray-Singer Torsion for Gamma=Dic k}}
\label{subsec:RSTGammaDick}

Now, we can move on to the case of interest $\Gamma=\text{Dic}_k$. Starting with \eqref{eqn:CheegerForm}, and using the first homology group from \tabref{tab:HomologyXY}, one finds:
\begin{align}
\label{eqn:DicTrivialTwist}
    \log\tau(S^3/ \text{Dic}_k, \triv)=\log\left\lbrack{V(S^3)/4k}\right\rbrack-{2}\log{2},
\end{align}
where we also used the fact that the order of Dic$_k$ is $4k$. Interestingly, this result can also be obtained directly from the results for Lens spaces, using the fact that Dic$_k$, as a set, can be represented as the union of certain cyclic subgroups. The log of the Ray-Singer torsion computes the regularized sum of the logs of eigenvalues of harmonic functions on $S^3$ that are invariant under $\Gamma$, and this sum can be recast in terms of sums of logs of eigenvalues of harmonic functions that are invariant only under subgroups of $\Gamma$, with added signs to achieve cancellations so that overall only eigenvalues of completely invariant harmonic functions will remain. Specifically, we can use the observation that Dic$_k$ can be represented as the union of a cyclic subgroup isomorphic to $\Z_{2k}$ and $k$ cyclic subgroups isomorphic to $\Z_4$ \cite{ikeda1995spectrum}. One needs to take into account the fact that some of their intersections are nontrivial and are isomorphic to $\Z_2$ subgroups. Overall, one finds \cite{ikeda1995spectrum,Dowker:2009in}:
\begin{align}
\nonumber
    \log\tau(S^3/ \text{Dic}_k, \triv)=\frac{1}{4k}\left\lbrack 
    2k\log\tau(S^3/ \Z_{2k}, \triv)+4k\log\tau(S^3/ \Z_{4}, \triv)-2k\log\tau(S^3/ \Z_{2}, \triv)\right\rbrack,
\end{align}
which recovers \eqref{eqn:DicTrivialTwist}.

For nontrivial representations of Dic$_k$, the calculation will be different depending on whether the representations are one- or two-dimensional. Again, we follow the method described in \cite{Dowker:2009in}. We start the discussion with conjugacy classes of Dic$_k$; there are $k+3$ of them, and using the generators $(r,s)$ they can be written as:
\begin{align}
\label{eqn:ConjClasses}
    \{e \}, \{r,r^{-1}\}, \{r^2, r^{-2}\},\dots \{r^{k-1},r^{-k+1}\},\{r^{k}\}, \{sr^{2m}\},\{sr^{2m+1}\},
\end{align}
where $m$ runs from $0$ to $2k-1$. Excluding the trivial one, the first $k$ conjugacy classes correspond to a rotation by an angle $\pi l/k$ where $l=1,\dots, k$. The last two are rotations by $\pi/2$ for all values of $m$. The torsion for a representation $\rho$ was explicitly computed in \cite{Dowker:2009in}, and we will now quote their result. Using $p$ to label the conjugacy classes \eqref{eqn:ConjClasses}, starting from $p=0$, and using $\theta(p)$ to denote the rotation angles that we just described above, the formula of \cite{Dowker:2009in} is
\begin{align}
\label{eqn:DowkerFormula}
    \log \tau(S^3/ \text{Dic}_k,R)=\frac{1}{4k}\sum_{p=0}^{k+2}\chi_R([p])c_p\sum_{q=1}^{2k-1}2\cos[q \theta(p)]\log\left[2\sin{\left(\frac{\pi q}{2k}\right)}\right],
\end{align}
where $c_p$ is the size of the $p^{th}$ conjugacy class, and $\chi_R([p])$ is the value of the character of the representation $R$ of $\text{Dic}_k$ on any element of the $p^{th}$ conjugacy class of $\text{Dic}_k$. 

We will now derive a closed formula for $\tau(S^3/ \text{Dic}_k,\NonAbRep{m})$, where we specialize to the two dimensional representations $R=\NonAbRep{m}$ that we need. Firstly, examining \eqref{eqn:NonARepsDic} we have the characters
\begin{align}
\label{eqn:DicCharacters}
    \chi_R(r^p)=2 \cos\left(\frac{\pi mp}{k}\right), \quad \chi_R(r^p s)=0,\,
    \qquad\text{for $R=\NonAbRep{m}$.}
\end{align}
for $m=1,3,\dots$ not exceeding $k-1$. Therefore, we can restrict the first sum of \eqref{eqn:DowkerFormula}, the one over $p$, to only the first $k$ conjugacy classes of \eqref{eqn:ConjClasses}. In this way, \eqref{eqn:DowkerFormula} simplifies to
\begin{align}
    \log \tau(S^3/ \text{Dic}_k,\NonAbRep{m})=\frac{1}{4k}\sum_{p=0}^k2c_p\cos\left(\frac{\pi m p}{k}\right)\sum_{q=1}^{2k-1}2\cos\left(\frac{\pi qp}{k} \right)\log\left[2\sin{\left(\frac{\pi q}{2k}\right)}\right].
\end{align}
Taking into account that $c_p=2$ for all $p$ except for $p=0$ and $p=k$, for which $c_p=1$, we can simplify the first sum to
\begin{align}
    \sum_{p=0}^kc_p\cos\left(\frac{\pi m p}{k}\right)\cos\left(\frac{\pi qp}{k} \right)=
    \left\{\begin{array}{ll}
    k & \text{for $q=m$ or $q=2k-m$,} \\
    0 & \text{otherwise.} 
    \end{array}\right.
\end{align}
Therefore, \eqref{eqn:DowkerFormula} simplifies significantly,
\begin{align}
\label{eqn:DicTorsion2d}
    \log \tau(S^3/ \text{Dic}_k,\NonAbRep{m})=2\log\left[2\sin{\left(\frac{\pi m}{2k}\right)}\right].
\end{align}
In fact, this formula is very reminiscent of the torsion formula for the Lens space \eqref{eqn:LensSpaceTorRep}, but it is not obvious to us why it should be so. As far as we know the closed form for this torsion was not presented in the literature before. 

Below, we list the values for the first few $k$'s. We also confirm results given in \cite{Dowker:2009in}, where cases $k=2,3,4$ were presented. For completion, we start first with results for one dimensional representations and then we follow with the two dimensional ones. 

The torsions associated to the one-dimensional representations of Dic$_k$ are
\begin{align}
\label{eqn:TauAb12}
    \tau(S^3/\Dic_k, \AbRep{2})&=4/k, \\
    \tau(S^3/\Dic_k, \AbRep{3})&=\tau(S^3/\Dic_k, \AbRep{4})= 2,\qquad (k=2,4,6,\dots),\\
    \tau(S^3/\Dic_k, \AbRep{3})&=\tau(S^3/\Dic_k, \AbRepCC{3})=4,\qquad (k=3,5,7,\dots).
\end{align}
It turns out that not all of these torsions are needed for our purposes, since as in the case of representations of $\Z_k$, we need to decompose the adjoint representation into one-dimensional representations. For both even and odd $k$, the abelian representations from \eqref{eqn:RepsDicEven} and \tabref{tab:RepsDicOdd} can be represented in the general diagonal form
$$
\lambda=\begin{pmatrix}
    \omega & 0 \\
    0 & \omega^{-1}
\end{pmatrix},
$$
therefore the adjoint action gives the same decomposition as the one we encountered before, $\gamma\cong\Sym^2\lambda\rightarrow \omega^{2}\oplus\triv\oplus\omega^{-2}$. Then, using the representations from \eqref{eqn:RepsDicEven}, we obtain for even $k$ 
\begin{align}
    \tau(S^3/\Dic_k,\Sym^2\AbRep{2})&=
        \tau(S^3/\Dic_k,\Sym^2\AbRep{3}) \nonumber \\
        &=
    \tau(S^3/\Dic_k, \Sym^2\AbRep{4})=\tau(S^3/\Dic_k, \triv)^3,
\end{align}
and for odd $k$ 
\begin{align}
    \tau(S^3/\Dic_k, \Sym^2\AbRep{2})&=\tau(S^3/\Dic_k, \triv)^3, \\
    \tau(S^3/\Dic_k, \Sym^2\AbRep{3})&=\tau(S^3/\Dic_k, \AbRep{2})^2\tau(S^3/\Dic_k, \triv).
\end{align}

For the two-dimensional representations, we list the results for the first $k=2,\dots,6$ values of $k$. We exclude $k=1$ since that case corresponds to the abelian $\text{Dic}_1\cong\Z_4$. As before since fields are transforming in the adjoint representation we find that
\begin{align}
    \gamma\cong\Sym^2\NonAbRep{m} \rightarrow \NonAbRep{2m}\oplus \AbRep{2},
\end{align}
therefore the torsion will be given by the product of torsions \eqref{eqn:DicTorsion2d} and by \eqref{eqn:TauAb12},
\begin{align}
\label{eqn:TorsionAdj} \nn
    \tau(S^3/ \text{Dic}_k, \gamma)&=\tau(S^3/ \text{Dic}_k, \Sym^2\NonAbRep{m}) \\ 
    &=\tau(S^3/ \text{Dic}_k, \NonAbRep{2m})\tau(S^3/ \text{Dic}_k, \AbRep{2})
=\frac{16}{k}\sin^2\left(\frac{\pi m}{k}\right).
\end{align}

For Dic$_2$, we have one possible value of $m=1$, hence, 
\begin{align}
    \tau(S^3/ \text{Dic}_2, \NonAbRep{2}\oplus \AbRep{2})=8.
\end{align}
For $k=3$ we have:
\begin{itemize}
    \item ${\tau(S^3/\Dic_3,\NonAbRep{2}\oplus \AbRep{2})}=4$.
\end{itemize}
For $k=4$ the torsion is:
\begin{itemize}
    \item ${\tau(S^3/\Dic_4, \NonAbRep{2}\oplus \AbRep{2})}=2$,
    \item ${\tau(S^3/\Dic_4, \NonAbRep{6}\oplus \AbRep{2})}=2$.
\end{itemize}
For $k=5$ the torsion is:
\begin{itemize}
    \item ${\tau(S^3/\Dic_5, \NonAbRep{2}\oplus \AbRep{2})}=\frac{16}{5}\left(\frac{5}{8}-\frac{\sqrt{5}}{8}\right)$,
    \item ${\tau(S^3/\Dic_5, \NonAbRep{6}\oplus \AbRep{2})}=\frac{16}{5}\left(\frac{5}{8}+\frac{\sqrt{5}}{8}\right)$.
\end{itemize}
Finally, for $k=6$ the torsion is:
\begin{itemize}
    \item ${\tau(S^3/\Dic_6, \NonAbRep{2}\oplus \AbRep{2})}=\frac{2}{3}$,
    \item ${\tau(S^3/\Dic_6, \NonAbRep{6}\oplus \AbRep{2})}=\frac{8}{3}$,
    \item ${\tau(S^3/\Dic_6, \NonAbRep{10}\oplus \AbRep{2})}=\frac{2}{3}$.
\end{itemize}

\section{Hilbert Spaces of the (2,0)-theory on Spherical Orbifolds}
\label{sec:HilbNoS}

We can now combine the results of the Ray-Singer torsion for various flat connections to explicitly compute the Witten indices of  spaces of the $(2,0)$-theory on $X_5=S^3/\Gamma\times S^2$, by evaluating \eqref{eqn:PartTorsion}. First we consider only gauge bundles with trivial Stiefel-Whitney classes. Later on, we extend the discussion to nontrivial Stiefel-Whitney classes as well. As we have seen in \secref{sec:physics}, because of the presence of $\rho$ and $\chi$ zero modes for all the abelian flat connections, only non-abelian flat connections contribute\footnote{In \cite{Bak:2017jwd}, an additional fermionic ``gauge symmetry'' was imposed whereby fermionic fields are shifted by terms proportional to harmonic forms, which allows for nonzero contributions from flat connections with zero modes (see Appendix D.2 of \cite{Bak:2017jwd}), but we do not impose this additional symmetry.} to the sum over $A$ in \eqref{eqn:PartTorsion}. For $\Gamma=\Z_k$, the Witten index therefore vanishes, because all flat connections are abelian. Thus, we restrict our attention to $\Gamma=\text{Dic}_k$. The non-abelian flat connections are in one-to-one correspondence with irreducible $2$-dimensional representations $R_{m}=\NonAbRep{1},\NonAbRep{3},\dots$ of $\text{Dic}_k$, and since our fields transform in the adjoint representation we find, using \eqref{eqn:symtensorp},
\begin{align} 
\label{eqn:PartTorsionDick}
Z(S^3 / \text{Dic}_k \times S^2) = \Norm \sum_{\begin{array}{c}\text{irrep $R_m$} \\ \dim R_m=2 \\ \end{array}} \frac{1}{2}\tau(\text{Dic}_k, R_{2m}\oplus \AbRep{2}).
\end{align}
where we write $\tau(\Gamma,R_{2m}\oplus \AbRep{2})$ as a shorthand for $\tau(S^3/\Gamma,A[R_{2m}\oplus \AbRep{2}])$, with $A[R_{2m}\oplus \AbRep{2}]$ being a flat connection whose holonomy corresponds to the adjoint representation $R_{2m}\oplus \AbRep{2}$. As we will argue in \secref{sec:DerCS}, this partition function can be also computed using Verlinde formula for Chern-Simons theory on $S^2$. From this observation we will find that $\Norm=1$. 
Combining the results of \secref{subsec:RSTGammaDick}, from \eqref{eqn:PartTorsionDick} we find
\begin{align}
\label{eqn:DimHBasic}
Z(S^3 / \text{Dic}_k \times S^2)=\left\{\begin{array}{ll}
4 & \text{for $k=2$;} \\
2 & \text{for $k\ge 3$.} \\
\end{array}\right.
\end{align}

For a nontrivial Stiefel-Whitney class, based on the discussion at the end of \secref{sec:geometry}, for odd $k$ the corresponding flat connections are all abelian, and hence admit fermionic zero modes and do not contribute to the partition function. For even $k$, we have one nontrivial Stiefel-Whitney class that supports a non-abelian flat connection associated with the representation \eqref{eqn:NonARepsDicSO}. It can be easily checked that for $k>2$ it gives the same partition function as the trivial Stiefel-Whitney class, and therefore the Witten index of space with $\vF=0$ in \eqref{eqn:wFZ} is double that of \eqref{eqn:DimHBasic}, and if $\vF\neq 0$ the Witten index vanishes due to cancellations between $\wF=0$ and $\wF\neq 0$ partition functions. For $k=2$, if $\wF\neq 0$ we get $\PFZ_\wF=0$, and therefore the Witten index \eqref{eqn:wFZ} is $4$ for all $\vF$.

\subsection{Hilbert Spaces with Surface Operator Insertions}
\label{subsec:operators}

We wish to generalize the results of \secref{sec:HilbNoS} by inserting external sources on $X_5=S^3/\text{Dic}_k\times S^2$. A possible external source for the $(2,0)$-theory comes in the form of a string along a closed one-dimensional curve $C\subset X_5$, and to compute the Witten index with the sources, we modify the partition function \eqref{eqn:PartTorsionDick} into a correlation function of BRST invariant surface operators along $C\times S^1$, for each of the curves $C$. The surface operators are labeled by irreducible representations $\RSurf$ of the gauge group SU$(2)$, and the curves $C$ are characterized by their homotopy, which corresponds to an element $\gamma\in\Gamma$. Thus, we can insert $\nC$ timelike surface operators corresponding to elements
\begin{align}
\gamma_1,\dots, \gamma_\nC\in\Gamma,
\end{align}
and irreducible representations 
\begin{align}
    \RSurf_1,\dots,\RSurf_\nC
\end{align}
of the 5d SYM's gauge group SU$(2)$.
Upon the compactification on $S^1$ the surface operators reduce to Wilson loops of the BRST invariant gauge field ${\mathcal A}$ in representations $\RSurf_1,\dots,\RSurf_\nC$. (See also  \cite{Rey:1998ik,Maldacena:1998im} for BPS Wilson loops.) For simplicity, we will take all the Wilson loops to be in the fundamental representation of SU$(2)$, and set $\RSurf_1=\cdots=\RSurf_\nC=\mathbf{2}$.

We denote the desired correlation function (in 5d SYM) by
\begin{align}
I(C_1,\dots,C_\nC)\stackrel{{\scriptstyle\rm def}}{=}
\left\langle
W(C_1)\cdots
W(C_\nC)
\right\rangle,\qquad
W(C)\stackrel{{\scriptstyle \rm def}}{=}
\tr\left[P \exp\left\{i\oint_{C} {\mathcal A}\right\}\right].
\end{align}
Similarly to the arguments in \secref{sec:HilbNoS}, the correlation function only receives contributions from flat connections for which there are no zero modes for $\rho$ or $\chi$. These are non-abelian flat connections, and they correspond to two dimensional representations of $\Gamma$ that are irreducible.
The expectation value is then
\begin{align}
\label{eqn:CorrFcnDef}
I(C_1,\dots,C_\nC)\stackrel{{\scriptstyle\rm def}}{=}
\left\langle
W(C_1)\cdots
W(C_\nC)
\right\rangle
=\sum_{\begin{array}{c}\text{irrep $R$} \\ \dim R=2 \\ \end{array}}\frac{1}{2}\tau(\Gamma,R)
\prod_{i=1}^\nC\chi_R(\gamma_i)
\end{align}
where $\tau(\Gamma,R)$ is the $1$-loop contribution which we computed using the Ray-Singer torsion in \secref{sec:ray} and $\chi_R(\gamma_i)$ is the character of the representation $R$ evaluated on the element $\gamma_i\in\Gamma$, we also set $\Norm=1$ based on the discussion above.

The expression $I(C_1,\dots,C_\nC)$ is the Witten index of the Hilbert space of the $\text{SU}(2)$ twisted $(2,0)$ theory on $(S^3/\Gamma)\times S^2$ with surface operator insertions that stretch in time and along $C_1,\dots,C_\nC$. This shows that $I(C_1,\dots,C_\nC)$ must be an integer, and below we will argue that this is indeed the case.
In general, the characters $\chi_R(\gamma_i)$ appearing in \eqref{eqn:CorrFcnDef} are not necessarily integers, and in fact are given by irrational numbers in many examples, as can be seen in \eqref{eqn:DicCharacters}. Without loss of generality, we can assume that all $\gamma_i$ in \eqref{eqn:CorrFcnDef} are of the form $r^{p_i}$, for some integers $p_i$, otherwise the characters and the index $I(C_1,\dots,C_\nC)$ vanish.
Therefore, using the torsion \eqref{eqn:TorsionAdj}, we find the Witten index with $\nC$ surface operator insertions labeled by $\pi_1(S^3/\Dic_k)$ elements $r^{p_1},\cdots, r^{p_\nC}$:
\bear
\lefteqn{
\text{Index}_{k,\nC}(p_1,\cdots,p_\nC) =
}\nn\\ &&
\frac{2^{\nC+3}}{k}\sum_{m=1,3,\dots}^{m\le k-1}
\sin^2\left(\frac{\pi m}{k}\right)
\cos\left(\frac{\pi m p_1}{k}\right)
\cos\left(\frac{\pi m p_2}{k}\right)
\cdots
\cos\left(\frac{\pi m p_{\nC}}{k}\right).
\label{eqn:NeedsToBeInZn}
\eear
Here we expanded \eqref{eqn:CorrFcnDef} and replaced $R$ with $\NonAbRep{2m}\oplus\AbRep{2}$.
We also used \eqref{eqn:TorsionAdj} to substitute
\be\label{eqn:RStorsionValue}
\tau(\text{Dic}_k,\NonAbRep{2m}\oplus \AbRep{2}) =
\frac{16}{k}\sin^2\left(\frac{\pi m}{k}\right).
\ee
That \eqref{eqn:NeedsToBeInZn} is an integer follows from the same statement for $\nC=1$,
\be
\label{eqn:NeedsToBeInZ1} 
\frac{16}{k}\sum_{m=1,3,\dots}^{m\le k-1}
\sin^2\left(\frac{\pi m}{k}\right)
\cos\left(\frac{\pi m p}{k}\right)
=\left\{\begin{array}{ll}
8 & \text{for $k=2$ and $p=0$,}\\
-8 & \text{for $k=2$ and $p=k$,}\\
4 & \text{for $k\ge 3$ and $p=0$,}\\
-4 & \text{for $k\ge 3$ and $p=k$,}\\
0 & \text{for $k=2$ and $p=1$,}\\
2 & \text{for $k=3$ and $p=1$,}\\
-2 & \text{for $k=3$ and $p=2$,}\\
0 & \text{for $k=4$ and $p=1,2,3$,}\\
-2 & \text{for $k\ge 5$ and $p=2$,}\\
2 & \text{for $k\ge 5$ and $p=k-2$,}\\
0 & \text{for $k\ge 5$ and $k\ge p\notin\{2,k-2\}$,}
\end{array}
\right.
\ee
which is always an integer.

In the case of nontrivial Stiefel-Whitney classes, for even $k$ we need to also add the contribution from the irreducible representations \eqref{eqn:NonARepsDicSO} that have even $m$. In this case, we can use the following sums:
\be\label{eqn:NeedsToBeInZ12ndRep}
\frac{16}{k}\sum_{m=0,2,\dots}^{m\le k-1}
\sin^2\left(\frac{\pi m}{k}\right)
\cos\left(\frac{\pi m p}{k}\right)
=\left\{\begin{array}{ll}
0 & \text{for $k=2$ and $p=0,1,2$,}\\
4 & \text{for $k\ge 4$ and $p=0,k$,}\\
-4 & \text{for $k=4$ and $p=2$,}\\
0 & \text{for $k=4$ and $p=1,3$,}\\
-2 & \text{for $k\ge 6$ and $p=2$,}\\
-2 & \text{for $k\ge 6$ and $p=k-2$,}\\
0 & \text{for $k\ge 6$ and $k\ge p\notin\{2,k-2\}$.}
\end{array}
\right.
\ee

We conclude the discussion of the Witten index of the Hilbert space of the $(2,0)$ theory with surface operator insertions by presenting a few examples with small $k$ and trivial Stiefel-Whitney class. As we argued above, the dimension is always given by a sum of products of torsion and characters \eqref{eqn:CorrFcnDef}, and we have shown that this number is necessarily an integer, as expected from the physics.

\subsection{Examples}
\label{sec:WittenWithSources}

In the examples below, we will take $\nC$ string sources each wound once along the $1$-cycle of $S^3/\Gamma$ associated to the generator $r^2\in\text{Dic}_k$. (We chose the element $r^2$ so as to get nontrivial results.)
We also write $\text{Index}_{k,\nC}$, short for $\text{Index}_{k,\nC}(p_1,\cdots,p_\nC)$ in \eqref{eqn:NeedsToBeInZn}.
Thus,
\be\label{eqn:IndexknX}
\text{Index}_{k,\nC}=
\frac{2^{\nC+3}}{k}\sum_{m=1,3,5,\dots}\cos^\nC\left(\frac{2\pi m}{k}\right)
\sin^2\left(\frac{\pi m}{k}\right),
\ee
and $m$ goes up to $k-1$ for even $k$ and up to $k-2$ for odd $k$.
Define
\be\label{eqn:defdeltak}
\delta_k(p)=\left\{\begin{array}{ll}
1 & \text{if $p=k N$ for some $N\in\Z$;} \\
0 & \text{otherwise.} \\
\end{array}
\right.
\ee
With basic algebraic manipulations such as a binomial expansion and summation of a geometric series, we can rewrite the index as
\bear
\text{Index}_{k,\nC} &=&
2\sum_{l=0}^{\nC}\binom{\nC}{l}(-1)^{2(\nC-2l)/k}\delta_k(2(\nC-2l))
\nn\\ &&
-\sum_{l=0}^{\nC+1}\binom{\nC+1}{l}(-1)^{2(\nC+1-2l)/k}\delta_k(2(\nC+1-2l)),
\label{eqn:AltIndexAnyk}
\eear
which for odd $k$ simplifies to
\be\label{eqn:AltIndexOddk}
\text{Index}_{k,\nC} 
=
2\sum_{l=0}^{\nC}\binom{\nC}{l}\delta_k(\nC-2l)
-\sum_{l=0}^{\nC+1}\binom{\nC+1}{l}\delta_k(\nC+1-2l)\qquad(k\in2\Z+1),
\ee
The expression for the index in \eqref{eqn:AltIndexAnyk} is manifestly an integer.
We will present additional manifestly integral formulas for specific values of $2\le k\le 6$ below.

\subsubsection*{Index for $\text{Dic}_2$ with $\nC$ strings}
$$
\text{Index}_{2,\nC}=
\frac{1}{2}\tau(\text{Dic}_2,\NonAbRep{2}\oplus\AbRep{2})\left\lbrack 2\cos\pi\right\rbrack^\nC
\nn\\
=(-2)^{\nC}4,
$$
for all $\nC$.
\subsubsection*{Index for $\text{Dic}_3$ with $\nC$ strings}
$$
\text{Index}_{3,\nC}=
\frac{1}{2}\tau(\text{Dic}_3,\NonAbRep{2}\oplus\AbRep{2})\left\lbrack 2\cos\tfrac{2\pi}{3}\right\rbrack^\nC
=(-1)^{\nC}2,
$$
for all $\nC$.

\subsubsection*{Index for $\text{Dic}_4$ with $\nC$ strings}
\begin{align}
\nn
\text{Index}_{4,\nC}=
\sum_{m=1,3}\frac{1}{2}\tau(\text{Dic}_4,\NonAbRep{2m}\oplus\AbRep{2})\left\lbrack 2\cos\tfrac{\pi m}{2}\right\rbrack^\nC=\begin{cases}
    2, &\quad \text{for $\nC=0$}, \\
    0, &\quad \text{otherwise} 
\end{cases}
\end{align}

\subsubsection*{Index for $\text{Dic}_5$ with $\nC$ strings}

\begin{align}
\text{Index}_{5,\nC}&=
\sum_{m=1,3}\frac{1}{2}\tau(\text{Dic}_5,\NonAbRep{2m}\oplus \AbRep{2})\left\lbrack 2\cos\tfrac{2 \pi m}{5}\right\rbrack^\nC
\nn\\
&=
\frac{1}{2^{\nC+2}\sqrt{5}}\left((\sqrt{5}-1)^{n+1}-(-1-\sqrt{5})^{n+1}\right) \nn \\ \nn
&\longrightarrow
2,-2,4,-6, 10,-16,26,\ldots, \quad \text{for $\nC\ge0$}
\nn
\end{align}
a Fibonacci-like sequence. The recurrence relation
$$
\text{Index}_{5,\nC+2} = -\text{Index}_{5,\nC+1} + \text{Index}_{5,\nC}
$$
suggests a fusion relation whereby two strings can fuse to a linear combination of one or zero strings. We derive these recurrence relations for general $k$ in \secref{subsec:Fusion} below. 

\subsubsection*{Index for $\text{Dic}_6$ with $\nC$ strings}

\begin{align}
\text{Index}_{6,\nC}&=
\sum_{m=1,3,5}\frac{1}{2}\tau(\text{Dic}_6,\NonAbRep{2m}\oplus \AbRep{2})\left\lbrack 2\cos\tfrac{\pi m}{3}\right\rbrack^\nC \\
&\rightarrow -2, 6,-10, 22, -42, 86, \ldots, \quad \text{for $\nC\ge1$}
\nn
\end{align}
with recurrence relation
$$
\text{Index}_{6,\nC+3} = 3\text{Index}_{6,\nC+1} - 2\text{Index}_{6,\nC}.
$$


\subsection{Fusion Rules for Surface Operators}
\label{subsec:Fusion}

The Witten indices that we found in \eqref{eqn:IndexknX} satisfy a linear recurrence relation of the form
$$
\text{Index}_{k,\nC}=
\sum_{i=1}^{\PolyDeg}\aRL_i\text{Index}_{k,\nC-i},\qquad
\PolyDeg = \left\lbrack\frac{k}{2}\right\rbrack,
$$
where the coefficients $\aRL_i$ are determined by expanding the polynomial whose roots are  $2\cos\frac{\pi m}{k}$:
\begin{align}
\label{eqn:ReccPol}
    \PolyP_k(\PolyX)=\PolyX^{\PolyDeg}-\sum\aRL_i\PolyX^{\PolyDeg-i} 
    =\prod_{m=1,3,5,\dots}^{2\PolyDeg-1}\left(\PolyX-2\cos\frac{2\pi m}{k}\right).
\end{align}
For organizational purposes, we will call these \textit{index polynomials}. These polynomials can be expressed in terms of Chebyshev polynomials, defined as follows \cite{abramowitz1968handbook}.
The Chebyshev polynomials of the first kind, $\ChebT_n$, can be defined recursively by
\begin{equation}
\begin{aligned}
T_0(x) & =1, \\
T_1(x) & =x, \\
T_{n+1}(x) & =2 x T_n(x)-T_{n-1}(x),
\end{aligned}
\end{equation}
and the Chebyshev polynomials of the second kind, $\ChebU_n$, can be defined by
\begin{equation}
\begin{aligned}
U_0(x) & =1, \\
U_1(x) & =2 x, \\
U_{n+1}(x) & =2 x U_n(x)-U_{n-1}(x) .
\end{aligned}
\end{equation}

The index polynomials can then be expressed in terms of the Chebyshev polynomials  $\ChebT_n$ and $\ChebU_n$ as
\begin{align}
\label{eqn:AllCheb}
    \PolyP_{2j}(\PolyX)=2\left\lbrack\ChebT_{j}\left(\frac{\PolyX}{2}\right)+1\right\rbrack, \qquad
    \PolyP_{2j+1}(\PolyX)=\ChebU_{j}\left(\frac{\PolyX}{2}\right)+\ChebU_{j-1}\left(\frac{\PolyX}{2}\right).
\end{align}
The index polynomials $\PolyP_j$ then satisfy recurrence relations,
\begin{align*}
    \PolyP_{2j+1}(\PolyX)&=\PolyP_{2j}(\PolyX)+\PolyP_{2j-1}(\PolyX)-2,\quad
\PolyP_{2j+2}(\PolyX)=\PolyP_{2j}(\PolyX)+(\PolyX-2)\PolyP_{2j+1}(\PolyX)
\end{align*}
with initial values
\begin{align*}
\PolyP_0(\PolyX)=4, \quad \PolyP_1(\PolyX)=1,
\end{align*}
which shows that the index polynomials have integer coefficients. The index polynomials can be encoded in a generating function
$$
\sum_{j=0}^\infty\PolyP_j(\PolyX)\PolyT^j=
\frac{t^3-\PolyX t^2+t+2}{1-\PolyT^2\PolyX+\PolyT^4}+\frac{2}{1-t^2}\,.
$$
The first few index polynomials are then
\begin{align}
\PolyP_1(X) &= 1,
\\
\PolyP_2(X) &= \PolyX+2,
\nn\\
\PolyP_3(X) &= \PolyX+1,
\nn\\
\PolyP_4(X) &= \PolyX^2,
\nn\\
\PolyP_5(X) &= \PolyX^2+\PolyX-1,
\nn\\
\PolyP_6(X) &= \PolyX^3-3\PolyX+2,
\nn\\
\PolyP_7(X) &= \PolyX^3+\PolyX^2-2\PolyX-1,
\nn\\
\PolyP_8(X) &= (\PolyX^2-2)^2,
\nn\\
\PolyP_9(X) &= \PolyX^4+\PolyX^3-3\PolyX^2-2\PolyX+1,
\nn\\
\PolyP_{10}(X) &= \PolyX^5-5\PolyX^3+5\PolyX+2.
\nn
\end{align}
The relations among Witten indices of different $n$'s suggest fusion relations for external string sources (implemented via the surface operator insertions) in the topological theory. For example, $\PolyP_{10}$ means that for $k=10$ we have a relation
$$
\text{Index}_{10,\nC+5}-5\text{Index}_{10,\nC+3}+5\text{Index}_{10,\nC+1}+2\text{Index}_{10,\nC}=0,
$$
which suggests that the operator $\OpS(r)$ that is a surface operator along the time direction and the $1$-cycle of $S^3/\text{Dic}_{10}$ associated with the $r^2$ generator of $\text{Dic}_{10}$ satisfies 
$\OpS^5-5\OpS^3+5\OpS+2=0$. One can check that these relations agree with the examples we presented in \secref{sec:WittenWithSources}.

In the next section, \secref{sec:DerCS}, we will see that the dimensions of Hilbert spaces that we studied can be computed as linear combinations of dimensions of Hilbert spaces of SU$(2)/\Z_2$ Chern-Simons theory on $S^2$ at a shifted level $k-2$, and with various numbers (between $0$ and $2\nC$) of quark sources inserted. Therefore, the fusion relations we found in terms of Chebyshev polynomials, are related to the fusion rules of Chern-Simons theory.

\section{Derivation of the index from Chern-Simons theory on $S^2$}
\label{sec:DerCS}

We will now show, from a physics perspective, how the expression \eqref{eqn:IndexknX} for the index with surface operator insertions can be recast as a linear combination of dimensions of Hilbert spaces of Chern-Simons theories on $S^2$ with $\nC$ quarks inserted. As we shall see, this involves a combination of two Chern-Simons theories with gauge groups (and levels)  $\text{SU}(2)_{k-2}$ and $\text{U}(1)_{2k}$. We will have to consider even and odd $k$ separately. The factor $\sin^2\left(\frac{\pi m}{k}\right)$, which in \eqref{eqn:IndexknX} arose from the analytic torsion on $S^3/\text{Dic}_k$, will appear in the present picture in two different ways: (i) it will be an element of the Verlinde matrix for gauge group $\text{SU}(2)_{k-2}$, and (ii) it will arise from the coupling of residual fermionic modes for gauge group $\text{U}(1)_{2k}$.

We begin with the fact that $S^3/\text{Dic}_k$ is a Hopf circle fibration over the real projective space $\mathbb{RP}^2$ (\figref{fig:S3DickHopf}), the structure group is composed of translations and orientation reversal (around any axis) of $S^1$. This allows a dimensional reduction to 5d SYM on $\R\times S^2\times\mathbb{RP}^2$ with an additional Chern-Simons term that arises from the nontriviality of the circle bundle. 
Such a dimensional reduction on a circle fibration has been applied in general for Seifert manifolds in \cite{Dimofte:2011ju, Dimofte:2011py, Gadde:2013sca, Chung:2014qpa, Gukov:2015sna}, and in \cite{Gukov:2017kmk} the $\hat{Z}$-invariant was derived from a consideration of the $(2,0)$-theory on $S^1\times D^2\times M_3$, where $M_3$ is a $3$-manifold and $D^2$ is a disk, and the $\hat{Z}$-invariant was computed for lens spaces in \cite{Gukov:2017kmk}, from which the superconformal index on $S^1\times S^2\times L(p,1)$ was derived. To the best of our knowledge the calculations we present below for $S^3/\text{Dic}_k$ have not been worked out before.

\begin{figure}
    \begin{center}
    \begin{tikzpicture}[scale=1, every node/.style={font=\small},>=Latex]
        \draw[thick] (180:2cm) arc (180:0:2cm);
        \draw[thick] (-2,0) arc (180:360:2cm and 0.5cm);
        \draw[dashed] (2,0) arc (0:180:2cm and 0.5cm);
        \node[black] at (0,-1) {$\phi\to\phi+\pi$};
        \coordinate (P) at (60:1.5cm);
        \fill (P) circle (2pt);
        \coordinate (F) at ($(P)+(1.5,0.7)$);
        \draw[-, blue, thick] (P) -- (F);
        \draw[blue, thick] (F) circle (0.5cm) node[right, xshift=10]  {$S^1_H$};
    \end{tikzpicture}
    \caption{$S^3/\text{Dic}_k$ represented as Hopf fibration:  $\mathbb{RP}^2\times S^1_H$}.
    \label{fig:S3DickHopf}
    \end{center}
\end{figure}

Next, we will dimensionally reduce from 5d to 3d by shrinking $\mathbb{RP}^2$, and we will need to specify which `t Hooft electric and magnetic fluxes are allowed in the 5d SYM theory. We will argue that for even $k$, the choices are specified by taking the gauge group to be $\text{SU}(2)/\Z_2$, which is the Langlands dual of the gauge group $\text{SU}(2)$ that we used in \secref{sec:HilbNoS}. For odd $k$, the situation is more complicated and we will address it in \secref{subsec:3dCS}. We will show that for nontrivial `t Hooft magnetic flux along $\mathbb{RP}^2$, the 3d Chern-Simons gauge group is $\text{SU}(2)$ for odd $k$ and $\text{SU}(2)/\Z_2$ for even $k$, while for trivial `t Hooft magnetic flux, the gauge group is a semidirect product $\Z_2\ltimes \text{U}(1)$.

The outline of the rest of this section is as follows. In \secref{subsec:FGeometry} we elaborate on the geometry of $S^3/\text{Dic}_k$ as a circle fibration. In \secref{subsec:FCHO} we discuss the $\text{SU}(2)$ gauge field and possible gauge bundles on $\mathbb{RP}^2$ by paying attention to the periodicity conditions along the nontrivial $1$-cycle of $\mathbb{RP}^2$ that arise after the dimensional reduction from $S^3/\text{Dic}_k$ to $\mathbb{RP}^2$. More details are shown in \secref{subsec:HOdef}, which can be skipped. In \secref{subsec:5dCS} we discuss the 5d Chern-Simons term, which is the only relevant term for the gauge field at low-energy. In \secref{subsec:3dCS} we will reduce the 5d Chern-Simons term to 3d, and discuss the 3d gauge groups and gauge bundles for even and odd $k$.
The surface operators of \secref{subsec:operators} reduce to a linear combination of Wilson loops in the adjoint and trivial representations, as we will show in \secref{subsec:3dSOtoW}. Next, we include the fermions in the discussion. In \secref{subsec:5df} we develop and write down the fermionic terms in the 5d SYM action, including the nontrivial periodicity conditions along $\mathbb{RP}^2$, and in \secref{subsec:3df} we reduce the fermionic action to 3d for the two gauge groups $\text{SU}(2)$ and $\Z_2\ltimes\text{U}(1)$. For gauge group $\text{SU}(2)$ the only effect of the fermions is a shift in the Chern-Simons level, with the details shown in \appref{app:LevelShift}. For gauge group $\Z_2\ltimes\text{U}(1)$, there is no level shift, but charged zero modes of fermions contribute as an additional effective Wilson loop.
Finally, we collect all the ingredients to reproduce the index \eqref{eqn:IndexknX} from the ``analytic torsion side" in \secref{subsec:Altogether}.

\subsection{$S^3/\text{Dic}_k$ as a Circle Fibration}
\label{subsec:FGeometry}

Since $\text{Dic}_k$ has a cyclic subgroup isomorphic to $\Z_{2k}$ and of index $2$ (as elaborated more in \secref{subsec:RSTGammaDick}), we can express $S^3/\text{Dic}_k$ as a $\Z_2$ quotient of the lens space $S^3/\Z_{2k}\cong L(2k,1)$, which is a (Hopf) $S^1$ fibration over $S^2$ as a base. We can take $\Z_{2k}\subset\text{Dic}_k$ to be generated by $r$, and $\Z_2$ to be generated by $s$, where $r,s$ were defined in \eqref{eqn:GenDic}. Concretely, we represent $S^3$ as the subset of $\C^2$ (with coordinates $z_1, z_2$), with $|z_1|^2+|z_2|^2=1$, and let $r,s\in\text{Dic}_k$ act as
$$
r:(z_1, z_2)\longrightarrow
(e^{\frac{\pi i}{k}}z_1, e^{\frac{\pi i}{k}}z_2),
\qquad
s:(z_1, z_2)\longrightarrow (-\bar{z}_2,\bar{z}_1).
$$
A translations by an arbitrary $\epsilon$ along the $S^1$ fiber (of the Hopf fibration) corresponds to
$$
(z_1, z_2)\longrightarrow
(e^{i\epsilon}z_1, e^{i\epsilon}z_2),
$$
and we can present the $S^2$ base of the fibration as $\CP^1$ with coordinate $z=z_1/z_2$ (in the patch $z_2\neq 0$). Note that $z$ is invariant under $r$. The generator $s$ then acts as 
$$
s: z\longrightarrow -\frac{1}{\bar{z}},
$$
and on the fiber it acts by an orientation reversal.
A fundamental region of the action of $s$ on $z$ is $|z|\le 1$.

We now parameterize $S^3/\Z_{2k}$ by $\theta,\Hphi,\Hchi$ through
\be\label{eqn:thetaHphiHchi}
z_1 = e^{i(\Hphi+\frac{\Hchi}{2k})}\sin\tfrac{\theta}{2},\qquad
z_2 = e^{i\frac{\Hchi}{2k}}\cos\tfrac{\theta}{2},\qquad
(0\le\Hphi, \Hchi<2\pi,\quad 0\le\theta\le\pi),
\ee
so that $z=e^{i\Hphi}\tan\tfrac{\theta}{2}$, and we can consider $\Hchi$ as a coordinate on the $S^1$ fiber. Then, $s$ acts as
\begin{align}
\label{eqn:AnglesIdentNew}
s:\Hphi\rightarrow\Hphi+\pi,\qquad
\Hchi\rightarrow -\Hchi-2k\Hphi,\qquad
\theta\rightarrow\pi-\theta.
\end{align}
Note that the action on the base ($\CP^1$) of the fibration reverses the orientation. The fundamental region $|z|\le 1$, with the identification $\theta\sim\theta+\pi$, becomes the real projective space $\RP^2$. The metric on $S^3/\text{Dic}_k$ can now be written as
\begin{align}
\label{eqn:S3metric}
ds^2 = \frac{1}{4}\left\{
d\theta^2+\sin^2\theta d\phi^2 + \frac{1}{k^2}\left[d\Hchi+2k\sin^2\left(\tfrac{\theta}{2}\right) d\phi\right]^2
\right\}.
\end{align}

\subsection{Gauge Bundles on $\mathbb{RP}^2$}
\label{subsec:FCHO}

To proceed with the plan of dimensionally reducing the $(2,0)$-theory on $S^3/\text{Dic}_k$ to 5d SYM on $\mathbb{RP}^2$, and then further shrinking $\mathbb{RP}^2$ to zero, we will now study flat $\text{SU(2)}$ connections on $\mathbb{RP}^2$. More precisely, by flat connections, we mean locally trivial $\text{SU(2)}$ gauge fields $A$ on the disc parameterized by $\theta\le\pi/2$ and $0\le\Hphi<2\pi$ [the coordinates defined in \eqref{eqn:thetaHphiHchi}], with suitable boundary conditions that identify antipodal points parameterized by $\Hphi$ and $\Hphi+\pi$ on the disc's boundary at $\theta=\pi/2$ [see \eqref{eqn:AnglesIdentNew}].

The orientation reversal of the Hopf fiber, which supplements the identification of antipoldal points in \eqref{eqn:AnglesIdentNew}, amounts to charge conjugation on fields, which acts as minus the transpose on hermitian matrices, like the components of the gauge field $A$. Thus, the components $A_\phi$ and $A_\theta$ of the 5d gauge field $A$ satisfy the following boundary conditions at $\theta=\pi/2$:
\be\label{eqn:AbcRP2TrivialBundleNew}
A_\Hphi(\tfrac{\pi}{2},\Hphi+\pi)=-A_\Hphi(\tfrac{\pi}{2},\Hphi)^t,
\qquad
A_\theta(\tfrac{\pi}{2},\Hphi+\pi)=A_\theta(\tfrac{\pi}{2},\Hphi)^t,
\ee
where the $(+)$ sign on the $A_\theta$ component is due to the combination of the $(-)$ sign in the charge conjugation $-(\cdots)^t$ and the $(-)$ sign in the orientation reversal of the $\theta$ direction in \eqref{eqn:AnglesIdentNew}. We can, in principle, include an additional $\Hphi$-dependent gauge transformation on the RHS of each equation in \eqref{eqn:AbcRP2TrivialBundleNew}. Such a gauge transformation can introduce a nontrivial `t Hooft magnetic flux, but other than that, its effects can be absorbed in a redefinition of the gauge field [we will expand on this point below \eqref{eqn:AbcRP2NonTrivialBundle}]. Which combination of `t Hooft magnetic fluxes is allowed will be addressed in \secref{subsec:3dCS}, but for the time being we will keep all the options. 

We will require $A_\Hphi$ and $A_\theta$ to be continuous for $0<\theta\le\pi/2$, but allow the holonomy around the origin ($\theta=0$) to be $\pm\Id\in\text{SU}(2)$, with the $(-)$ sign corresponding to nontrivial `t Hooft magnetic flux. We will show below that, on the disc, such locally trivial gauge fields that satisfy the boundary conditions \eqref{eqn:AbcRP2TrivialBundleNew} are characterized by a constant element $\HO\in\text{SU}(2)$  which satisfies
\be\label{eqn:HOHOstNew}
\HO^\star\HO = (-1)^{\tHm}\Id,
\ee
where the integers $\tHm\in\{0,1\}$ are known as the `t Hooft magnetic flux along $\RP^2$, and $\Id$ is the identity matrix. Roughly speaking, $\HO$ is the holonomy of a gauge field along a loop that represents the nontrivial element of $\pi_1(\RP^2)\cong \mathbb{Z}_2 $, and represents the leftover global data that cannot be gauged away. A precise definition will be given in \eqref{eqn:HOdefNew}. 

Let $\AdjField(\theta,\phi)$ be any adjoint-valued field in the theory. We can choose a gauge that locally trivializes the gauge field along $\RP^2$, and in this gauge the components of the covariant derivative $\partial_\mu\AdjField-i[A_\mu,\AdjField]$ reduce to $\partial_\mu\AdjField$, for $\mu=\theta$ or $\mu=\Hphi$. In this gauge, the field $\AdjField$ satisfies the boundary condition
\be\label{eqn:HOAphi}
\HO^{-1}\AdjField(\tfrac{\pi}{2},\phi+\pi)\HO = \pm\AdjField(\tfrac{\pi}{2},\phi)^t.
\ee
The $(\pm)$ sign is determined by the action of the orientation reversal of the Hopf fiber on the field and by the action of orientation reversal of the $\theta$ direction. For example, for the components of the gauge field that are not along $\theta$, the sign is $(-)$; while for $A_\theta$, the sign is $(+)$. Equation \eqref{eqn:HOAphi} can be taken as an implicit definition of $\HO$ for now, and more details will be presented in \secref{subsec:HOdef}.

We can consider solutions $\HO_1,\HO_2\in\text{SU}(2)$ of \eqref{eqn:HOHOstNew} as gauge equivalent if there exists $\Lambda\in\text{SU}(2)$ such that
\be\label{eqn:HOequivNew}
\HO_2=\Lambda^{-1}\HO_1\Lambda^\star.
\ee
Furthermore, when the gauge group is $\text{SU}(2)/\Z_2$, we will additionally consider $\HO$ and $-\HO$ as equivalent.

We may investigate the solutions to \eqref{eqn:HOHOstNew} by choosing a Lie algebra basis given by the standard Pauli matrices $\{\sigma_{1},\sigma_2,\sigma_3\}$. For $\tHm=0$, all solutions to \eqref{eqn:HOHOstNew} are equivalent, by \eqref{eqn:HOequivNew}, to one of the two solutions $\HO=\pm\Id$.
For $\tHm=1$, all solutions are equivalent to $\pm i$ multiple of a Pauli matrix, say, $\HO=\pm i\GGPauli_2$.
For gauge group $\text{SU}(2)/\Z_2$, the $\pm$ signs can be dropped, and there is one solution up to equivalence for each `t Hooft magnetic flux.

We will now define $\HO$ more precisely, and derive \eqref{eqn:HOHOstNew} and \eqref{eqn:HOequivNew}. These details are not needed for the rest of the paper, and readers who accept \eqref{eqn:HOHOstNew} and \eqref{eqn:HOequivNew} can skip to \secref{subsec:5dCS}.

\subsection{Definition of $\HO$ and Its Properties}
\label{subsec:HOdef}

We begin with the boundary conditions \eqref{eqn:AbcRP2TrivialBundleNew}. If $A$ is a flat connection on $\RP^2$, we can find a local trivialization such that 
\be\label{eqn:AYgKicakNew}
A_\Hphi = i\Yg^{-1}\partial_\Hphi\Yg,\qquad 
\Yg(\theta,\Hphi)\in\text{SU}(2),\qquad
0\le\theta\le\tfrac{\pi}{2},\qquad 0\le\Hphi<2\pi,
\ee
where $\Yg$ is the gauge transformation that trivializes $A$. To allow for the possibility of a nontrivial `t Hooft magnetic flux along $\RP^2$, we remove $\theta=0$ and modify the condition of $2\pi$ periodicity in $\Hphi$ to
\be\label{eqn:tHooftYgNew}
\Yg(\theta,\phi)=(-1)^{\tHm}\Yg(\theta,\phi+2\pi),\qquad\tHm\in\{0,1\},
\ee
where $\tHm$ is the magnetic flux.
From now on we consider $\Yg(\theta,\phi)$ only at $\theta=\pi/2$ and omit the argument $\theta$, for simplicity.
From \eqref{eqn:AbcRP2TrivialBundleNew}, we find
\bear
\lefteqn{
\partial_\Hphi\left\lbrack
\Yg(\Hphi+\pi)\Yg(\Hphi)^t
\right\rbrack
=\partial_\Hphi\Yg(\Hphi+\pi)\Yg(\Hphi)^t
+\Yg(\Hphi+\pi)\partial_\Hphi\Yg(\Hphi)^t
}\nn\\
&=&
-i\Yg(\Hphi+\pi)A_\Hphi(\Hphi+\pi)\Yg(\Hphi)^t-i\Yg(\Hphi+\pi)A_\Hphi(\Hphi)^t\Yg(\Hphi)^t=0,
\nn
\eear
and therefore $\Yg(\Hphi+\pi)\Yg(\Hphi)^t$ is independent of $\phi$ and we set
\be\label{eqn:HOdefNew}
\HO=\Yg(\tfrac{\pi}{2},\Hphi+\pi)\Yg(\tfrac{\pi}{2},\Hphi)^t.
\ee
Since $\HO$ is a constant element of $\text{SU}(2)$, we have at $\theta=\pi/2$:
$$
\HO=\Yg(\Hphi+\pi)\Yg(\Hphi)^t=\Yg(\Hphi+2\pi)\Yg(\Hphi+\pi)^t=(-1)^{\tHm}\Yg(\phi)\Yg(\Hphi+\pi)^t,
$$
where we have used \eqref{eqn:tHooftYgNew}.
Therefore,
\be\label{eqn:HOHOst}
\HO^\star\HO = (-1)^{\tHm}\Yg(\Hphi+\pi)^\star\Yg(\Hphi)^\dagger\Yg(\Hphi)\Yg(\Hphi+\pi)^t=(-1)^{\tHm}\Id,
\ee
which proves \eqref{eqn:HOHOstNew}.
Note that \eqref{eqn:AYgKicakNew} does not require $A$ to be a flat connection, since we only used the $\Hphi$ component, and $\tHm$ is a topological quantum number of the field configuration.

We can modify \eqref{eqn:AbcRP2TrivialBundleNew} by allowing another gauge transformation $\Pg$ defined in the vicinity of $\theta=\frac{\pi}{2}$,
\be\label{eqn:AbcRP2NonTrivialBundle}
\Pg(\Hphi+\pi)^{-1}A_\Hphi(\tfrac{\pi}{2},\Hphi+\pi)\Pg(\Hphi+\pi)+i\Pg(\Hphi+\pi)^{-1}\partial_\Hphi\Pg(\Hphi+\pi)=-A_\Hphi(\tfrac{\pi}{2},\Hphi)^t,
\ee
and in the case where the gauge group is $\text{SU}(2)/\Z_2$, we will allow
\be\label{eqn:Pgbc}
\Pg(\Hphi+2\pi)=(-1)^{\tHPg}\Pg(\Hphi),
\ee
for some $\tHPg\in\{0,1\}$, but we will see below \eqref{eqn:PgPibc} that $\tHPg=0$ is required.
From \eqref{eqn:AbcRP2NonTrivialBundle} we also find
\be\label{eqn:AbcRP2Shift}
\Pg(\Hphi)^{-1}A_\Hphi(\tfrac{\pi}{2},\Hphi)\Pg(\Hphi)+i\Pg(\Hphi)^{-1}\partial_\Hphi\Pg(\Hphi)=-A_\Hphi(\tfrac{\pi}{2},\Hphi-\pi)^t,
\ee
and plugging this back into \eqref{eqn:AbcRP2NonTrivialBundle} we get
\begin{align}
&\Pg(\Hphi+\pi)^{-1}A_\Hphi(\tfrac{\pi}{2},\Hphi+\pi)\Pg(\Hphi+\pi)+i\Pg(\Hphi+\pi)^{-1}\partial_\Hphi\Pg(\Hphi+\pi)
=-A_\Hphi(\tfrac{\pi}{2},\Hphi)^t
\nn\\
&=
-\left\lbrack
-\Pg(\Hphi)A_\Hphi(\tfrac{\pi}{2},\Hphi-\pi)^t\Pg(\Hphi)^{-1}
-i\partial_\Hphi\Pg(\Hphi)\Pg(\Hphi)^{-1}
\right\rbrack^t \nn \\
&=
\Pg(\Hphi)^\star A_\Hphi(\tfrac{\pi}{2},\Hphi+\pi)\Pg(\Hphi)^t
+i\Pg(\Hphi)^\star\partial_\Hphi\Pg(\Hphi)^t.
\label{eqn:Aconsistent}
\end{align}
If this is to hold for an arbitrary gauge field $A_\Hphi(\tfrac{\pi}{2},\Hphi+\pi)$, we must require
\be\label{eqn:PgPibc}
\Pg(\Hphi+\pi)=(-1)^{\tHX}\Pg(\Hphi)^t,\qquad\text{for some $\tHX\in\{0,1\}$},
\ee
so that $\Pg(\Hphi+\pi)^{-1}=(-1)^{\tHX}\Pg(\Hphi)^\star$.
Equation \eqref{eqn:PgPibc} now implies that $\tHPg=0$ in \eqref{eqn:Pgbc}.

We now trivialize $A$ again, locally, as in \eqref{eqn:AYgKicakNew}.
Now, from \eqref{eqn:AbcRP2NonTrivialBundle} we find that $\Yg(\Hphi+\pi)\Pg(\Hphi+\pi)\Yg(\Hphi)^t$ is independent of $\Hphi$ and we set
\be\label{eqn:HOdef}
\HO=\Yg(\tfrac{\pi}{2},\Hphi+\pi)\Pg(\Hphi+\pi)\Yg(\tfrac{\pi}{2},\Hphi)^t.
\ee
We will now compute $\HO^\star\HO$. Since $\HO$ is a constant element of $\text{SU}(2)$, we have at $\theta=\pi/2$:
$$
\HO=\Yg(\Hphi+\pi)\Pg(\Hphi+\pi)\Yg(\Hphi)^t=\Yg(\Hphi+2\pi)\Pg(\Hphi+2\pi)\Yg(\Hphi+\pi)^t=(-1)^{\tHm}\Yg(\Hphi)\Pg(\Hphi)\Yg(\Hphi+\pi)^t,
$$
and therefore
\be\label{eqn:HOHOstHm}
\HO^\star\HO = (-1)^{\tHm}\Yg(\Hphi+\pi)^\star\Pg(\Hphi+\pi)^\star\Yg(\Hphi)^\dagger\Yg(\Hphi)\Pg(\Hphi)\Yg(\Hphi+\pi)^t=(-1)^{\tHm+\tHX}\Id.
\ee
Let $\AdjField(\theta,\Hphi)$ be any adjoint-valued field in the theory, which is required to satisfy twisted boundary conditions similar to \eqref{eqn:AbcRP2NonTrivialBundle}
\be\label{eqn:AdjFbcNonT}
\Pg(\Hphi+\pi)^{-1}\AdjField(\tfrac{\pi}{2},\Hphi+\pi)\Pg(\Hphi+\pi)=\AdjField(\tfrac{\pi}{2},\Hphi)^t,
\ee
and after the gauge transformation by $\Yg$, these are modified to
\be \nn
\Pg(\Hphi+\pi)^{-1}\Yg(\tfrac{\pi}{2},\Hphi+\pi)^{-1}\AdjField(\tfrac{\pi}{2},\Hphi+\pi)\Yg(\tfrac{\pi}{2},\Hphi+\pi)\Pg(\Hphi+\pi)=
\Yg(\tfrac{\pi}{2},\Hphi)^t\AdjField(\tfrac{\pi}{2},\Hphi)^t\Yg(\tfrac{\pi}{2},\Hphi)^\star,
\ee
which reduces to
\be\label{eqn:HOAphiX}
\HO^{-1}\AdjField(\tfrac{\pi}{2},\Hphi+\pi)\HO = \AdjField(\tfrac{\pi}{2},\Hphi)^t.
\ee
[Generally, when we allow a nontrivial action of orientation reversal on the fields, we get \eqref{eqn:HOAphi}.]

Overall, we have seen that a flat $\text{SU}(2)$ connection $A$ defines a constant $\HO\in\text{SU}(2)$ via \eqref{eqn:HOdef} [in terms of $\Yg$ defined in \eqref{eqn:AYgKicakNew}], and for gauge group $\text{SU}(2)/\Z_2$ we should project $\HO$ to $\text{SU}(2)/\Z_2$.
For a constant $\Lambda\in\text{SU}(2)$ [or $\text{SU}(2)/\Z_2$], $\Yg$ and $\Lambda\Yg$ define the same $A$ in \eqref{eqn:AYgKicakNew}, and so we can consider $\HO_1,\HO_2\in\text{SU}(2)$ as gauge equivalent, in this context, if
\be\label{eqn:HOequiv}
\HO_2=\Lambda^{-1}\HO_1\Lambda^\star.
\ee

We now show that, up to gauge equivalence, $\HO\in\text{SU}(2)$ uniquely defines a flat connection $A$ that satisfies the boundary conditions \eqref{eqn:AbcRP2TrivialBundleNew}, where uniqueness is up to the obvious fact that $\pm\HO$ define the same flat connection. What we need to show is that there exists a continuous $\Yg(\theta,\Hphi)$ that satisfies \eqref{eqn:HOdefNew} and for which \eqref{eqn:AYgKicakNew} satisfies the boundary conditions \eqref{eqn:AbcRP2TrivialBundleNew}. We set $\Yg(\pi/2,0)=\Id$ and $\Yg(\pi/2,\pi)=\HO$, which satisfies \eqref{eqn:HOdefNew} at $\phi=0$. We then choose an arbitrary $\Yg(\pi/2,\Hphi)$ in the range $0\le\Hphi\le\pi$ that equals $\Id$ at $\Hphi=0$ and equals $\HO$ at $\Hphi=\pi$, and we extend it to the range $\pi\le\Hphi\le2\pi$ by using \eqref{eqn:HOdefNew}, setting $\Yg(\pi/2,\Hphi)=\HO\Yg(\pi/2,\Hphi-\pi)^\star$.
Using \eqref{eqn:HOHOstNew}, at $\Hphi=2\pi$ we find that $\Yg(\pi/2,2\pi)=(-1)^{\tHm}\Id$. If $\tHm=0$ we can now extend $\Yg$ continuously to the entire disk $0\le\theta<\pi/2$, using the fact that $\text{SU}(2)$ is simply connected. And if $\tHm=1$ we can extend $\Yg$ to $0<\theta<\pi/2$ keeping the boundary condition $\Yg(\theta,2\pi)=-\Yg(\theta,0)$, which corresponds to a connection $A$ with nontrivial `t Hooft magnetic flux. Similarly, since $\text{SU}(2)$ is simply connected, $\Pg(\pi/2,\Hphi)$ that satisfies \eqref{eqn:Pgbc} with $\tHPg=0$, can be extended continuously to $0\le\theta<\pi/2$. We can then gauge transform all the fields by this continuously defined $\Pg$ on the disk. Since $\Pg$ also satisfies \eqref{eqn:PgPibc}, such a gauge transformation will eliminate $\Pg$ entirely from \eqref{eqn:AbcRP2NonTrivialBundle}.
Thus, the flat connections are characterized by the constant $\HO\in\text{SU}(2)$ up to gauge equivalence and $\HO\sim-\HO$.

\subsection{The 5d Chern-Simons Term}
\label{subsec:5dCS}

In general, the $A_1$ $(2,0)$-theory compactified on a circle fibration over a base space $Y_5$ which is a Riemannian manifold equipped with a curvature $\omega$ [which is a $2$-form on $Y_5$ so that the first Chern class is $c_1=[\omega]\in H^2(Y_5,\Z)$] reduces to $\mathrm{SU}(2)$ SYM on $Y_5$ with an extra Chern-Simons term proportional to \cite{Douglas:1995bn}
\be\label{eqn:CSY5New}
\frac{1}{4\pi}
\int_{Y_5}\omega\wedge\tr(A\wedge dA+\tfrac{2}{3}A\wedge A\wedge A).
\ee
As a quick warmup, we will investigate the form of this Chern-Simons term in the case of $\R\times S^2\times (S^3/\Z_{2k})$.
Since the lens space $S^3/\Z_{2k}\cong L(2k,1)$ is a circle fibration over $S^2$, we take the base manifold to be $Y_5=\R\times S^2\times S^2$, where the second $S^2$ factor is the base of the Hopf fibration $L(2k,1)\rightarrow S^2$.
Let $\AreaFormS$ be the area $2$-form on the second $S^2$ factor, normalized so that $\int_{S^2}\AreaFormS=1$. Thus $\omega=2k\AreaFormS$.

All this was for $S^3/\Z_{2k}$, but to get to $S^3/\text{Dic}_k$ we need to further quotient by $\Z_2$ that is generated by the inversion $s$ in \eqref{eqn:AnglesIdentNew}. This quotient does not introduce any uncontrollable fixed points and replaces $Y_5$ in \eqref{eqn:CSY5New} with $Y_5/\Z_2\cong\R\times S^2\times (S^2/\Z_2)$. We can identify $S^2/\Z_2$ with the disc $0\le\theta\le\pi/2$, so that the Chern-Simons term becomes
\be\label{eqn:CSY5Z2}
\frac{k}{2\pi}
\int_{Y_5/\Z_2}\AreaFormS\wedge\tr(A\wedge dA+\tfrac{2}{3}A\wedge A\wedge A)
\,.
\ee
Note that $\omega$ is invariant under the action of $s$, because the first Chern class of the fibration changes sign under orientation reversal of the fiber, and also $\AreaFormS$ changes sign under orientation reversal of the base. In \eqref{eqn:CSY5Z2}, we have slightly abused the notation, since the $2$-form $\AreaFormS$ cannot be integrated over the non-orientable $S^2/\Z_2\cong\RP^2$, and what we actually mean by the integral in \eqref{eqn:CSY5Z2} is an integral over $\R\times S^2$ times the disc $\theta<\pi/2$, so that $\int\AreaFormS$ over the disc evaluates to $1/2$.

Alternatively, the Chern-Simons term \eqref{eqn:CSY5Z2} arises from the effect of the shift by $-2k\Hphi$ in the action of $s$ on $\Hchi$ in \eqref{eqn:AnglesIdentNew}. In general, a translation along the Hopf fiber corresponds in 5d SYM to the symmetry associated with the instanton number as the conserved charge \cite{Douglas:1995bn,Vafa:1995bm,Seiberg:1996bd, Berkooz:1997cq, Kallen:2012cs}, and so when we construct $\RP^2$ from the disk $\theta<\pi/2$ by gluing the arc of the boundary with, say, $0\le\Hphi<\pi$ to the arc with $\pi\le\Hphi<2\pi$, we must insert a term
\be\label{eqn:dtheta}
\frac{k}{4\pi^2}\int_0^\pi d\Hphi\wedge\tr\left(A\wedge dA+\frac{2}{3}A\wedge A\wedge A\right),
\ee
along the cut.

\subsection{3d Chern-Simons and Gauge Bundles}
\label{subsec:3dCS}

We now shrink $S^2/\Z_2\cong\RP^2$ to zero and reduce the Chern-Simons term \eqref{eqn:dtheta} from 5d to 3d.
Let $A$ now denote only the 3d part of the gauge field, along $\R\times S^2$, and we assume that it is independent of the point in $\RP^2$ i.e. we keep the zero mode.
According to the discussion below \eqref{eqn:HOAphi}, this means that 
\be\label{eqn:A3dHO}
\HO^{-1}A\HO^\star = -A^t.
\ee
We will now substitute the solutions of $\HO$ from \secref{subsec:FCHO}.
For $\HO=\pm i\GGPauli_2$, every element of the $\mathfrak{su}(2)$ Lie algebra satisfies an equation similar to \eqref{eqn:A3dHO}, and therefore the 3d gauge algebra is the full Lie algebra $\mathfrak{su}(2)$.
For $\HO=\pm\Id$, on the other hand, the subalgebra of $\mathfrak{su}(2)$ that solves \eqref{eqn:A3dHO} is the C
artan subalgebra generated by $\GGPauli_2$, and the 3d gauge algebra is therefore $\mathfrak{u}(1)$.

The 5d term \eqref{eqn:dtheta} reduces to a 3d Chern-Simons action, 
\be\label{eqn:3dUnshifted}
\frac{k}{4\pi}\int\tr\left(A\wedge dA+\frac{2}{3}A\wedge A\wedge A\right),
\ee
where $\tr$ is in the $2$-dimensional representation. We will now proceed separately with the two cases $\HO=\pm i\GGPauli_2$ and $\HO=\pm\Id$, starting with the former.

The term \eqref{eqn:dtheta}, for $\HO=\pm i\GGPauli_2$, describes 3d $\mathfrak{su}(2)$ Chern-Simons at level $k$, but the level is shifted because of a well-known $1$-loop effect due to the fermionic degrees of freedom. Massive gluinos join the gauge fields of the level $k$ Chern-Simons term to form a supersymmetric Chern-Simons theory with $\NS=2$ SUSY. For $\mathfrak{su}(2)$ gauge group, after integrating out the gluinos at low-energy, the supersymmetric theory at level $k$ becomes pure Chern-Simons theory at level $(k-2)$ \cite{Niemi:1983rq, Redlich:1983kn, Redlich:1983dv, Kao:1995gf,Tong:2000ky,Gukov:2015sna}. The shift in level occurs because of a mismatch in the number of positive and negative eigenvalues of the Dirac operator. For completeness, we will show the calculation of the index of the Dirac operator explicitly in \appref{app:LevelShift}, which can be read after the fermions are discussed in \secref{subsec:5df}.

For the case $\HO=\pm\Id$, the gauge group is $\text{U}(1)$, and the $\tr$ in \eqref{eqn:3dUnshifted} gives an extra factor of $2$. This time there is no $1$-loop contribution to the level, and so the total level is $2k$. We will parameterize the low-energy 3d gauge field by
\be\label{eqn:AGdef}
A = \aG\GGPauli_2.
\ee
So far, we discussed the Lie algebras, and now we will discuss the Lie groups. Based on \eqref{eqn:HOequivNew}, the question is: for a given $\HO$ what is the group of $\Lambda\in\text{SU}(2)$ that satisfies
\be\label{eqn:LHOL}
\Lambda^{-1}\HO\Lambda^\star=\pm\HO,
\ee
where we included a $(\pm)$ to allow for $\text{SU}(2)/\Z_2$ gauge group?
For $\HO=\pm i\GGPauli_2$, every $\Lambda\in\text{SU}(2)$ satisfies $\Lambda^{-1}\HO\Lambda^\star=\HO$, and the 3d gauge group is the full $\text{SU}(2)$. (We will discuss below the `t Hooft electric and magnetic flux restrictions.)

For $\HO=\pm\Id$,
$$
\Lambda= \Lambda_1(\alpha)\equiv\Id\cos\alpha+i\GGPauli_2\sin\alpha,
$$
for any $\alpha$, satisfies \eqref{eqn:LHOL} with the $(+)$ sign, while for the $(-)$ sign in \eqref{eqn:LHOL} we have
\be\label{eqn:Lambda2}
\Lambda=\Lambda_2(\alpha)\equiv i\GGPauli_1\cos\alpha+i\sin\alpha\GGPauli_3.
\ee
If the 5d gauge group is $\text{SU}(2)$ then the 3d gauge group is the set of $\Lambda_1(\alpha)$'s, which is isomorphic to $\text{U}(1)$. In this case, the 3d reduction of the gauge sector is simply $\text{U}(1)_{2k}$ Chern-Simons theory. If the 5d gauge group is instead $\text{SU}(2)/\Z_2$, then the 3d gauge group is the union of the $\Lambda_1(\alpha)$'s and $\Lambda_2(\alpha)$'s, which is equivalent to the a $\Z_2$ quotient of the semidirect product $\Z_4\ltimes\text{U}(1)$. The $\Z_4$ factor in the last expression can be taken as the element $\Lambda_2(0)=i\GGPauli_1$. A constant gauge transformation $\Lambda=\Lambda_2(0)$ acts on the 3d $\text{U}(1)$ gauge field $\aG$ in \eqref{eqn:AGdef} as $\aG\rightarrow \Lambda^{-1}\aG\Lambda=-\aG$, and if $\text{U}(1)$ is taken to be a Cartan subgroup of $\text{SU}(2)$, then we can identify the action of $\Lambda_2(0)$ with the nontrivial element of the Weyl group. An additional $\Z_2$ quotient is required to account for the relation $\Lambda_2(0)^2=\Lambda_1(\pi)$, and the resulting gauge group is therefore $[\Z_4\ltimes\text{U}(1)]/\Z_2\cong\text{Pin}^{-}(2)$.

What is left to discuss is which combination of electric and magnetic fluxes are allowed.
For this purpose it is useful to think about the 6d space $S^1\times S^2\times S^3/\text{Dic}_k$ as a $T^2$ fibered over $S^2\times\mathbb{RP}^2$, where $T^2\cong S^1_t\times S^1_H$ is the product of the time circle that we now temporarily denote by $S^1_t$, and the Hopf fiber that we temporarily denote by $S^1_H$. Then, the $(2,0)$-theory reduces locally to 4d $\NS=4$ SYM, and the difference between the dimensional reduction of \secref{sec:HilbNoS} (where the analytic torsion was used) and the dimensional reduction of the present section is in the order in which we reduce on $S^1_t$ and $S^1_H$ -- in \secref{sec:HilbNoS} we reduced on $S^1_t$ first, and now $S^1_H$ is first. In terms of gauge groups, the difference is therefore a Langlands duality, and if the gauge group was $\text{SU}(2)$ in \secref{sec:HilbNoS}, it should be $\text{SU}(2)/\Z_2$ presently. This is indeed the case for even $k$, and therefore for even $k$ the 3d gauge groups are $\text{SU}(2)/\Z_2\cong\text{SO}(3)$ for $\HO=\pm i\GGPauli_2$ and $\text{Pin}^{-}(2)/\Z_2\cong\Z_2\ltimes\text{U}(1)\cong\text{O}(2)$ for $\HO=\pm\Id$. 

For odd $k$, there is an additional complication. SO$(3)$ Chern-Simons theory is not well-defined for odd $k$, and indeed the geometry of the dicyclic orbifold requires a more delicate combination of magnetic fluxes. From \eqref{eqn:H2Z2} we see that $H^2(S^3/\text{Dic}_k,\Z_2)$ has one less generator for odd $k$. Whereas, it is $\Z_2^2$ for even $k$, it is only $\Z_2$ for odd $k$. The absence of an additional $\Z_2$ factor for odd $k$ means that we cannot define nontrivial `t Hooft magnetic flux on $\mathbb{RP}^2$ in the $S^1_t$-first approach. To see this, note that for odd $k$ the generator of $H^2(S^3/\text{Dic}_k,\Z_2)\cong H_1(S^3/\text{Dic}_k,\Z_2)$ corresponds to the nontrivial $1$-cycle associated with $s$, while unlike the case of even $k$, there is no generator in $H_1(S^3/\text{Dic}_k,\Z_2)$ that corresponds to the $1$-cycle associated with $r$.
In string theory, replacing $S^2$ with $\R^{1,1}$ for the argument, we can take two D$4$-branes on $\R^{1,1}\times(S^3/\text{Dic}_k)$, and then magnetic flux along $\mathbb{RP}^2$ would correspond to a D$2$-brane charge, with the D$2$-brane wrapped around the Hopf fiber $S^1_H$ (and along $\R\subset\R^{1,1}$), but the Hopf fiber is trivial in $H_1(S^3/\text{Dic}_k,\Z_2)$ for odd $k$.
Thus, for odd $k$, in the $S^1_t$-first approach of \secref{sec:HilbNoS}, we allow `t Hooft magnetic flux along $S^2$ but not on $\mathbb{RP}^2$. In the dual $S^1_H$-first approach of the present section we therefore allow magnetic flux along $\mathbb{RP}^2$, but not along $S^2$, for odd $k$. This means that we can consider solutions $\HO=\pm i\GGPauli_2$ for $\HO\HO^\star=-\Id$ [\eqref{eqn:HOHOstNew} with nontrivial flux], which we identify within $\text{SU}(2)/\Z_2$, but since for odd $k$ we do not allow magnetic flux along $S^2$ the gauge group is $\text{SU}(2)$. In summary, when $\HO=\pm i\GGPauli_2$, for even $k$ we have SU$(2)/\Z_2$ gauge group and for odd $k$ the gauge group is SU$(2)$.

For $\HO=\pm\Id$, when we plug \eqref{eqn:AGdef} into \eqref{eqn:3dUnshifted}, the action becomes
$$
S_\aG= \frac{k}{2\pi}\int\aG\wedge d\aG,
$$
which is that of $\text{U}(1)_{2k}$ for $\aG$, as mentioned above. But if our 5d gauge group is $\text{SU}(2)/\Z_2$, and not $\text{SU}(2)$, the periodicity of a gauge parameter $\lambda$ in a gauge transformation $\aG\rightarrow\aG+d\lambda$ is $\pi$ and not the standard $2\pi$. This can be fixed by rescaling the gauge field and define
$$
\aGp=2\aG.
$$
The action is now
$$
S_{\aGp}= \frac{k/2}{4\pi}\int\aGp\wedge d\aGp,
$$
which is that of $\text{U}(1)_{k/2}$, and in particular requires $k$ to be even. However, we will see in \secref{subsec:Altogether} that because of an unusual fermion number $(-)^F$ assignment to the sector with nontrivial `t Hooft magnetic flux along $S^2$, the effective theory on $S^2\times S^1$ is not quite $\text{U}(1)_{k/2}$ but rather a combination of sectors of $\text{U}(1)_k$ with `t Hooft magnetic flux associated with $\Z_2\subset\text{U}(1)$.
For convenience, we can also rotate the embedding in $\text{SU}(2)$ to be $A=i\GGPauli_3\aG$, so that the $\text{SU}(2)$ gauge field is
\be\label{eqn:AaGp}
A = \begin{pmatrix} \aG & 0 \\ 0 & -\aG \\ \end{pmatrix}.
\ee
If $k$ is odd, we forbid magnetic flux on the $S^2$ factor, and so effectively the 3d Chern-Simons level remains $2k$.

Altogether, for even $k$ we have two sectors to consider with 3d Chern-Simons gauge groups and levels $\text{SU}(2)_{k-2}/\Z_2$ and $\text{U}(1)_{k}$. For odd $k$, we also have two sectors, one with gauge group $\text{SU}(2)_{k-2}$ and a restriction of no magnetic flux on $S^2$, and the other is $\text{Pin}^{-}(2)$, with a similar restriction on magnetic flux on $S^2$.

\subsection{Reduction of Surface Operators to 3d Wilson Loops}
\label{subsec:3dSOtoW}

Our next task is to reduce the 6d Surface operators introduced in \secref{subsec:operators} in $S^1\times S^2\times S^3/\text{Dic}_k$ to 5d SYM on $S^1\times S^2\times\mathbb{RP}^2$. Since our goal is to reproduce \eqref{eqn:IndexknX}, for which the surface operators are all along $S^1$ times the $r^2$ cycle of $S^3/\text{Dic}_k$, we will restrict to this topology. 
As we are now dimensionally reducing along the $r$-cycle, which is the homotopy class of the fiber of the circle fibration $S^3/\text{Dic}_k\rightarrow\RP^2$ (up to conjugation), our surface operators should reduce to Wilson loops along $S^1$. In fact, they should compute the trace of the square of the gauge holonomy around $S^1$. Concretely, let $M\in\text{SU}(2)$ correspond to the holonomy of the Chern-Simons gauge field around $S^1$ (the ``time'' direction). As $M$ is an $\text{SU}(2)$ matrix, we have
\be\label{eqn:trM2}
\tr(M^2)=(\tr M)^2-2.
\ee
In \eqref{eqn:trM2}, $\tr(M^2)$ represents a Wilson loop that is wound twice around the ``time'' circle $S^1$, and to compare with \eqref{eqn:IndexknX}, we need to compute the correlation function of $\nC$ such Wilson loops.

 If we are in the $\HO=\pm\Id$ sector, then insertion of $\nC$ surface operators reduces to an insertion of
\begin{align}
\label{eqn:CosInIndex}
    \left(2\cos 2\oint\aG\right)^\nC,
\end{align}
in the path integral, where we have used \eqref{eqn:AaGp}, and set $M=\exp(i\oint A)$ in \eqref{eqn:trM2}. The integrals $\oint$ are along the ``time'' circle $S^1$.

\subsection{5d Fermions}
\label{subsec:5df}

Next, we study the fermionic sector of the effective 3d field theory on $S^1\times S^2$ in two steps.
In the present subsection we develop the 5d effective action, and in \secref{subsec:3df} we will reduce the action to 3d on $\mathbb{RP}^2$ with either $\HO=\pm i\GGPauli_2$ or $\HO=\pm\Id$. The only effect of the fermionic fields that will be relevant for us is (i) a shift of the Chern-Simons level for $\HO=\pm i\GGPauli_2$, and (ii) a nontrivial charge of the ground states of the fermionic sector that has the combined effect of an additional insertion of a Wilson loop operator for $\HO=\pm\Id$.

Let us first reduce a free chiral 6d spinor $\Psi$ on $\R\times S^2\times S^3/\text{Dic}_k$ to 5d, on the Hopf fiber. 
The R-symmetry twist, as in \secref{sec:physics}, is a bit unusual in this formulation, because the space is six-dimensional while the R-symmetry is only Spin$(5)$. We therefore single out the $\R$ (time) direction, and only perform a twist in the spatial directions, i.e., along $S^2\times S^3/\text{Dic}_k$. The spinor is chiral in 6d, and we would first like to derive the resulting Dirac equation in 5d on $S^1\times S^2\times\RP^2$.

Let $\DiracVI^I$ ($I=0,\dots,5$) be the 6d Dirac matrices, and let $\DiracV^\mu$ ($\mu=1,\dots,5$) be 5d Dirac matrices, excluding the timelike $S^1$, which does not participate in the R-symmetry twist.
The 6d spinor satisfies the chirality condition $\DiracVI^{012345}\Psi=\Psi$, and the 5d Dirac matrices act on $\Psi$ as $\DiracV^\mu\Psi=\DiracVI^{0\mu}\Psi$ and satisfy $\DiracV^{12345}=1$. Now we twist so that the R-symmetry background gauge field will match the spin connection in directions $12345$, so the component $\Psi_\alpha^a$, with $\alpha$ a chiral spinor index and $a$ an R-symmetry index, becomes $\Psi_\alpha^\beta$, which can be interpreted as a $4\times 4$ matrix in the same basis as the $\DiracV^\mu$'s, so we can expand
$$
\Psi = \Xrho\Id +\Xpsi_\mu\DiracV^\mu+\Xchi_{\mu\nu}\DiracV^{\mu\nu},
$$
where $\Xrho$, $\Xpsi_\mu$, and $\Xchi_{\mu\nu}$ are scalar, vector, and tensor 6d fermionic fields, but their components $\mu,\nu=1,\dots,5$ are missing the time direction.
Note that in 5d, $\DiracV^{\mu\nu\sigma}=-\tfrac{1}{2}\epsilon^{\mu\nu\sigma\alpha\beta}\DiracV_{\alpha\beta}$, so there is no need for a $3$-form component in $\Psi$.
We translate the 6d massless Dirac equation $0=\partial_0\Psi+\DiracV^\mu\partial_\mu\Psi$ to the fields $\Xrho,\Xpsi_\mu,\Xchi_{\mu\nu}$ and get the Dirac-K\"ahler equations
\bear
\partial_0\Xrho &=& -\partial^\mu\Xpsi_\mu
\,,\label{eqn:Dirac5DXrho}\\
\partial_0\Xpsi_\mu &=& -\partial_\mu\Xrho -2\partial^\alpha\Xchi_{\alpha\mu}
\,,\label{eqn:Dirac5DXpsi}\\
\partial_0\Xchi_{\mu\nu} &=& 
-\frac{1}{2}{\epsilon^{\alpha\beta\gamma}}_{\mu\nu}\partial_\alpha\Xchi_{\beta\gamma}
-\partial_{[\mu}\Xpsi_{\nu]}
\,.\label{eqn:Dirac5DXchi}
\eear
These arise from the Lagrangian density
\begin{align}
\mathcal{L} &= 
\tfrac{1}{2}\Xrho\partial_0\Xrho
+\tfrac{1}{2}\Xpsi^\mu\partial_0\Xpsi_\mu
+\Xchi^{\mu\nu}\partial_0\Xchi_{\mu\nu}
+\Xrho\partial^\mu\Xpsi_\mu
+\tfrac{1}{2}\epsilon^{\alpha\beta\gamma\mu\nu}\Xchi_{\mu\nu}\partial_\alpha\Xchi_{\beta\gamma}
+2\Xchi^{\mu\nu}\partial_{\mu}\Xpsi_{\nu}
\,.\label{eqn:LDen}
\end{align}
On a curved space, we can simply replace the derivatives with covariant derivatives and one can check  that there is no 5d covariant ways to couple the fields quadratically to the curvature tensor and hence no additional curvature terms are introduced in the Lagrangian density.
 
Now reduce on the Hopf circle, i.e., the fiber of the Hopf fibration $S^3/\text{Dic}_k\rightarrow\mathbb{RP}^2$, with $\Hchi$ denoting the coordinate on the fiber circle, as in \secref{subsec:FGeometry}. Then, the 5d fields are
$$
\Xrho, \Xpsi_i, \Xpsi_\Hchi, \Xchi_{ij}, \Xchi_{i\Hchi},
$$
with the index $i$ running over the coordinates of $S^2\times\RP^2$, and we assume that the fields are independent of the coordinate $\Hchi$ along the Hopf circle. We now shrink the Hopf fiber by introducing a radius parameter $\HopfRadius$ in \eqref{eqn:S3metric}, which we will take to be small.
The metric is now
$$
d s^2= d\theta^2+\sin^2\theta d\phi^2 + \frac{\HopfRadius^2}{k^2}\left[d\Hchi+2k\sin^2\left(\tfrac{\theta}{2}\right) d\phi\right]^2.
$$
We denote by $h_{ij}$ the 4d metric on $S^2\times\RP^2$:
$$
ds^2 = h_{ij}dx^i dx^j = d\Omega_2^2 + d \theta^2+\sin ^2 \theta d \phi^2,
$$
with $d\Omega_2^2$ being the metric on $S^2$.
The 5d Lagrangian is then
\begin{align}
\label{eqn:5dActionExp}
\mathcal{L} &=
\tfrac{1}{2}\Xrho\partial_0\Xrho
+\tfrac{1}{2}\Xpsi^i\partial_0\Xpsi_i
+\frac{k^2}{2}\left[\frac{1}{\HopfRadius^2}+\tan^2\left(\tfrac{\theta}{2}\right)\right]\Xpsi_\Hchi\partial_0\Xpsi_\Hchi
-\frac{k}{2}\sec^2\left(\tfrac{\theta}{2}\right)\Xpsi_\Hchi\partial_0\Xpsi_\phi \\ \nn
&+\Xchi^{ij}\partial_0\Xchi_{ij}
+k^2\left[\frac{1}{\HopfRadius^2}+\tan^2\left(\tfrac{\theta}{2}\right)\right]{\Xchi_\Hchi}{}^j\partial_0\Xchi_{\Hchi j}
-k\sec^2\left(\tfrac{\theta}{2}\right){\Xchi_\Hchi}{}^j\partial_0\Xchi_{\phi j}
+\Xrho\nabla^i\Xpsi_i \\ \nn
&+\frac{2k}{\HopfRadius\sqrt{h}}\epsilon^{ijkl}\Xchi_{\Hchi i}\partial_j\Xchi_{kl}
+2\Xchi^{ij}\partial_{i}\Xpsi_{j}
+2k^2\left\lbrack\frac{1}{\HopfRadius^2}+\tan^2\left(\tfrac{\theta}{2}\right)\right]{\Xchi_\Hchi}{}^i\partial_i\Xpsi_\Hchi \\ \nn
&
-k\sec^2\left(\tfrac{\theta}{2}\right){\Xchi_\Hphi}{}^i\partial_i\Xpsi_\Hchi-k\sec^2\left(\tfrac{\theta}{2}\right){\Xchi_\Hchi}{}^i\partial_i\Xpsi_\Hphi
\,. 
\end{align}

On the fermionic fields we need to impose the boundary conditions \eqref{eqn:HOAphi}. 
Let $p$ and $\antip$ be two antipodal points on the circle $\theta=\pi/2$, with $\Hphi$ coordinates that differ by $\pi$. The boundary conditions then take the form $\HO^{-1}\AdjField(p)\HO=\pm\AdjField^t(\antip)$, where the $\pm$ sign depends on whether the component field has an even or odd total number of $\Hchi$ and $\theta$ indices.
Thus
\begin{align}
\label{eqn:bcwithminus}
\HO^{-1}\Xchi_{\Hchi i}(p)\HO &= -\Xchi_{\Hchi i}^t(\antip),
\quad
\HO^{-1}\Xchi_{\theta i}(p)\HO = -\Xchi_{\theta i}^t(\antip),
\\ \nn
\HO^{-1}\Xpsi_{\Hchi}(p)\HO &= -\Xpsi_{\Hchi}^t(\antip),
\quad
\HO^{-1}\Xpsi_{\theta}(p)\HO = -\Xpsi_{\theta}^t(\antip),
\qquad
(i=1,2,\Hphi),
\end{align}
and
\begin{align}\label{eqn:bcwithplus}
\HO^{-1}\Xchi_{ij}(p)\HO &= \Xchi_{ij}^t(\antip),
\quad
\HO^{-1}\Xchi_{\Hchi\theta}(p)\HO = \Xchi_{\Hchi\theta}^t(\antip), \\ \nn
\HO^{-1}\Xpsi_{i}(p)\HO &= \Xpsi_{i}^t(\antip),
\quad
\HO^{-1}\Xrho(p)\HO = \Xrho^t(\antip), 
\qquad
(i,j=1,2,\Hphi),
\end{align}
where $i,j=1,2$ are the $S^2$ directions.

\subsection{3d Fermions}
\label{subsec:3df}

We now reduce from 5d to 3d by shrinking $\mathbb{RP}^2$, and we need to separate the discussion of $\HO=\pm i\GGPauli_2$ [with 3d gauge group $\text{SU}(2)$ or $\text{SU}(2)/\Z_2$] from the discussion of $\HO=\pm\Id$ [with 3d gauge group $\text{Pin}^{-}(2)$ or $\text{O}(2)$].

For $\HO=\pm i\GGPauli_2$, it can be shown that the equations of motion of the Lagrangian \eqref{eqn:5dActionExp} with boundary conditions \eqref{eqn:bcwithminus}-\eqref{eqn:bcwithplus} have no zero modes, but there is a mismatch between Kaluza-Klein modes with positive and negative masses that shift the Chern-Simons level from $k$ to $(k-2)$. The details are left for \appref{app:LevelShift}.

For $\HO=\pm\Id$, there is no level shift, but there is a pair of fermionic zero modes. The Witten index, however, does not vanish because the modes interact with the Chern-Simons gauge fields, as we will see below.
We begin by finding the modes of the fermionic fields on $S^1\times S^2\times\RP^2$, with boundary conditions \eqref{eqn:bcwithminus}-\eqref{eqn:bcwithplus}, which survive the low-energy reduction to $S^1\times S^2$.
We separate the components to those proportional to $\GGPauli_2$ and those proportional to a linear combination of $\GGPauli_1$ and $\GGPauli_3$. 
Applied to the field $\Xrho$, for example, the components proportional to $\GGPauli_2$ are antisymmetric matrices and satisfy the boundary conditions $\Xrho(p)=-\Xrho(\antip)$. They, therefore, do not survive the reduction to $S^1\times S^2$. The components that do survive are proportional to a linear combination of $\GGPauli_1$ and $\GGPauli_3$, which does not commute with the 3d $\text{U}(1)$ gauge field (which is proportional to $\GGPauli_2$). The field $\Xrho$ therefore reduces to a pair of charged fermionic scalars of charges $\pm 2$. Similarly, the components of $\Xchi_{ij}$ and the components of $\Xpsi_i$ with $i,j$ along $S^2$ also reduce to charged fermions. The remaining components $\Xchi_{\Hchi i}$, $\Xchi_{\theta i}$, $\Xchi_{\Hchi\theta}$, $\Xchi_{\Hphi i}$, $\Xpsi_\Hchi$, $\Xpsi_\Hphi$, $\Xpsi_\theta$ can be lifted to components of $1$-forms or $2$-forms on $S^3/\Z_{2k}$ and are therefore massive after the Kaluza-Klein reduction to $S^1\times S^2$, whether they are symmetric or antisymmetric matrices.
Thus, after the Kaluza-Klein reduction, what remains of \eqref{eqn:5dActionExp} is
\be\label{eqn:5dActionExpKK3d}
\mathcal{L}_{3d} =
\tfrac{1}{2}\Xrho D_0\Xrho
+\tfrac{1}{2}\Xpsi^i D_0\Xpsi_i
+\Xchi^{ij}D_0\Xchi_{ij}
+\Xrho D^i\Xpsi_i
+2\Xchi^{ij}D_{i}\Xpsi_{j}
\,,
\ee
where $i,j=1,2$ are along $S^2$, and where $D_0, D_i$ are components of the covariant derivative with respect to the $\text{U}(1)$ gauge field. Reducing further to zero modes on $S^2$, and assuming the gauge bundle is trivial on $S^2$, we find that only the zero modes of $\Xrho$ and $\Xchi_{12}$ survive. We denote them by $\qXrho_{\pm\pm}$ and $\qXchi_{\pm\pm}$.
Consider first the case $\nC=0$, i.e., without the surface operator insertions.
The relevant Lagrangian reduces to
$$
\mathcal{L}_{1d} =
\qXrho_{++} D_0\qXrho_{--}
+\qXchi_{++}D_0\qXchi_{--}
\,,
$$
on $S^1$.
Quantizing either the $\qXrho_{\pm\pm}$ system or the $\qXchi_{\pm\pm}$ system gives a pair of ground states $\ket{\pm}$ of opposite fermion numbers and opposite charges, and the total system has four states
\be\label{eqn:kets}
\ket{++},\, \ket{+-},\, \ket{-+},\, \ket{--},
\ee
but only $\ket{\pm\mp}$ are neutral, and hence gauge invariant. The fermion number $(-)^F$ is opposite for $\ket{\pm\pm}$ and $\ket{\pm\mp}$, and for consistency with the results of \secref{sec:HilbNoS}, we must assign a $(-1)^\tHmS$ eigenvalue for $\ket{\pm\mp}$, where $\tHmS$ is the `t Hooft magnetic flux on $S^2$, as we will see below in \secref{subsec:Altogether}.

But [for 5d gauge group $\text{SU}(2)/\Z_2$] after the compactification, the 3d gauge group is $\Z_2\ltimes\text{U}(1)$, and we need to mod out by the $\Z_2$ (Weyl group) as well. The Weyl group is generated by $\Lambda_2(0)=i\GGPauli_1$ in \eqref{eqn:Lambda2}.
The Weyl group acts as $\ket{+}\leftrightarrow\ket{-}$, leaving only one Weyl-invariant $\text{U}(1)$-neutral state, namely $\frac{1}{\sqrt{2}}(\ket{+-}+\ket{-+})$. The contribution of $\HO=\Id$ to the Witten index for $\nC=0$ is therefore $1$, because for $\nC=0$ the $\text{U}(1)$ Chern-Simons sector has only one state with $\tHmS=0$.

Adding the single ground state of the $\HO=\pm i\GGPauli_2$ sector, which is the single state of $\text{SU}(2)_{k-2}/\Z_2$ on $S^2$, we find the total Witten index of $2$, in agreement with \eqref{eqn:DimHBasic} (for $k>2$).

Now, we discuss the case of general $\nC$. Let $\aG$ be the $\text{U}(1)$ gauge field, as in \eqref{eqn:AGdef}. In the sector of the Hilbert space where the fermionic system is in the $\ket{++}$ state of \eqref{eqn:kets}, the charge $+2$ introduces an additional Wilson loop $\exp\left(2i\oint\aG\right)$ along the time direction $S^1$, and since $\ket{\pm\pm}$ are fermionic states, while $\ket{\pm\mp}$ are bosonic, the fermions can be ``integrated out'' and replaced with the combination
\be\label{eqn:WilsonsU1SU2}
2-e^{2i\oint\aG}-e^{-2i\oint\aG}=4\sin^2\oint\aG,
\ee
of Wilson lines along $S^1$. In \eqref{eqn:WilsonsU1SU2} we have still not projected to the Weyl invariant subspace. The space has a basis:
$$
\frac{1}{\sqrt{2}}(\ket{+-}+\ket{-+}),
\qquad
\frac{1}{\sqrt{2}}(\ket{++}+\ket{--}).
$$
Inserting the $\text{SU}(2)$ Wilson loop associated to the gauge field \eqref{eqn:AaGp} we find that the fermions can be integrated out to give half of \eqref{eqn:WilsonsU1SU2}, namely
\be\label{eqn:WilsonsU1SU2h}
2\sin^2\oint\aG.
\ee

We will also need to consider cases where the gauge bundle on $S^2$ is not trivial, originating from a nontrivial `t Hooft magnetic flux for $\text{SU}(2)/\Z_2$ gauge group. In this case the zero modes that survive the dimensional reduction of \eqref{eqn:5dActionExpKK3d} on $S^2$ are not $\Xrho$ and $\Xchi_{ij}$ but rather belong to the charged components of $\Xpsi_{j}$. To see this, we note that the covariant derivative of a gauge field with one unit of magnetic flux acts on charged components with charge $\pm 2$ as the spin connection of $S^2$ acts on vector fields. Thus, $\Xrho$ becomes a $1$-form, with no zero modes on $S^2$, and similarly $\Xchi_{ij}$ becomes a tensor of rank $3$. On the other hand,  $\Xpsi_j$ becomes a tensor with two zero modes associated with components that are proportional to $\delta_{ij}$ and to $\epsilon_{ij}$ on $S^2$. Overall, we end up with the same system as in \eqref{eqn:kets}, which is captured by the insertion of \eqref{eqn:WilsonsU1SU2h} in the Chern-Simons effective action on $S^2\times S^1$.

\subsection{The Total Index}
\label{subsec:Altogether}

We now have all the ingredients to compute the total Witten index. The total is the sum of two sectors, each described by Chern-Simons theory on $S^1\times S^2$ with $\nC$ insertions of linear combinations of Wilson lines along $S^1$. The details are different for even and odd $k$ and we calculate the index separately for each case below. For clarity, the results of these calculations are summarized in \tabref{tab:Index}, which reader can use to compare with the analytic torsion results.

\begin{table}[ht]
    \centering
    \resizebox{\textwidth}{!}{
    \begin{tabular}{|c|c|c|c|c|c|c|} \hline
         & \multicolumn{2}{|c|}{Torsion side} & \multicolumn{2}{|c|}{CS $\Omega=\pm i\GGPauli_2$} & \multicolumn{2}{|c|}{CS $\Omega=\pm\Id$}  \\ \hline 
         & Gauge group & Index & Gauge group & Index & Gauge group & Index \\ \hline 
         odd $k$ & SU$(2)$ & $I_{k,n}$ \eqref{eqn:IndexknX} & SU$(2)_{k-2}$ & $\frac{1}{2}I_{k,n}$ \eqref{eqn:IndexknXCSoddk} & $\text{Pin}^{-}(2)_{2k}$ & $\frac{1}{2}I_{k,n}$ \eqref{eqn:IndexKoddSector2}  \\ \hline 
         even $k$ & SU$(2)$ & $I_{k,n}$ \eqref{eqn:IndexknX} & SU$(2)_{k-2}/\Z_2$ & $\frac{1}{2}I_{k,n}$ \eqref{eqn:IndexknXCS} & U$(1)_{k}$ & $\frac{1}{2}I_{k,n}$ \eqref{eqn:CSIndexEvenk} \\ \hline 
    \end{tabular}
    }
    \caption{Summary of the index calculation from both sides of the duality.}
    \label{tab:Index}
\end{table}

\subsubsection*{Odd $k$}

The $\HO=\pm i\GGPauli_2$ sector reduces to $\text{SU}(2)_{k-2}$ Chern-Simons theory on $S^1\times S^2$. The dimension of the Hilbert space of Chern-Simons theory at level $(k-2)$ with $\tilde{\nC}$ quarks in the fundamental representation is given by \cite{Witten:1988hf}:
\be\label{eqn:CSkminus2}
\dim\Hilb^{\text{SU}(2)}_{k-2,\tilde{\nC}}=
\frac{2^{\nC+1}}{k}\sum_{m=1}^{k-1}
\cos^{\tilde{\nC}}\left(\frac{m\pi}{k}\right)
\sin^2\left(\frac{m\pi}{k}\right).
\ee
But as discussed in \secref{subsec:3dSOtoW}, we do not insert simple $\nC$ Wilson loops in fundamental representations, but rather Wilson
loops that wind twice around $S^1$, and this modification of \eqref{eqn:CSkminus2} can be achieved, according to \eqref{eqn:trM2}, by replacing
\begin{align}
\label{eqn:ModWilsonLine}
2\cos\left(\frac{m\pi}{k}\right)\longrightarrow
\left\lbrack 2\cos\left(\frac{m\pi}{k}\right)\right\rbrack^2-2
=2\cos\left(\frac{2\pi m}{k}\right),
\end{align}
leading to (on the Chern-Simons side we denote the Witten index by $\text{I}$ to distinguish it from the analytic torsion one $I$)
\begin{align}
\nn
\text{I}_{k,\nC}^{\text{SU}(2)}&=
\frac{2^{\nC+1}}{k}\sum_{m=0}^{m\le k-1}\cos^\nC\left(\frac{2\pi m}{k}\right)
\sin^2\left(\frac{\pi m}{k}\right) \\ \label{eqn:IndexknXCSoddk}
&=
\frac{2^{\nC+2}}{k}\sum_{m=1,3,5,\dots}^{m\le k-2}\cos^\nC\left(\frac{2\pi m}{k}\right)
\sin^2\left(\frac{\pi m}{k}\right)=\frac{1}{2}I_{k,\nC},
\end{align}
where in the second line we used the fact that the summand is invariant under $m\rightarrow k-m$, so for odd $k$ one can replace the sum over all $m$ with the doubled sum over odd $m$ only. This way we obtain exactly a half of the index computed using analytic torsion \eqref{eqn:IndexknX}.

To compute the contribution of the $\HO=\pm\Id$ sector, we first need to insert \eqref{eqn:CosInIndex}:
$
\left( 2\cos 2\oint\aG\right)^\nC,
$
in the $\text{U}(1)_{2k}$ Chern-Simons path integral, where $\oint\aG$ is the $\text{U}(1)$ Wilson loop along $S^1$, and the factor of $2$ was explained in \eqref{eqn:trM2}.
We also insert the contribution of the fermionic zero modes \eqref{eqn:WilsonsU1SU2h}, and altogether we have to calculate
\be\label{eqn:U1aGCorrelator}
\left\langle\left( 2\cos 2\oint\aG\right)^\nC\left(2\sin^2\oint\aG\right)\right\rangle
=\left\langle\left(2\cos 2\oint\aG\right)^\nC\right\rangle
-\left\langle\left(2\cos 2\oint\aG\right)^{\nC+1}\right\rangle.
\ee
This is easily done as in \cite{Witten:1988hf} by decomposing $S^2\times S^1$ as two solid tori, each corresponding to a product of $S^1$ and one of the two hemispheres of $S^2$. On the boundary of the solid torus we choose a basis of states of Chern-Simons theory for which $\exp\left(i\oint\aG\right)$ is diagonal, with eigenvalues $\exp(\pi i p/k)$, and $p=0,\dots,2k-1$.
The result of the contribution of this $\text{Pin}^{-}(2)$ sector to the index is
\bear
\label{eqn:IndexKoddSector2}
\text{I}_{k,\nC}^{\text{Pin}^{-}(2)} &=&
2^{\nC}\left\langle\left(\cos 2\oint\aG\right)^\nC\right\rangle_{\text{U}(1)_{2k}}
-2^{\nC}\left\langle\left(\cos 2\oint\aG\right)^{\nC+1}\right\rangle_{\text{U}(1)_{2k}}
\nn\\
&=&
\frac{2^{\nC-1}}{k}\sum_{p=0}^{2k-1}\cos^\nC\left(\frac{2\pi p}{k}\right)
-\frac{2^{\nC-1}}{k}\sum_{p=0}^{2k-1}\cos^{\nC+1}\left(\frac{2\pi p}{k}\right)
\nn\\
&=&
\frac{2^{\nC+1}}{k}\sum_{p\; \text{odd}}^{k-2}\cos^\nC\left(\frac{2\pi p}{k}\right)
-\frac{2^{\nC+1}}{k}\sum_{p\; \text{odd}}^{k-2}\cos^{\nC+1}\left(\frac{2\pi p}{k}\right)
\nn\\
&=&
\sum_{l=0}^\nC\binom{\nC}{l}\delta_k(2(\nC-2l))
-\frac{1}{2}\sum_{l=0}^{\nC+1}\binom{\nC+1}{l}\delta_k(2(\nC+1-2l))
\,,
\eear
where in the second and third lines we used the fact that $p$, $p+k$, $k-p$, and $2k-p$ give the same summand.
For odd $k$ we have $\delta_k(2(\nC-2l))=\delta_k(\nC-2l)$, and so the last formula is exactly $1/2$ of \eqref{eqn:AltIndexOddk}.
Altogether, the contribution of both sectors \eqref{eqn:IndexknXCSoddk} and \eqref{eqn:IndexKoddSector2}  recovers the Witten index that was calculated by the analytic torsion approach for even $k$.

\subsubsection*{Even $k$}

The $\HO=\pm i\GGPauli_2$ sector reduces to $\text{SU}(2)_{k-2}/\Z_2$ Chern-Simons theory on $S^1\times S^2$. The dimension of the Hilbert is again given by \eqref{eqn:CSkminus2}. This formula is for a gauge group SU$(2)$, but the modification for SU$(2)/\Z_2$ is simple. We must add the contributions of both values $\mF=0,1$ of the `t Hooft magnetic flux on the space $S^2$. 
The partition function with trivial magnetic flux, $\mF=0$, is as in \eqref{eqn:CSkminus2}, while the partition function with $\mF=1$ can be computed by decomposing $S^2$ into two hemispheres, and is given by
\be\label{eqn:CSkminus2MagFlux}
\dim\Hilb^{\text{SU}(2)}_{k-2,\tilde{\nC},\mF=1}=
\frac{2^{\nC+1}}{k}\sum_{m=1}^{k-1}(-1)^{m-1}
\cos^{\tilde{\nC}}\left(\frac{m\pi}{k}\right)
\sin^2\left(\frac{m\pi}{k}\right).
\ee
The sum of the contributions of $\mF=0$ and $\mF=1$ becomes $\sum_m[1-(-1)^m](\cdots)=2\sum_{m=1,3,\dots}(\cdots)$, and we get
\be\label{eqn:CSkminus2All}
\dim\Hilb^{\text{SU}(2)/\Z_2}_{k-2,\tilde{\nC}}=
\frac{2^{\nC+2}}{k}\sum_{m=1,3,5,\dots}^{k-1}
\cos^{\tilde{\nC}}\left(\frac{m\pi}{k}\right)
\sin^2\left(\frac{m\pi}{k}\right).
\ee
Using again the modification \eqref{eqn:ModWilsonLine} for the contribution of Wilson lines, we obtain
\be\label{eqn:IndexknXCS}
\text{I}_{k,\nC}^{\text{SU}(2)/\Z_2}=
\frac{2^{\nC+2}}{k}\sum_{m=1,3,5,\dots}^{m\le k-1}\cos^\nC\left(\frac{2\pi m}{k}\right)
\sin^2\left(\frac{\pi m}{k}\right)=\frac{1}{2}I_{k,\nC}.
\ee 

For $\HO=\pm\Id$ and gauge group $\mathrm{O}(2)$, we again insert the combination of Wilson loop operators \eqref{eqn:U1aGCorrelator} into the partition function of $\text{U}(1)_{2k}$ on $S^1\times S^2$, but we now have two sectors with `t Hooft magnetic flux $\tHmS=0$ and $\tHmS=1$ along $S^2$. We proceed as before by decomposing $S^2\times S^1$ as two solid tori, with a basis of states of Chern-Simons theory for which $\exp\left(i\oint\aG\right)$ is diagonal, with eigenvalues $\exp(\pi i p/k)$, and $p=0,\dots,2k-1$. We now need to mod out by the $\Z_2$ factor that originated as the center of $\text{SU}(2)$. [We have already applied invariance under the $\Z_2$ Weyl group by using contribution of the zero modes in \eqref{eqn:WilsonsU1SU2h}.] Modding out by the center $\Z_2$ amounts to requiring invariance under large gauge transformations with holonomy $\Z_2$ either along the ``time'' $S^1$ or along the boundary of a hemisphere. The requirement of invariance under a large gauge transformation along the boundary of the hemisphere, i.e., the equator of $S^2$, is implemented by adding the $\tHmS=0$ and $\tHmS=1$ contributions.

We will implement the remaining required invariance under large gauge transformations along $S^1$ by noting that such a gauge transformation acts as $\ket{p}\rightarrow\ket{k+p}$, where $p$ is considered modulo $2k$. In the approach we follow, in the $\tHmS=0$ sector we need to calculate the expectation value of \eqref{eqn:U1aGCorrelator} in the state
\be\label{eqn:ketPsi}
\ket{\Psi}=\frac{1}{\sqrt{2k}}\sum_{p=0}^{2k-1}\ket{p},
\ee
and in the $\tHmS=1$ sector we need to insert a large gauge transformation along the equator of $S^2$, which acts as $\ket{p}\rightarrow(-1)^p\ket{p}$. The state \eqref{eqn:ketPsi} is invariant under the large gauge transformation $\ket{p}\rightarrow\ket{k+p}$, but not under $\ket{p}\rightarrow(-1)^p\ket{p}$.
The result of the contribution of the $\tHmS=0$ sector to the index is therefore
\bear
\nn
\text{I}_{k,\nC}^{\text{O}(2),\tHmS=0} &=&
2^{\nC}\left\langle\left(\cos 2\oint\aG\right)^\nC\right\rangle_{\text{U}(1)_{2k}}
-2^{\nC}\left\langle\left(\cos 2\oint\aG\right)^{\nC+1}\right\rangle_{\text{U}(1)_{2k}}
\nn\\
&=&
\frac{2^{\nC-1}}{k}\sum_{p=0}^{2k-1}\cos^\nC\left(\frac{2\pi p}{k}\right)
-\frac{2^{\nC-1}}{k}\sum_{p=0}^{2k-1}\cos^{\nC+1}\left(\frac{2\pi p}{k}\right),
\nn\\ \label{eqn:IndexKoddSector2v0}
&=&
\frac{2^{\nC}}{k}\sum_{p=0}^{k-1}\cos^\nC\left(\frac{2\pi p}{k}\right)
-\frac{2^{\nC}}{k}\sum_{p=0}^{k-1}\cos^{\nC+1}\left(\frac{2\pi p}{k}\right)
\eear
and for $\tHmS=1$ we get
\bear
\label{eqn:IndexKoddSector2v1}
\text{I}_{k,\nC}^{\text{O}(2),\tHmS=1} &=&
\frac{2^{\nC}}{k}\sum_{p=0}^{k-1}(-1)^p\cos^\nC\left(\frac{2\pi p}{k}\right)
-\frac{2^{\nC}}{k}\sum_{p=0}^{k-1}(-1)^p\cos^{\nC+1}\left(\frac{2\pi p}{k}\right).
\eear

If we add \eqref{eqn:IndexKoddSector2v0} and \eqref{eqn:IndexKoddSector2v1} we get a sum with $[1+(-1)^p]$ inserted, which projects onto even $p$ and does not reproduce the result for the index \eqref{eqn:IndexknX}. We therefore conclude that the $\tHmS=1$ must be added with a $(-)$ sign to the partition function, as we alluded to at the end of \secref{subsec:3dCS}. Now we get
\begin{align} \nn
    \text{I}_{k,\nC}^{\text{O}(2),\tHmS=0}-\text{I}_{k,\nC}^{\text{O}(2),\tHmS=1}&=
\frac{2^{\nC}}{k}\sum_{p=0}^{k-1}[1-(-1)^p]\cos^\nC\left(\frac{2\pi p}{k}\right)
-\frac{2^{\nC}}{k}\sum_{p=0}^{k-1}[1-(-1)^p]\cos^{\nC+1}\left(\frac{2\pi p}{k}\right)
\\ \nn
&=
\frac{2^{\nC+1}}{k}\sum_{p=1,3,\dots}^{k-1}\cos^\nC\left(\frac{2\pi p}{k}\right)
-\frac{2^{\nC+1}}{k}\sum_{p=1,3,\dots}^{k-1}\cos^{\nC+1}\left(\frac{2\pi p}{k}\right)
\\ \label{eqn:CSIndexEvenk}
&=
\frac{2^{\nC+2}}{k}\sum_{p=1,3,\dots}^{k-1}\cos^\nC\left(\frac{2\pi p}{k}\right)\sin^2\left(\frac{\pi p}{k}\right)
=\frac{1}{2}I_{k,\nC}.
\end{align}
Adding the contribution of the sector $\HO=\pm i\GGPauli_2$ from \eqref{eqn:IndexknXCS}, we get
\begin{align}
\label{eqn:MinusSign}
\text{I}_{k,\nC}^{\text{SU}(2)/\Z_2}+
\text{I}_{k,\nC}^{\text{O}(2),\tHmS=0}-\text{I}_{k,\nC}^{\text{O}(2),\tHmS=1}=I_{k,\nC}.
\end{align}
We do not have an independent argument for why the $\tHmS=1$ sector of $\text{O}(2)$ should have a $(-)$ sign, or alternatively, why the $(-)^F$ fermion number assignments in that sector, with $S^1$ taken as ``time'', should be flipped, but this seems to us to be the only plausible way to recover the results of the analytic torsion.

\section{SU$(2)/\Z_2$ Gauge Group}
\label{sec:SU2Z2}

Since the partition function of the $(2,0)$-theory is not unique \cite{Aharony:1998qu,Tachikawa:2013hya}, we can also consider a different reduction to 5d where the gauge group is taken to be $\text{SU}(2)/\Z_2$ on the analytic torsion side of the calculation, and on the Chern-Simons side it is taken to be $\text{SU}(2)$. 
On the analytic torsion side with gauge group $\text{SU}(2)/\Z_2$, the calculation in \secref{sec:HilbNoS} is modified in three ways:
\begin{enumerate}
    \item 
    For $m=\frac{k}{2}$, the order of the isotropy group, i.e., the factor $|H_A|$, is equal to $4$, as explained below; 
    \item 
    The representation $\NonAbRep{m}$ in \eqref{eqn:NonARepsDic} is equivalent to $\NonAbRep{k-m}$;
    \item 
    The binary parameters $\sF_1,\sF_2$ in \eqref{eqn:GenDicVF} may be nontrivial, but since the $\sF_1=1$ solutions \eqref{eqn:sF1} are reducible, only $\sF_1=0$ can contribute to the Witten index.
\end{enumerate}
We start with explaining the enhancement of the isotropy group. If $k$ is even for $m=k/2$ in \eqref{eqn:NonARepsDic}, the group $H_A$ of $\Lambda\in\text{SU}(2)/\Z_2$ for which $\Lambda^{-1}g_m(\gamma)\Lambda=\pm g_m(\gamma)$ for all elements $\gamma\in\text{Dic}_k$ is equivalent to $\Z_2\times \Z_2$ and is generated by $\pm i\GGPauli_1,\pm i\GGPauli_2$. Therefore, $|H_A|=4$ for $m=k/2$. For $1\le m<k/2$ the center is $\Z_2$ and is generated by $i\GGPauli_3$ so the isotropy group is as before.
For even $k$, we find instead of \eqref{eqn:IndexknX}:
\be\label{eqn:IndexknXSU2Z2Evenk}
\text{Index}_{k,\nC}^{\text{SU}(2)/\Z_2}=
\frac{2^{3+\nC}}{k}\sum_{m=0}^{\frac{k}{2}-1}\cos^\nC\left(\frac{2\pi m}{k}\right)
\sin^2\left(\frac{\pi m}{k}\right)
+\frac{2^{2+\nC}}{k}(-1)^\nC,\qquad\text{(even $k$).}
\ee
Here, we added to \eqref{eqn:IndexknXSU2Z2Evenk} the sum the even values of $m$, for $\sF_2=1$, and the rightmost term is the contribution of $m=k/2$. It can be checked that $\text{Index}_{k,\nC}^{\text{SU}(2)/Z_2}$ is indeed an integer, even though individual terms on the RHS are not. Since the summand in the first term on the RHS of \eqref{eqn:IndexknXSU2Z2Evenk} does not change if we replace $m$ with $(k-m)$, we can extend the range of the sum up to $m=k-1$, and divide by $2$. We then rewrite \eqref{eqn:IndexknXSU2Z2Evenk} as
\bear
\text{Index}_{k,\nC}^{\text{SU}(2)/\Z_2} &=&
\frac{2^{2+\nC}}{k}\sum_{m=1}^{k-1}\cos^\nC\left(\frac{2\pi m}{k}\right)
\sin^2\left(\frac{\pi m}{k}\right)
\nn\\
&=&
2\sum_{l=0}^\nC\binom{\nC}{l}\delta_{k}(\nC-2l)
-\sum_{l=0}^{\nC+1}\binom{\nC+1}{l}\delta_{k}(\nC+1-2l).
\label{eqn:IndexknXSU2Z2New}
\eear
where we dropped the $m=0$ and $m=k$ terms for which the summand vanishes.

For odd $k$, we have seen at the end of \secref{sec:geometry} that for nontrivial $\sF_1,\sF_2$ the representations are abelian (and hence do not contribute to the Witten index because of zero modes). Also, since $(k-m)$ is even for odd $m$, the representation $\NonAbRep{k-m}$ is not in $\text{SU}(2)$, and so we get the same result as for $\text{SU}(2)$, namely,
\be\label{eqn:IndexknXSU2Z2Oddk}
\text{Index}_{k,\nC}^{\text{SU}(2)/\Z_2}=
\frac{2^{3+\nC}}{k}\sum_{m=1,3,\dots}^{k-2}\cos^\nC\left(\frac{2\pi m}{k}\right)
\sin^2\left(\frac{\pi m}{k}\right),\qquad\text{(odd $k$).}
\ee
In the sum over $m$ above, the summand does not change when we replace $m$ with $(k-m)$, and so we can add the terms with $m=k-1,k-3,\dots,2$ to the sum and divide by $2$. Thus, we get the same formula for the index as \eqref{eqn:IndexknXSU2Z2New}.

How do we derive these results from Chern-Simons theory on $S^2$?
Since now the gauge group is $\text{SU}(2)$, there is no possibility of a `t Hooft magnetic flux, and so only $\HO\HO^\star=\Id$ solutions are allowed in \eqref{eqn:HOHOst}. The solutions $\HO=\pm i\GGPauli_2$ are now absent, and so is the contribution of $\text{SU}(2)_{k-2}$ Chern-Simons theory. Only $\text{U}(1)$ Chern-Simons theory remains. Calculating \eqref{eqn:U1aGCorrelator} explicitly, for $\aG$ a $\text{U}(1)_{2k}$ Chern-Simons gauge field, we get
\begin{align}
    \text{I}_{k,\nC}^{\text{U}(1)}=\frac{2^{\nC+2}}{2k}\sum_{m=0}^{2k-1}\cos^{\tilde{\nC}}\left(\frac{2m\pi}{k}\right) \sin^2\left(\frac{m\pi}{k}\right),
\end{align}
and noting that the summand is the same for $m$ and $(k+m)$, we can reduce the range to $m=0,\dots,k-1$ and we reproduce \eqref{eqn:IndexknXSU2Z2New}, which is another example of the duality introduced in \secref{sec:DerCS}. Below we list the value of the index for $k=1,2,\dots,6$ with $\nC$ surface operators inserted.

\subsubsection*{$\text{SU}(2)/\Z_2$ index for $\text{Dic}_2$ with $\nC$ strings}
$$
\text{Index}^{\text{SU}(2)/\Z_2}_{2,\nC}=(-1)^\nC2^{\nC+1}.
$$
\subsubsection*{$\text{SU}(2)/\Z_2$ index for $\text{Dic}_3$ with $\nC$ strings}
$$
\text{Index}^{\text{SU}(2)/\Z_2}_{3,\nC}=(-1)^\nC2.
$$

\subsubsection*{$\text{SU}(2)/\Z_2$ index for $\text{Dic}_4$ with $\nC$ strings}
$$
\text{Index}^{\text{SU}(2)/\Z_2}_{4,\nC}=
\left\{
\begin{array}{ll}
2 & \text{for $\nC=0$}, \\
(-1)^\nC2^\nC & \text{for $\nC>0$}. \\
\end{array}
\right.
$$

\subsubsection*{$\text{SU}(2)/\Z_2$ index for $\text{Dic}_5$ with $\nC$ strings}

\begin{align*}    
\text{Index}^{\text{SU}(2)/\Z_2}_{5,\nC}&=\frac{1}{5}\left(\left(5+\sqrt{5}\right)\left(\frac{1}{2}(-1-\sqrt{5})\right)^\nC-\left(-5+\sqrt{5}\right)\left(\frac{1}{2}(-1+\sqrt{5})\right)^\nC\right) \\
&=2,-2,4, -6, 10, -16,\dots.
\end{align*}

\subsubsection*{$\text{SU}(2)/\Z_2$ index for $\text{Dic}_6$ with $\nC$ strings}

$$
\text{Index}^{\text{SU}(2)/\Z_2}_{6,\nC}=\frac{1}{3}+(-1)^\nC+\frac{1}{3}(-1)^\nC2^{\nC+1}=2, -2, 4, -6, 12, -22, \ldots.
$$

\section{Conclusions and Further Directions}
\label{sec:conclusions}

In this paper, we built on the work of Bak and Gustavsson \cite{Bak:2017jwd}, combined with results of Chang and Dowker \cite{Dowker:2009in} on the Ray-Singer torsion of spherical orbifolds, to compute the Witten index of various systems obtained from the 6d (2,0)-theory associated to gauge group SU$(2)$. We compactified the 6d theory on 5-manifolds $X_5$ constructed from spherical orbifolds with a SUSY preserving twist, and with possible insertions of surface operators playing the role of external sources. A connection between the values of the Ray-Singer torsion and the characters of representations of the orbifold groups, in terms of which the Witten index is expressed, ensures that the result is always an integer.

Our main example is for the manifold $X_5 =S^3/\text{Dic}_k \times S^2$, where $\text{Dic}_k$ is the dicyclic group (of order $4k$), and in this case the results for the Witten index also allow us to postulate conjectures regarding fusion relations among external string-like sources, represented by timelike surface operators. We also discussed $S^5/\Z_k$ and $S^3/\Z_k\times S^2$, where the Witten index is zero, because of fermionic zero modes. Extensions to $S^5/\Gamma$ and $S^3/\Gamma \times S^2$ with other (non-abelian) finite orbifold groups $\Gamma$ are possible future developments of this work.

Our initial strategy, which follows \cite{Bak:2017jwd}, was to reduce the problem to a partition function of topologically twisted 5d SYM on $X_5$, containing gauge fields, bosonic and Grassmann odd 0-forms, 1-forms and 2-forms which participate in a 5d ``Fermionic Chern Simons'' term in the action. Supersymmetry ensures that the partition function localizes on gauge field configurations that are flat connections, and a study of fermionic zero modes shows that only irreducible non-abelian flat connections contribute. The simplest nontrivial case is then $X_5=S^3/\text{Dic}_k\times S^2$, since for $\Z_k$ quotient, all representations are abelian. Using expressions developed in \cite{Bak:2017jwd} for the one-loop contribution in terms of the Ray-Singer analytic torsion, we recast the partition function as a sum \eqref{eqn:PartTorsion} over non-abelian representations of Dic$_k$ torsions (associated with a flat connection) times products of characters of group elements. We simplified the torsion results of \cite{Dowker:2009in} for $S^3/\text{Dic}_k$ to a compact expression \eqref{eqn:DicTorsion2d}, and we used it to confirm that the Witten index is indeed given by an integer. 
Recurrence relations among the Witten indices for various numbers of timelike surface operator insertions lead to conjectural fusion rules for the external string sources associated with those operators. These relations are expressed as combinations of Chebyshev polynomials.

Moreover, we were able to rederive the same expressions for the Witten indices from a dual perspective that led to 3d Chern-Simons theory with two different sectors, labeled by an SU$(2)$ element $\HO$ that determines boundary conditions on fields, and gauge groups that depend on the parity of $k$. Thus, the results obtained in this work provide a physical explanation for the existence  of ``nice'' closed form expressions for the Ray-Singer analytic torsion for $S^3/\text{Dic}_k$, and they follow from the dimensions of Hilbert spaces of Chern-Simons theory on $S^2$ with quark sources, as computed in \cite{Witten:1988hf}. The fusion relations among surface operators then must follow from the basic fusion relations of Wilson loops in Chern-Simons theory, and can be expressed in terms of the $N_{ijk}$ fusion coefficients and Verlinde's formula \cite{Verlinde:1988sn,Witten:1988hf}. This connection between Chern-Simons theory and analytic torsion on $S^3/\text{Dic}_k$ has, as far as we know, not been reported before in the literature.

It is interesting to reflect on how the expression \eqref{eqn:IndexknX}, derived from analytic torsion in \secref{sec:WittenWithSources}, and the derivation of the same formula from Chern-Simons theory in \secref{sec:DerCS} are connected. The factor $\cos\left(\frac{2m\pi}{k}\right)$ in \eqref{eqn:IndexknX} arises from a certain character of $\text{Dic}_k$ in a representation labeled by $m$, while the same factor in \secref{sec:DerCS} can be traced back, via \cite{Witten:1988hf}, to an eigenvalue of a Wilson loop operator acting on ground states of $\text{SU}(2)$ Chern-Simons theory on $T^2$. We therefore see that the connection hinges on the observation by Vafa and Witten \cite{Vafa:1994tf} about the connection between the partition function of $\NS=4$ Super-Yang-Mills theory on $\R^4/\Gamma$ and characters of an affine Lie algebra associated to the Dynkin diagram of $\Gamma$, which in turn was obtained through earlier results of Nakajima \cite{Nakajima:1993,Nakajima:1994}. In particular, the number of ground states of Chern-Simons theory on $T^2$ is related to the number of representations of $\Gamma$. (See also a recent proof by Kojima and Tachikawa \cite{Kojima:2025hzr} of the equality of the number of embeddings of $\Gamma$ in $\text{SU}(n)$ and in its Langlands dual group $\text{SU}(n)/\Z_n$, as well as other Langlands dual pairs.)

In the process of matching the results that were obtained in \secref{sec:HilbNoS} via analytic torsion with the results that were obtained in \secref{sec:DerCS} Chern-Simons theory, we were led to assume that a certain sector of the low-energy $\text{U}(1)$ Chern-Simons theory with nontrivial `t Hooft magnetic flux should be counted with a $(-)$ sign in the total Witten index [see \eqref{eqn:MinusSign}]. We do not have an independent argument for this predicted effect. Within 5d SYM, there is no way to continuously deform the fields of the sector with `t Hooft magnetic flux $\tHmS=1$ on $S^2$ to the sector without it, and there is no way to continuously deform the fields of the $\tHm=0$ [for which the solutions of \eqref{eqn:HOHOstNew} lead to a $\text{U}(1)$ gauge group] to the fields of $\tHm=1$, and so the explanation for this effect seems to originate in 6d.

Looking ahead, our analysis may be naturally extended in several directions. In particular, one could study the smooth orbifolds $X_5=S^5/\Gamma$ for the non-abelian $\Gamma$'s that appear in \eqref{eqn:BobbyGroup}. Such an extension is expected to involve similar techniques, namely the computation of the Ray-Singer torsion in the background of flat connections associated to representations of $\Gamma$. Our results can also be examined in the context of 4d $\mathcal{N}=2$ $\ClassS$ theories \cite{Gaiotto:2009we, Gaiotto:2009hg}, by first compactifying on $S^2$ (with a topological twist), leading to a gapped 4d theory. In that case, the Witten index gives information about the nontrivial topological nature of this gapped system. 

We have worked with an SU$(2)$ gauge group, because for SU$(N)$ with $N\ge 3$ the fermionic 0-forms will always have zero modes. This is because there are no irreducible representations of $\text{Dic}_k$ of dimension higher than $2$, and therefore for any representation $\mathbf{R}$ of $\text{Dic}_k$ there are nontrivial elements of the Lie algebra $\mathfrak{su}(N)$ (in which the 0-form takes values) that are invariant under the adjoint action of $\text{Dic}_k$ (embedded in SU$(N)$ via $\mathbf{R}$). 

As another possible extension, one may consider more general surface operators \cite{Drukker:2020bes}. It would be interesting to find explicit operators that have nontrivial matrix elements between different ground states, and a natural candidate would be a surface operator wrapped on $S^2$. It would also be interesting to explore the effect of the braiding group on the Wilson loops of the effective 3d theory, although the structure might not be stable under coupling constant deformation, and it might not be visible in the dual picture with the analytic torsion. In general, such studies could reveal a richer structure in the spectrum of BPS states and their associated indices.

\acknowledgments
We wish to thank Sergey Cherkis, Vi Hong, Petr Ho\v{r}ava and Jesus Sanchez Jr. for helpful discussions. This work has been supported by the Leinweber Institute for Theoretical Physics at UC Berkeley.


\begin{appendix}
\section{Chern-Simons Level Shift from Integration of Massive Gluinos}
\label{app:LevelShift}

In \secref{subsec:3dCS} we noted that the level of the effective $\text{SU}(2)$ Chern-Simons theory that we obtained in 3d, for the case $\HO=\pm i\GGPauli_2$, is shifted from its classical value of $k$ to $(k-2)$ because of a mismatch between positive and negative fermionic Kaluza-Klein masses. We will show this explicitly below, and in the process, we will also show that there are no massless fermionic zero modes.

Luckily, for $\HO=\pm i\GGPauli_2$ we do not actually have to solve the full equations of motion obtained from the fermionic action \eqref{eqn:5dActionExp}. This is because in that case the boundary conditions \eqref{eqn:HOAphi}, i.e., $\HO^{-1}\AdjField(p)\HO=\pm\AdjField(\tilde{p})^t$ [where we wrote $p$ and $\antip$ for the antipodal points $(\frac{\pi}{2},\phi+\pi)$ and $(\frac{\pi}{2},\phi)$ in \eqref{eqn:HOAphi}] combined with the fact that $\HO^{-1}\Phi(p)\HO=-\Phi(p)^t$ for every $\mathfrak{su}(2)$ Lie-algebra valued matrix, implies that the SU$(2)$ indices of fields decouple from the geometry. There is only an overall $\pm$ sign in the boundary conditions:
\begin{align}
\label{eqn:bcwithplusNoHO}
\Xchi_{\Hchi i}(p) &= \Xchi_{\Hchi i}(\antip),
\quad
\Xchi_{\theta i}(p) = \Xchi_{\theta i}(\antip),
\\ \nn
\Xpsi_{\Hchi}(p) &= \Xpsi_{\Hchi}(\antip),
\quad
\Xpsi_{\theta}(p)= \Xpsi_{\theta}(\antip),
\qquad
(i=1,2,\Hphi),
\end{align}
and
\begin{align}\label{eqn:bcwithminusNoHO}
\Xchi_{ij}(p) &= -\Xchi_{ij}(\antip),
\quad
\Xchi_{\Hchi\theta}(p) = -\Xchi_{\Hchi\theta}(\antip), \\ \nn
\Xpsi_{i}(p) &= -\Xpsi_{i}(\antip),
\quad
\Xrho(p) = -\Xrho(\antip), 
\qquad
(i,j=1,2,\Hphi).
\end{align}

Thus, we can lift the problem back to the Dirac equation on $S^3/\Z_{2k}$, but with the extra condition that the modes are odd or even (depending on the $\mp$ sign) under the $\Z_2$ isometry that flips the orientation of the fiber. We are therefore back to the Dirac-K\"ahler equations \eqref{eqn:Dirac5DXrho}-\eqref{eqn:Dirac5DXchi} on $\R\times S^2\times S^3/\Z_{2k}$, with the extra conditions that the fields $\Xrho, \Xpsi_\mu,\Xchi_{\mu\nu}$ are
\begin{enumerate}
\item
invariant under translations in the direction of the fiber, i.e., along the Killing vector $\partial/\partial\Hchi$, and
\item
odd under the discrete isometry given by the action \eqref{eqn:AnglesIdentNew} of $s$ on $S^3$.
\end{enumerate}

We need to find the Kaluza-Klein modes of the reduction of such a twisted spinor from 6d to a low-energy 3d twisted spinor on $\R\times S^2$. Firstly, we note that a 3d spinor has two components, and if it is also a doublet of an SU$(2)$ R-symmetry, we can twist it by identifying the R-symmetry with the holonomy group of $S^2$, thus converting the spinor into two scalars and a vector along $S^2$. Note that in this formalism full covariance along $\R\times S^2$ is lost, and only covariance along the directions of $S^2$ is maintained. With $\xi,\xj$ labeling directions along $S^2$, after this twist, the Kaluza-Klein modes of $\Xrho, \varepsilon^{\xi\xj}\Xchi_{\xi\xj}$ and $\Xpsi_\xi$ form a multiplet. Similarly, with $\xi,\xj$ still labeling directions along $S^2$ and $\yl,\ym,\yn$ labeling directions along $S^3/\Z_{2k}$, we find that Kaluza-Klein modes of $\psi_\yl, {\varepsilon_\yl}^{\ym\yn}\Xchi_{\ym\yn}$ and $\Xchi_{\xi\yl}$ form a multiplet. This grouping into multiplets will become explicit in \eqref{eqn:Dirac5DXrhoM}-\eqref{eqn:Dirac5DXchiM3}.

To describe the Kaluza-Klein reduction more concretely, we look for modes of the fields that factorize as products of a (bosonic) field on $S^3/\Z_{2k}$ and a (fermionic) field on $\R\times S^2$. Let $\yCo^\ym$ be coordinates on $S^3/\Z_{2k}$ and let $\xCo^\xi$ be coordinates on $\R\times S^2$.
We find that modes with a given Kaluza-Klein mass $\lambda$ that factorize come in two species:
\begin{enumerate}
    \item 
    modes for which the nonzero fields are proportional to components of solutions of $\star d\psiF=\lambda\psiF$ on $S^3/\Z_{2k}$, where $\psiF$ is a 1-form, and
    \item 
    modes for which the nonzero fields are proportional to a solution $\Xrho$ of $d\star d\rho=-\lambda^2\star\rho$ on $S^3/\Z_{2k}$, or one of its derivatives $\partial_\ym\Xrho$, where $\rho$ is a 0-form.
\end{enumerate}
For both kinds of modes, we can first rewrite \eqref{eqn:Dirac5DXrho}-\eqref{eqn:Dirac5DXchi} by separating the derivatives along $\R\times S^2$ (i.e., with respect to $\xCo_0$ and $\xCo^\xi$) and along $S^3/\Z_{2k}$ (i.e., with respect to $\yCo^\ym$). For factorization, we assume that the parts of the equations with derivatives along $S^3/\Z_{2k}$ satisfy the following eigenvalue conditions:
\bear
\lambda\varepsilon^{\xi\xj}\Xchi_{\xi\xj} &=& \nabla^\ym\Xpsi_\ym
\,,\label{eqn:Dirac5DXrhoL}\\
\lambda\varepsilon_{\xi\xj}\Xpsi^\xj &=& 2\nabla^\ym\Xchi_{\ym\xi}
\,,\label{eqn:Dirac5DXpsiL1}\\
\lambda\varepsilon_{\ym\yn\yl}\Xchi^{\yn\yl} &=& -\partial_\ym\Xrho -2\nabla^\yl\Xchi_{\yl\ym}
\,,\label{eqn:Dirac5DXpsiL2}\\
\lambda\Xrho &=& 
-\varepsilon^{\yl\ym\yn}\partial_\yl\Xchi_{\ym\yn}
\,,\label{eqn:Dirac5DXchiL1}\\
\lambda\varepsilon_{\xi\xj}{\Xchi_\ym}{}^{\xj} &=& 
\varepsilon_{\xi\xj}\varepsilon_{\yl\ym\yn}\nabla^\yl\Xchi^{\yn\xj}
-\frac{1}{2}\partial_\ym\Xpsi_\xi
\,,\label{eqn:Dirac5DXchiL2}\\
\lambda\Xpsi^\yl &=& 
-\frac{1}{2}\varepsilon^{\xi\xj}\partial^\yl\Xchi_{\xi\xj}
-\varepsilon^{\yl\ym\yn}\partial_\ym\Xpsi_\yn
\,.\label{eqn:Dirac5DXchiL3}
\eear
Then, \eqref{eqn:Dirac5DXrho}-\eqref{eqn:Dirac5DXchi} can be rewritten along $\R\times S^2$ as the following massive twisted 3d Dirac equations:
\bear
\lambda\varepsilon^{\xi\xj}\Xchi_{\xi\xj} &=& -\partial_0\Xrho -\partial^\xi\Xpsi_\xi
\,,\label{eqn:Dirac5DXrhoM}\\
\lambda\varepsilon_{\xi\xj}\Xpsi^\xj &=& -\partial_0\Xpsi_\xi -2\partial^\xj\Xchi_{\xj\xi}
\,,\label{eqn:Dirac5DXpsiM1}\\
\lambda\varepsilon_{\ym\yn\yl}\Xchi^{\yn\yl} &=& -\partial_0\Xpsi_\ym-2\partial^\xi\Xchi_{\xi\ym}
\,,\label{eqn:Dirac5DXpsiM2}\\
\lambda\varepsilon_{\xi\xj}\Xrho &=& 
-\partial_0\Xchi_{\xi\xj}
-\frac{1}{2}\varepsilon^{\yl\ym\yn}\varepsilon_{\xk\xj}\partial^\xk\Xchi_{\ym\yn}
\,,\label{eqn:Dirac5DXchiM1}\\
\lambda\varepsilon_{\xi\xj}{\Xchi_\ym}{}^{\xj} &=& 
-\partial_0\Xchi_{\ym\xi}
+\varepsilon_{\xi\xk}\varepsilon_{\yl\ym\yn}\partial^\xk\Xchi^{\yn\yl}
+\frac{1}{2}\partial_\xi\Xpsi_\ym
\,,\label{eqn:Dirac5DXchiM2}\\
\lambda\Xpsi^\yl &=& 
-\varepsilon^{\yl\ym\yn}\partial_0\Xchi_{\ym\yn}
-\frac{1}{2}\varepsilon_{\xi\xk}\partial^\xk\Xchi^{\xi\yl}
\,.\label{eqn:Dirac5DXchiM3}
\eear
The 6 equations \eqref{eqn:Dirac5DXrhoL}-\eqref{eqn:Dirac5DXchiL3} are actually 3 decoupled pairs: \eqref{eqn:Dirac5DXrhoL} and \eqref{eqn:Dirac5DXchiL3} only involve $\Xchi_{\xi\xj}$ and $\Xpsi_\ym$, \eqref{eqn:Dirac5DXpsiL1} and \eqref{eqn:Dirac5DXchiL2} only involve $\Xchi_{\ym\xi}$ and $\Xpsi_\xi$, and \eqref{eqn:Dirac5DXpsiL2} and \eqref{eqn:Dirac5DXchiL1} only involve $\Xrho$ and $\Xchi_{\ym\yn}$.
We will begin by analyzing the solutions to the first pair, \eqref{eqn:Dirac5DXrhoL} and \eqref{eqn:Dirac5DXchiL3}.

For a fixed point on $S^2\times\R$ (labeled by $\xCo^\xi$), if we take $\Xpsi_\ym d\yCo^\ym$ to be proportional to an eigenform $\psiF$ of $\star d$, satisfying $\star d\psiF=-\lambda\psiF$, and $\lambda\neq 0$, then $\partial^\ym\Xpsi_\ym$ in \eqref{eqn:Dirac5DXrhoL} vanishes, since it is proportional to $d\star (\star d\psiF) = 0$. \eqref{eqn:Dirac5DXchiL3} will then immediately follow from $\star d\psiF=\lambda\psiF$. We can set all other fields to zero, and such mode will therefore have
$$
\Xrho=0,\quad
\Xchi_{\xi\xj}=0, \quad
\Xchi_{\ym\xi}=0,\quad
\Xchi_{\ym\yn}=0,\quad
\Xpsi_\xi=0,\quad
\Xpsi_\ym\sim\psiF_\ym.
$$
On the other hand, if $\lambda=0$, and $\star d\psiF=-\lambda\psiF=0$, then $\psiF$ is an exact 1-form, since $H^1(S^3/\Z_{2k},\R)=0$. Then \eqref{eqn:Dirac5DXrhoL} shows that $\psiF$ is also coexact, and hence vanishes. \eqref{eqn:Dirac5DXchiL3} shows that $\varepsilon^{\xi\xj}\partial^\yl\Xchi_{\xi\xj}=0$, and hence $\varepsilon^{\xi\xj}\Xchi_{\xi\xj}$ is a constant on $S^3/\Z_{2k}$ (at fixed $\xCo^\xi$). Thanks to the additional conditions \eqref{eqn:bcwithminusNoHO}, we find that $\varepsilon^{\xi\xj}\Xchi_{\xi\xj}=0$, and hence $\Xchi_{\xi\xj}$ must vanish as well. Thus there are no modes of this kind with $\lambda=0$, i.e., no zero modes, as argued in \secref{sec:physics}. 

Now assume that $\varepsilon^{\xi\xj}\Xchi_{\xi\xj}\neq 0$ and set $\EpChS=\frac{1}{2}\varepsilon^{\xi\xj}\Xchi_{\xi\xj}$. (We are still at a fixed point on $S^2\times\R$.) Then, \eqref{eqn:Dirac5DXchiL3} becomes:
\be\label{eqn:lpsiF}
-\lambda\psiF=d\EpChS+\star d\psiF,
\ee
and \eqref{eqn:Dirac5DXrhoL} becomes
\be\label{eqn:lEpChS}
\lambda\EpChS=\star d\star\psiF.
\ee
Plugging \eqref{eqn:lpsiF} into \eqref{eqn:lEpChS} we find $\lambda^2\EpChS=\star d\star d\EpChS$, and so $\EpChS$ is an eigenfunction of the Laplacian on $S^3/\Z_{2k}$ with eigenvalue $-\lambda^2$.
Because we also need to require $\EpChS$ to be invariant under translations along $\partial/\partial\Hchi$, i.e., along the Hopf fiber, we see that $\EpChS$ is the pullback to $S^3/\Z_{2k}$ of a spherical harmonic $Y_{lm}(\theta,\phi)$ on the base of the Hopf fibration, and therefore $\lambda=\pm\sqrt{l(l+1)}$. Moreover, by \eqref{eqn:bcwithminusNoHO} only odd $l$'s are allowed, because if $p$ is parameterized by $(\theta,\phi)$ then $\antip$ is parameterized by $(\pi-\theta,\phi+\pi)$, and $Y_{lm}(\pi-\theta,\phi+\pi)=(-1)^l Y_{lm}(\theta,\phi)$. Thus, there are no zero modes, and the $\lambda$'s come in pairs of positive and negative masses. From \eqref{eqn:lpsiF} we get $\lambda\star d\psiF=\star d\star d\psiF$, and plugging this back to the RHS of \eqref{eqn:lpsiF}, and using \eqref{eqn:lEpChS} for $\EpChS$ we find
$$
-\lambda^2\psiF=\lambda d\EpChS+\star d\star d\psiF = d\star d\star\psiF+\star d\star d\psiF
=\triangle_1\psiF,
$$
where $\triangle_1=d\star d\star+\star d\star d$.

The Hodge decomposition of $\psiF$ into exact and coexact 1-forms, which incidentally is given by \eqref{eqn:lpsiF} for $\lambda\neq 0$, separates $\psiF$ into exact and coexact eigenforms of $\triangle_1$, and applying $\star d$ to \eqref{eqn:lpsiF} we get $-\lambda(\star d\psiF)=\star d\star d\psiF$, which shows that $\star d\psiF$ is an eigenform of $\star d$ with eigenvalue $-\lambda$. But, as we will soon see below, the eigenvalues of $\star d$ are integers while $\pm\sqrt{l(l+1)}$ is never an integer for integer $l\ge 1$. Therefore, we must have $\star d\psiF=0$ or $\lambda=0$. If $\star d\psiF=0$ and $\lambda\neq 0$, then from \eqref{eqn:lpsiF} again we find $\psiF =-(1/\lambda)d\EpChS$ and so the solution to \eqref{eqn:Dirac5DXrhoL} and \eqref{eqn:Dirac5DXchiL3} with eigenvalue $\lambda=\pm\sqrt{l(l+1)}$ and $l\ge 1$ takes the form
\begin{align}
\Xrho&=0,\quad
\Xchi_{\xi\xj}=\epsilon_{\xi\xj}\EpChS, \quad
\Xchi_{\ym\xi}=0,\quad
\Xchi_{\ym\yn}=0,\quad
\Xpsi_\xi=0,
\nn\\
\Xpsi_\ym&=-\frac{1}{\lambda}\partial_\ym\EpChS,\quad
\EpChS\propto Y_{lm}(\theta,\phi),\quad
\lambda=\pm\sqrt{l(l+1)}
\end{align}
If $\lambda=0$ then \eqref{eqn:lpsiF} and \eqref{eqn:lEpChS} imply that $\EpChS$ is harmonic, and hence constant, and it follows that $\psiF$ is both exact and coexact, and hence vanishes. We also saw above that if $\EpChS$ is constant, it must vanish by \eqref{eqn:bcwithminusNoHO}.

We will now show that the eigenvalues $\lambda$ of $\lambda\psiF=\star d\psiF$ are odd integers.
We set
$$
\psiF=f(\theta,\phi)(d\chi+2k\sin^2\left(\tfrac{\theta}{2}\right)d\phi)
+\Wa_\theta(\theta,\phi)d\theta+\Wa_\phi(\theta,\phi)d\phi.
$$
which enforces invariance under translations along fiber $\partial/\partial\chi$.
Here $f$ is (the pullback to $S^3/\Z_{2k}$ of) a globally defined $0$-form on the base $S^2$ of the Hopf fibration $S^3/\Z_{2k}\rightarrow S^2$, and $\Wa=\Wa_\theta d\theta + \Wa_\phi d\phi$ is  (the pullback to $S^3/\Z_{2k}$ of) a globally defined $1$-form on the same $S^2$. We now need to solve
\be\label{eqn:dfWa}
\star df = \lambda\Wa,\qquad \star d\Wa = -(\lambda+1) f,
\ee
where the $\star$ in \eqref{eqn:dfWa} is the 2d Hodge dual (along the base $S^2$ of the Hopf fibration).
These imply $\star d\star df=-\lambda(\lambda+1)f$, whose solutions are spherical harmonics $Y_{lm}$ with $\lambda=l$ or $\lambda=-l-1$. The additional requirement of being even under $s$ is that $f(\pi-\theta,\phi+\pi)=f(\theta,\phi)$, which requires $l$ to be even. For $l>0$, the modes come in pairs of even and odd eigevalues $l>0$ and $-l-1<0$, and those add up to a net shift of $0$ to the low-energy effective Chern-Simons level. For $l=0$ we get a constant $f$ and $d\Wa$ is proportional to $\sin\theta d\theta\wedge d\phi$. But $\Wa$ has to be globally defined and $\sin\theta d\theta\wedge d\phi$ is not exact, so there is no solution for $\Wa$ with $l=0$ and the mode with $\lambda=-1$ is unpaired.

The solution of \eqref{eqn:Dirac5DXpsiL2} and \eqref{eqn:Dirac5DXchiL1} is similar to the solution we discussed above of \eqref{eqn:Dirac5DXrhoL} and \eqref{eqn:Dirac5DXchiL3}, with $\Xrho$ playing the role of $\varepsilon^{\xi\xj}\Xchi_{\xi\xj}$, and $\varepsilon_{\ym\yn\yl}\Xchi^{\yn\yl}$ playing the role of $\Xpsi_\ym$. The only remaining pair of equations to solve is therefore \eqref{eqn:Dirac5DXpsiL1} and \eqref{eqn:Dirac5DXchiL2}. This is again a similar pair of equations, with a vector index $\xi$ (on $S^2$ added), so that for $\xi$ fixed $\varepsilon_{\xi\xj}\Xpsi^\xj$ now plays the role of $\varepsilon^{\xk\xj}\Xchi_{\xk\xj}$, and $\Xchi_{\ym\xi}$ now plays the role of $\Xpsi_\ym$. Altogether, there is exactly one unpaired 3d fermion mass, leading to a shift of the Chern-Simons level to $k-2$ as claimed.

\end{appendix}

\bibliographystyle{JHEP}
\bibliography{references.bib}

\providecommand{\href}[2]{#2}\begingroup\raggedright\begin{thebibliography}{10}

\bibitem{Verlinde:1988sn}
E.P.~Verlinde, \emph{{Fusion Rules and Modular Transformations in 2D Conformal Field Theory}}, \href{https://doi.org/10.1016/0550-3213(88)90603-7}{\emph{Nucl. Phys. B} {\bfseries 300} (1988) 360}.

\bibitem{Witten:1988hf}
E.~Witten, \emph{{Quantum Field Theory and the Jones Polynomial}}, \href{https://doi.org/10.1007/BF01217730}{\emph{Commun. Math. Phys.} {\bfseries 121} (1989) 351}.

\bibitem{Witten:1995zh}
E.~Witten, \emph{{Some comments on string dynamics}},  in \emph{{STRINGS 95: Future Perspectives in String Theory}}, pp.~501--523, 7, 1995 [\href{https://arxiv.org/abs/hep-th/9507121}{{\ttfamily hep-th/9507121}}].

\bibitem{Strominger:1995ac}
A.~Strominger, \emph{{Open p-branes}}, \href{https://doi.org/10.1201/9781482268737-13}{\emph{Phys. Lett. B} {\bfseries 383} (1996) 44} [\href{https://arxiv.org/abs/hep-th/9512059}{{\ttfamily hep-th/9512059}}].

\bibitem{Moore2012}
G.~Moore, ``Applications of the six-dimensional (2,0) theories to physical mathematics.'' Felix Klein Lectures, Bonn, 2012.

\bibitem{Heckman:2018jxk}
J.J.~Heckman and T.~Rudelius, \emph{{Top Down Approach to 6D SCFTs}}, \href{https://doi.org/10.1088/1751-8121/aafc81}{\emph{J. Phys. A} {\bfseries 52} (2019) 093001} [\href{https://arxiv.org/abs/1805.06467}{{\ttfamily 1805.06467}}].

\bibitem{Aharony:1997th}
O.~Aharony, M.~Berkooz, S.~Kachru, N.~Seiberg and E.~Silverstein, \emph{{Matrix description of interacting theories in six-dimensions}}, \href{https://doi.org/10.4310/ATMP.1997.v1.n1.a5}{\emph{Adv. Theor. Math. Phys.} {\bfseries 1} (1998) 148} [\href{https://arxiv.org/abs/hep-th/9707079}{{\ttfamily hep-th/9707079}}].

\bibitem{Banks:1996vh}
T.~Banks, W.~Fischler, S.H.~Shenker and L.~Susskind, \emph{{M theory as a matrix model: A conjecture}}, \href{https://doi.org/10.1201/9781482268737-37}{\emph{Phys. Rev. D} {\bfseries 55} (1997) 5112} [\href{https://arxiv.org/abs/hep-th/9610043}{{\ttfamily hep-th/9610043}}].

\bibitem{Arkani-Hamed:2001wsh}
N.~Arkani-Hamed, A.G.~Cohen, D.B.~Kaplan, A.~Karch and L.~Motl, \emph{{Deconstructing (2,0) and little string theories}}, \href{https://doi.org/10.1088/1126-6708/2003/01/083}{\emph{JHEP} {\bfseries 01} (2003) 083} [\href{https://arxiv.org/abs/hep-th/0110146}{{\ttfamily hep-th/0110146}}].

\bibitem{Douglas:2010iu}
M.R.~Douglas, \emph{{On D=5 super Yang-Mills theory and (2,0) theory}}, \href{https://doi.org/10.1007/JHEP02(2011)011}{\emph{JHEP} {\bfseries 02} (2011) 011} [\href{https://arxiv.org/abs/1012.2880}{{\ttfamily 1012.2880}}].

\bibitem{Lambert:2010iw}
N.~Lambert, C.~Papageorgakis and M.~Schmidt-Sommerfeld, \emph{{M5-Branes, D4-Branes and Quantum 5D super-Yang-Mills}}, \href{https://doi.org/10.1007/JHEP01(2011)083}{\emph{JHEP} {\bfseries 01} (2011) 083} [\href{https://arxiv.org/abs/1012.2882}{{\ttfamily 1012.2882}}].

\bibitem{Lambert:2012qy}
N.~Lambert, C.~Papageorgakis and M.~Schmidt-Sommerfeld, \emph{{Deconstructing (2,0) Proposals}}, \href{https://doi.org/10.1103/PhysRevD.88.026007}{\emph{Phys. Rev. D} {\bfseries 88} (2013) 026007} [\href{https://arxiv.org/abs/1212.3337}{{\ttfamily 1212.3337}}].

\bibitem{Seiberg:1997ax}
N.~Seiberg, \emph{{Notes on theories with 16 supercharges}}, \href{https://doi.org/10.1016/S0920-5632(98)00128-5}{\emph{Nucl. Phys. B Proc. Suppl.} {\bfseries 67} (1998) 158} [\href{https://arxiv.org/abs/hep-th/9705117}{{\ttfamily hep-th/9705117}}].

\bibitem{Lambert:2010wm}
N.~Lambert and C.~Papageorgakis, \emph{{Nonabelian (2,0) Tensor Multiplets and 3-algebras}}, \href{https://doi.org/10.1007/JHEP08(2010)083}{\emph{JHEP} {\bfseries 08} (2010) 083} [\href{https://arxiv.org/abs/1007.2982}{{\ttfamily 1007.2982}}].

\bibitem{Bagger:2006sk}
J.~Bagger and N.~Lambert, \emph{{Modeling Multiple M2's}}, \href{https://doi.org/10.1103/PhysRevD.75.045020}{\emph{Phys. Rev. D} {\bfseries 75} (2007) 045020} [\href{https://arxiv.org/abs/hep-th/0611108}{{\ttfamily hep-th/0611108}}].

\bibitem{Gustavsson:2007vu}
A.~Gustavsson, \emph{{Algebraic structures on parallel M2-branes}}, \href{https://doi.org/10.1016/j.nuclphysb.2008.11.014}{\emph{Nucl. Phys. B} {\bfseries 811} (2009) 66} [\href{https://arxiv.org/abs/0709.1260}{{\ttfamily 0709.1260}}].

\bibitem{Saemann:2011nb}
C.~Saemann and M.~Wolf, \emph{{On Twistors and Conformal Field Theories from Six Dimensions}}, \href{https://doi.org/10.1063/1.4769410}{\emph{J. Math. Phys.} {\bfseries 54} (2013) 013507} [\href{https://arxiv.org/abs/1111.2539}{{\ttfamily 1111.2539}}].

\bibitem{Kallen:2012cs}
J.~K\"all\'en and M.~Zabzine, \emph{{Twisted supersymmetric 5D Yang-Mills theory and contact geometry}}, \href{https://doi.org/10.1007/JHEP05(2012)125}{\emph{JHEP} {\bfseries 05} (2012) 125} [\href{https://arxiv.org/abs/1202.1956}{{\ttfamily 1202.1956}}].

\bibitem{Kim:2012ava}
H.-C.~Kim and S.~Kim, \emph{{M5-branes from gauge theories on the 5-sphere}}, \href{https://doi.org/10.1007/JHEP05(2013)144}{\emph{JHEP} {\bfseries 05} (2013) 144} [\href{https://arxiv.org/abs/1206.6339}{{\ttfamily 1206.6339}}].

\bibitem{Kim:2013nva}
H.-C.~Kim, S.~Kim, S.-S.~Kim and K.~Lee, \emph{{The general M5-brane superconformal index}},  \href{https://arxiv.org/abs/1307.7660}{{\ttfamily 1307.7660}}.

\bibitem{Cheung:1998te}
Y.-K.E.~Cheung, O.J.~Ganor and M.~Krogh, \emph{{On the twisted (2,0) and little string theories}}, \href{https://doi.org/10.1016/S0550-3213(98)00583-5}{\emph{Nucl. Phys. B} {\bfseries 536} (1998) 175} [\href{https://arxiv.org/abs/hep-th/9805045}{{\ttfamily hep-th/9805045}}].

\bibitem{Cheung:1998wj}
Y.-K.E.~Cheung, O.J.~Ganor, M.~Krogh and A.Y.~Mikhailov, \emph{{Instantons on a noncommutative T**4 from twisted (2,0) and little string theories}}, \href{https://doi.org/10.1016/S0550-3213(99)00539-8}{\emph{Nucl. Phys. B} {\bfseries 564} (2000) 259} [\href{https://arxiv.org/abs/hep-th/9812172}{{\ttfamily hep-th/9812172}}].

\bibitem{Gaiotto:2009we}
D.~Gaiotto, \emph{{N=2 dualities}}, \href{https://doi.org/10.1007/JHEP08(2012)034}{\emph{JHEP} {\bfseries 08} (2012) 034} [\href{https://arxiv.org/abs/0904.2715}{{\ttfamily 0904.2715}}].

\bibitem{Gaiotto:2009hg}
D.~Gaiotto, G.W.~Moore and A.~Neitzke, \emph{{Wall-crossing, Hitchin systems, and the WKB approximation}}, \href{https://doi.org/10.1016/j.aim.2012.09.027}{\emph{Adv. Math.} {\bfseries 234} (2013) 239} [\href{https://arxiv.org/abs/0907.3987}{{\ttfamily 0907.3987}}].

\bibitem{Alday:2009aq}
L.F.~Alday, D.~Gaiotto and Y.~Tachikawa, \emph{{Liouville Correlation Functions from Four-dimensional Gauge Theories}}, \href{https://doi.org/10.1007/s11005-010-0369-5}{\emph{Lett. Math. Phys.} {\bfseries 91} (2010) 167} [\href{https://arxiv.org/abs/0906.3219}{{\ttfamily 0906.3219}}].

\bibitem{Nekrasov:2002qd}
N.A.~Nekrasov, \emph{{Seiberg-Witten prepotential from instanton counting}}, \href{https://doi.org/10.4310/ATMP.2003.v7.n5.a4}{\emph{Adv. Theor. Math. Phys.} {\bfseries 7} (2003) 831} [\href{https://arxiv.org/abs/hep-th/0206161}{{\ttfamily hep-th/0206161}}].

\bibitem{Dimofte:2011ju}
T.~Dimofte, D.~Gaiotto and S.~Gukov, \emph{{Gauge Theories Labelled by Three-Manifolds}}, \href{https://doi.org/10.1007/s00220-013-1863-2}{\emph{Commun. Math. Phys.} {\bfseries 325} (2014) 367} [\href{https://arxiv.org/abs/1108.4389}{{\ttfamily 1108.4389}}].

\bibitem{Gukov:2015sna}
S.~Gukov and D.~Pei, \emph{{Equivariant Verlinde formula from fivebranes and vortices}}, \href{https://doi.org/10.1007/s00220-017-2931-9}{\emph{Commun. Math. Phys.} {\bfseries 355} (2017) 1} [\href{https://arxiv.org/abs/1501.01310}{{\ttfamily 1501.01310}}].

\bibitem{Gadde:2009kb}
A.~Gadde, E.~Pomoni, L.~Rastelli and S.S.~Razamat, \emph{{S-duality and 2d Topological QFT}}, \href{https://doi.org/10.1007/JHEP03(2010)032}{\emph{JHEP} {\bfseries 03} (2010) 032} [\href{https://arxiv.org/abs/0910.2225}{{\ttfamily 0910.2225}}].

\bibitem{Gadde:2010te}
A.~Gadde, L.~Rastelli, S.S.~Razamat and W.~Yan, \emph{{The Superconformal Index of the $E_{6}$ SCFT}}, \href{https://doi.org/10.1007/JHEP08(2010)107}{\emph{JHEP} {\bfseries 08} (2010) 107} [\href{https://arxiv.org/abs/1003.4244}{{\ttfamily 1003.4244}}].

\bibitem{Gadde:2011ik}
A.~Gadde, L.~Rastelli, S.S.~Razamat and W.~Yan, \emph{{The 4d Superconformal Index from q-deformed 2d Yang-Mills}}, \href{https://doi.org/10.1103/PhysRevLett.106.241602}{\emph{Phys. Rev. Lett.} {\bfseries 106} (2011) 241602} [\href{https://arxiv.org/abs/1104.3850}{{\ttfamily 1104.3850}}].

\bibitem{Witten:1982df}
E.~Witten, \emph{{Constraints on Supersymmetry Breaking}}, \href{https://doi.org/10.1016/0550-3213(82)90071-2}{\emph{Nucl. Phys. B} {\bfseries 202} (1982) 253}.

\bibitem{Aharony:1998qu}
O.~Aharony and E.~Witten, \emph{{Anti-de Sitter space and the center of the gauge group}}, \href{https://doi.org/10.1088/1126-6708/1998/11/018}{\emph{JHEP} {\bfseries 11} (1998) 018} [\href{https://arxiv.org/abs/hep-th/9807205}{{\ttfamily hep-th/9807205}}].

\bibitem{Henningson:2010rc}
M.~Henningson, \emph{{The partition bundle of type A$_{N-1}$ (2, 0) theory}}, \href{https://doi.org/10.1007/JHEP04(2011)090}{\emph{JHEP} {\bfseries 04} (2011) 090} [\href{https://arxiv.org/abs/1012.4299}{{\ttfamily 1012.4299}}].

\bibitem{Tachikawa:2013hya}
Y.~Tachikawa, \emph{{On the 6d origin of discrete additional data of 4d gauge theories}}, \href{https://doi.org/10.1007/JHEP05(2014)020}{\emph{JHEP} {\bfseries 05} (2014) 020} [\href{https://arxiv.org/abs/1309.0697}{{\ttfamily 1309.0697}}].

\bibitem{Ganor:1996nf}
O.J.~Ganor, \emph{{Six-dimensional tensionless strings in the large N limit}}, \href{https://doi.org/10.1016/S0550-3213(96)00702-X}{\emph{Nucl. Phys. B} {\bfseries 489} (1997) 95} [\href{https://arxiv.org/abs/hep-th/9605201}{{\ttfamily hep-th/9605201}}].

\bibitem{Maldacena:1998im}
J.~Maldacena, \emph{{Wilson loops in large N field theories}}, \href{https://doi.org/10.1103/PhysRevLett.80.4859}{\emph{Phys. Rev. Lett.} {\bfseries 80} (1998) 4859} [\href{https://arxiv.org/abs/hep-th/9803002}{{\ttfamily hep-th/9803002}}].

\bibitem{Drukker:1999zq}
N.~Drukker, D.J.~Gross and H.~Ooguri, \emph{{Wilson loops and minimal surfaces}}, \href{https://doi.org/10.1103/PhysRevD.60.125006}{\emph{Phys. Rev. D} {\bfseries 60} (1999) 125006} [\href{https://arxiv.org/abs/hep-th/9904191}{{\ttfamily hep-th/9904191}}].

\bibitem{Witten:1994cg}
E.~Witten, \emph{{Monopoles and four manifolds}}, \href{https://doi.org/10.4310/MRL.1994.v1.n6.a13}{\emph{Math. Res. Lett.} {\bfseries 1} (1994) 769} [\href{https://arxiv.org/abs/hep-th/9411102}{{\ttfamily hep-th/9411102}}].

\bibitem{Vafa:1994tf}
C.~Vafa and E.~Witten, \emph{{A Strong coupling test of S duality}}, \href{https://doi.org/10.1016/0550-3213(94)90097-3}{\emph{Nucl. Phys. B} {\bfseries 431} (1994) 3} [\href{https://arxiv.org/abs/hep-th/9408074}{{\ttfamily hep-th/9408074}}].

\bibitem{Bershadsky:1995qy}
M.~Bershadsky, C.~Vafa and V.~Sadov, \emph{{D-branes and topological field theories}}, \href{https://doi.org/10.1016/0550-3213(96)00026-0}{\emph{Nucl. Phys. B} {\bfseries 463} (1996) 420} [\href{https://arxiv.org/abs/hep-th/9511222}{{\ttfamily hep-th/9511222}}].

\bibitem{Geyer:2002gy}
B.~Geyer and D.~Mulsch, \emph{{Higher dimensional analog of the Blau-Thompson model and N(T) = 8, D = 2 Hodge type cohomological gauge theories}}, \href{https://doi.org/10.1016/S0550-3213(03)00260-8}{\emph{Nucl. Phys. B} {\bfseries 662} (2003) 531} [\href{https://arxiv.org/abs/hep-th/0211061}{{\ttfamily hep-th/0211061}}].

\bibitem{Bak:2015hba}
D.~Bak and A.~Gustavsson, \emph{{The geometric Langlands twist in five and six dimensions}}, \href{https://doi.org/10.1007/JHEP07(2015)013}{\emph{JHEP} {\bfseries 07} (2015) 013} [\href{https://arxiv.org/abs/1504.00099}{{\ttfamily 1504.00099}}].

\bibitem{Bak:2017jwd}
D.~Bak and A.~Gustavsson, \emph{{Five-dimensional fermionic Chern-Simons theory}}, \href{https://doi.org/10.1007/JHEP02(2018)037}{\emph{JHEP} {\bfseries 02} (2018) 037} [\href{https://arxiv.org/abs/1710.02841}{{\ttfamily 1710.02841}}].

\bibitem{ray1970reidemeister}
D.~Ray, \emph{Reidemeister torsion and the laplacian on lens spaces}, {\emph{Advances in Mathematics} {\bfseries 4} (1970) 109}.

\bibitem{RAY1971145}
D.~Ray and I.~Singer, \emph{R-torsion and the laplacian on riemannian manifolds}, \href{https://doi.org/https://doi.org/10.1016/0001-8708(71)90045-4}{\emph{Advances in Mathematics} {\bfseries 7} (1971) 145}.

\bibitem{mckay1981graphs}
J.~McKay, \emph{Graphs, singularities, and finite groups},  in \emph{Proceedings of Symposia in Pure Mathematics}, pp.~183--186, American Mathematical Society, 1981.

\bibitem{wolf1972spaces}
J.A.~Wolf, \emph{Spaces of constant curvature}, vol.~372, American Mathematical Soc. (1972).

\bibitem{WIP}
E.~Albrychiewicz, A.~Franco~Valiente and O.J.~Ganor, ``{Work in Progress}.'' 2025.

\bibitem{GanorJu}
O.~Ganor and C.~Ju, ``{Action of S-duality on Ground States of $N=4$ Super-Yang-Mills on $S^3/\Gamma$ Orbifold (unpublished)}.'' 2023.

\bibitem{Ju:2023umb}
C.~Ju, \emph{{Chern-Simons theory, Ehrhart polynomials, and representation theory}}, \href{https://doi.org/10.1007/JHEP01(2024)052}{\emph{JHEP} {\bfseries 01} (2024) 052} [\href{https://arxiv.org/abs/2304.11830}{{\ttfamily 2304.11830}}].

\bibitem{Albrychiewicz:2024fkr}
E.~Albrychiewicz, A.~Franco~Valiente, O.J.~Ganor and C.~Ju, \emph{{Ground states of Class $ \mathcal{S} $ theory on ADE singularities and dual Chern-Simons theory}}, \href{https://doi.org/10.1007/JHEP10(2024)219}{\emph{JHEP} {\bfseries 10} (2024) 219} [\href{https://arxiv.org/abs/2404.12446}{{\ttfamily 2404.12446}}].

\bibitem{Wolf:1974}
J.A.~Wolf, \emph{{Spaces of constant curvature}}, AMS Chelsea Publishing (1974).

\bibitem{Acharya:1998db}
B.S.~Acharya, J.M.~Figueroa-O'Farrill, C.M.~Hull and B.J.~Spence, \emph{{Branes at conical singularities and holography}}, \href{https://doi.org/10.4310/ATMP.1998.v2.n6.a2}{\emph{Adv. Theor. Math. Phys.} {\bfseries 2} (1999) 1249} [\href{https://arxiv.org/abs/hep-th/9808014}{{\ttfamily hep-th/9808014}}].

\bibitem{Tsimpis:2022orc}
D.~Tsimpis, \emph{{Relative scale separation in orbifolds of S$^{2}$ and S$^{5}$}}, \href{https://doi.org/10.1007/JHEP03(2022)169}{\emph{JHEP} {\bfseries 03} (2022) 169} [\href{https://arxiv.org/abs/2201.10916}{{\ttfamily 2201.10916}}].

\bibitem{tHooft:1977nqb}
G.~'t~Hooft, \emph{{On the Phase Transition Towards Permanent Quark Confinement}}, \href{https://doi.org/10.1016/0550-3213(78)90153-0}{\emph{Nucl. Phys. B} {\bfseries 138} (1978) 1}.

\bibitem{Witten:1996hc}
E.~Witten, \emph{{Five-brane effective action in M theory}}, \href{https://doi.org/10.1016/S0393-0440(97)80160-X}{\emph{J. Geom. Phys.} {\bfseries 22} (1997) 103} [\href{https://arxiv.org/abs/hep-th/9610234}{{\ttfamily hep-th/9610234}}].

\bibitem{Freed:2006yc}
D.S.~Freed, G.W.~Moore and G.~Segal, \emph{{Heisenberg Groups and Noncommutative Fluxes}}, \href{https://doi.org/10.1016/j.aop.2006.07.014}{\emph{Annals Phys.} {\bfseries 322} (2007) 236} [\href{https://arxiv.org/abs/hep-th/0605200}{{\ttfamily hep-th/0605200}}].

\bibitem{Elliott:2020ecf}
C.~Elliott, P.~Safronov and B.R.~Williams, \emph{{A taxonomy of twists of supersymmetric Yang\textendash{}Mills theory}}, \href{https://doi.org/10.1007/s00029-022-00786-y}{\emph{Selecta Math.} {\bfseries 28} (2022) 73} [\href{https://arxiv.org/abs/2002.10517}{{\ttfamily 2002.10517}}].

\bibitem{Witten:1990bs}
E.~Witten, \emph{{Introduction to cohomological field theories}}, \href{https://doi.org/10.1142/S0217751X91001350}{\emph{Int. J. Mod. Phys. A} {\bfseries 6} (1991) 2775}.

\bibitem{Freed:1991wd}
D.S.~Freed and R.E.~Gompf, \emph{{Computer calculation of Witten's three manifold invariant}}, \href{https://doi.org/10.1007/BF02100006}{\emph{Commun. Math. Phys.} {\bfseries 141} (1991) 79}.

\bibitem{Rozansky:1993zx}
L.~Rozansky, \emph{{A Large k asymptotics of Witten's invariant of Seifert manifolds}}, \href{https://doi.org/10.1007/BF02099272}{\emph{Commun. Math. Phys.} {\bfseries 171} (1995) 279} [\href{https://arxiv.org/abs/hep-th/9303099}{{\ttfamily hep-th/9303099}}].

\bibitem{Adams:1995np}
D.H.~Adams and S.~Sen, \emph{{Phase and scaling properties of determinants arising in topological field theories}}, \href{https://doi.org/10.1016/0370-2693(95)00590-H}{\emph{Phys. Lett. B} {\bfseries 353} (1995) 495} [\href{https://arxiv.org/abs/hep-th/9506079}{{\ttfamily hep-th/9506079}}].

\bibitem{Adams:1996hi}
D.H.~Adams, \emph{{A Note on the Faddeev-Popov determinant and Chern-Simons perturbation theory}}, \href{https://doi.org/10.1023/A:1007442121759}{\emph{Lett. Math. Phys.} {\bfseries 42} (1997) 205} [\href{https://arxiv.org/abs/hep-th/9704159}{{\ttfamily hep-th/9704159}}].

\bibitem{Marino:2002fk}
M.~Marino, \emph{{Chern-Simons theory, matrix integrals, and perturbative three manifold invariants}}, \href{https://doi.org/10.1007/s00220-004-1194-4}{\emph{Commun. Math. Phys.} {\bfseries 253} (2004) 25} [\href{https://arxiv.org/abs/hep-th/0207096}{{\ttfamily hep-th/0207096}}].

\bibitem{Marino:2011nm}
M.~Marino, \emph{{Lectures on localization and matrix models in supersymmetric Chern-Simons-matter theories}}, \href{https://doi.org/10.1088/1751-8113/44/46/463001}{\emph{J. Phys. A} {\bfseries 44} (2011) 463001} [\href{https://arxiv.org/abs/1104.0783}{{\ttfamily 1104.0783}}].

\bibitem{Witten:1991mk}
E.~Witten, \emph{{The N matrix model and gauged WZW models}}, \href{https://doi.org/10.1016/0550-3213(92)90235-4}{\emph{Nucl. Phys. B} {\bfseries 371} (1992) 191}.

\bibitem{Kapustin:2009kz}
A.~Kapustin, B.~Willett and I.~Yaakov, \emph{{Exact Results for Wilson Loops in Superconformal Chern-Simons Theories with Matter}}, \href{https://doi.org/10.1007/JHEP03(2010)089}{\emph{JHEP} {\bfseries 03} (2010) 089} [\href{https://arxiv.org/abs/0909.4559}{{\ttfamily 0909.4559}}].

\bibitem{rosenberg1997laplacian}
S.~Rosenberg, \emph{The Laplacian on a Riemannian manifold: an introduction to analysis on manifolds}, no.~31, Cambridge University Press (1997).

\bibitem{Schwarz:1978cn}
A.S.~Schwarz, \emph{{The Partition Function of Degenerate Quadratic Functional and Ray-Singer Invariants}}, \href{https://doi.org/10.1007/BF00406412}{\emph{Lett. Math. Phys.} {\bfseries 2} (1978) 247}.

\bibitem{Witten:1991we}
E.~Witten, \emph{{On quantum gauge theories in two-dimensions}}, \href{https://doi.org/10.1007/BF02100009}{\emph{Commun. Math. Phys.} {\bfseries 141} (1991) 153}.

\bibitem{Friedmann:2002ty}
T.~Friedmann and E.~Witten, \emph{{Unification scale, proton decay, and manifolds of G(2) holonomy}}, \href{https://doi.org/10.4310/ATMP.2003.v7.n4.a1}{\emph{Adv. Theor. Math. Phys.} {\bfseries 7} (2003) 577} [\href{https://arxiv.org/abs/hep-th/0211269}{{\ttfamily hep-th/0211269}}].

\bibitem{Murthy:2025ioh}
S.~Murthy and E.~Witten, \emph{{Localization of strings on group manifolds}},  \href{https://arxiv.org/abs/2506.20028}{{\ttfamily 2506.20028}}.

\bibitem{Blau:2022odi}
M.~Blau, M.~Kakona and G.~Thompson, \emph{{Massive Ray-Singer torsion and path integrals}}, \href{https://doi.org/10.1007/JHEP08(2022)230}{\emph{JHEP} {\bfseries 08} (2022) 230} [\href{https://arxiv.org/abs/2206.12268}{{\ttfamily 2206.12268}}].

\bibitem{cheeger1979analytic}
J.~Cheeger, \emph{Analytic torsion and the heat equation}, {\emph{Annals of Mathematics} {\bfseries 109} (1979) 259}.

\bibitem{muller1978analytic}
W.~M{\"u}ller, \emph{Analytic torsion and r-torsion of riemannian manifolds}, {\emph{Advances in Mathematics} {\bfseries 28} (1978) 233}.

\bibitem{Nash:1992sf}
C.~Nash and D.J.~O'Connor, \emph{{Determinants of Laplacians, the Ray-Singer torsion on lens spaces and the Riemann zeta function}}, \href{https://doi.org/10.1063/1.531134}{\emph{J. Math. Phys.} {\bfseries 36} (1995) 1462} [\href{https://arxiv.org/abs/hep-th/9212022}{{\ttfamily hep-th/9212022}}].

\bibitem{Dowker:2009in}
J.S.~Dowker and P.~Chang, \emph{{Analytic torsion on spherical factors and tessellations}},  \href{https://arxiv.org/abs/0904.0744}{{\ttfamily 0904.0744}}.

\bibitem{ikeda1995spectrum}
A.~Ikeda, \emph{On the spectrum of homogeneous spherical space forms}, {\emph{Kodai Mathematical Journal} {\bfseries 18} (1995) 57}.

\bibitem{Rey:1998ik}
S.-J.~Rey and J.-T.~Yee, \emph{{Macroscopic strings as heavy quarks in large N gauge theory and anti-de Sitter supergravity}}, \href{https://doi.org/10.1007/s100520100799}{\emph{Eur. Phys. J. C} {\bfseries 22} (2001) 379} [\href{https://arxiv.org/abs/hep-th/9803001}{{\ttfamily hep-th/9803001}}].

\bibitem{abramowitz1968handbook}
M.~Abramowitz and I.A.~Stegun, \emph{Handbook of mathematical functions with formulas, graphs, and mathematical tables}, vol.~55, US Government printing office (1968).

\bibitem{Dimofte:2011py}
T.~Dimofte, D.~Gaiotto and S.~Gukov, \emph{{3-Manifolds and 3d Indices}}, \href{https://doi.org/10.4310/ATMP.2013.v17.n5.a3}{\emph{Adv. Theor. Math. Phys.} {\bfseries 17} (2013) 975} [\href{https://arxiv.org/abs/1112.5179}{{\ttfamily 1112.5179}}].

\bibitem{Gadde:2013sca}
A.~Gadde, S.~Gukov and P.~Putrov, \emph{{Fivebranes and 4-manifolds}}, \href{https://doi.org/10.1007/978-3-319-43648-7_7}{\emph{Prog. Math.} {\bfseries 319} (2016) 155} [\href{https://arxiv.org/abs/1306.4320}{{\ttfamily 1306.4320}}].

\bibitem{Chung:2014qpa}
H.-J.~Chung, T.~Dimofte, S.~Gukov and P.~Su\l{}kowski, \emph{{3d-3d Correspondence Revisited}}, \href{https://doi.org/10.1007/JHEP04(2016)140}{\emph{JHEP} {\bfseries 04} (2016) 140} [\href{https://arxiv.org/abs/1405.3663}{{\ttfamily 1405.3663}}].

\bibitem{Gukov:2017kmk}
S.~Gukov, D.~Pei, P.~Putrov and C.~Vafa, \emph{{BPS spectra and 3-manifold invariants}}, \href{https://doi.org/10.1142/S0218216520400039}{\emph{J. Knot Theor. Ramifications} {\bfseries 29} (2020) 2040003} [\href{https://arxiv.org/abs/1701.06567}{{\ttfamily 1701.06567}}].

\bibitem{Douglas:1995bn}
M.R.~Douglas, \emph{{Branes within branes}}, {\emph{NATO Sci. Ser. C} {\bfseries 520} (1999) 267} [\href{https://arxiv.org/abs/hep-th/9512077}{{\ttfamily hep-th/9512077}}].

\bibitem{Vafa:1995bm}
C.~Vafa, \emph{{Instantons on D-branes}}, \href{https://doi.org/10.1016/0550-3213(96)00075-2}{\emph{Nucl. Phys. B} {\bfseries 463} (1996) 435} [\href{https://arxiv.org/abs/hep-th/9512078}{{\ttfamily hep-th/9512078}}].

\bibitem{Seiberg:1996bd}
N.~Seiberg, \emph{{Five-dimensional SUSY field theories, nontrivial fixed points and string dynamics}}, \href{https://doi.org/10.1016/S0370-2693(96)01215-4}{\emph{Phys. Lett. B} {\bfseries 388} (1996) 753} [\href{https://arxiv.org/abs/hep-th/9608111}{{\ttfamily hep-th/9608111}}].

\bibitem{Berkooz:1997cq}
M.~Berkooz, M.~Rozali and N.~Seiberg, \emph{{Matrix description of M theory on T**4 and T**5}}, \href{https://doi.org/10.1016/S0370-2693(97)00800-9}{\emph{Phys. Lett. B} {\bfseries 408} (1997) 105} [\href{https://arxiv.org/abs/hep-th/9704089}{{\ttfamily hep-th/9704089}}].

\bibitem{Niemi:1983rq}
A.J.~Niemi and G.W.~Semenoff, \emph{{Axial Anomaly Induced Fermion Fractionization and Effective Gauge Theory Actions in Odd Dimensional Space-Times}}, \href{https://doi.org/10.1103/PhysRevLett.51.2077}{\emph{Phys. Rev. Lett.} {\bfseries 51} (1983) 2077}.

\bibitem{Redlich:1983kn}
A.N.~Redlich, \emph{{Gauge Noninvariance and Parity Violation of Three-Dimensional Fermions}}, \href{https://doi.org/10.1103/PhysRevLett.52.18}{\emph{Phys. Rev. Lett.} {\bfseries 52} (1984) 18}.

\bibitem{Redlich:1983dv}
A.N.~Redlich, \emph{{Parity Violation and Gauge Noninvariance of the Effective Gauge Field Action in Three-Dimensions}}, \href{https://doi.org/10.1103/PhysRevD.29.2366}{\emph{Phys. Rev. D} {\bfseries 29} (1984) 2366}.

\bibitem{Kao:1995gf}
H.-C.~Kao, K.-M.~Lee and T.~Lee, \emph{{The Chern-Simons coefficient in supersymmetric Yang-Mills Chern-Simons theories}}, \href{https://doi.org/10.1016/0370-2693(96)00119-0}{\emph{Phys. Lett. B} {\bfseries 373} (1996) 94} [\href{https://arxiv.org/abs/hep-th/9506170}{{\ttfamily hep-th/9506170}}].

\bibitem{Tong:2000ky}
D.~Tong, \emph{{Dynamics of N=2 supersymmetric Chern-Simons theories}}, \href{https://doi.org/10.1088/1126-6708/2000/07/019}{\emph{JHEP} {\bfseries 07} (2000) 019} [\href{https://arxiv.org/abs/hep-th/0005186}{{\ttfamily hep-th/0005186}}].

\bibitem{Nakajima:1993}
H.~Nakajima, ``{Instantons on ALE spaces, quiver varieties, and Kac-Moody Algebras}.'' 1993.

\bibitem{Nakajima:1994}
H.~Nakajima, \emph{{Gauge Theory on Resolutions of Simple Singularities and Simple Lie Algebras}}, {\emph{{Int. Math. Res. Notices}} {\bfseries {}} (1994) 61}.

\bibitem{Kojima:2025hzr}
Y.~Kojima and Y.~Tachikawa, \emph{{On homomorphisms from finite subgroups of $SU(2)$ to Langlands dual pairs of groups}},  \href{https://arxiv.org/abs/2505.01253}{{\ttfamily 2505.01253}}.

\bibitem{Drukker:2020bes}
N.~Drukker and M.~Tr\'epanier, \emph{{Observations on BPS observables in 6d}}, \href{https://doi.org/10.1088/1751-8121/abf38d}{\emph{J. Phys. A} {\bfseries 54} (2021) 20} [\href{https://arxiv.org/abs/2012.11087}{{\ttfamily 2012.11087}}].

\end{thebibliography}\endgroup

\end{document}